%File information: C:\TEXTS\VARELA08\VARELA2.TEX 8    1400   31/03/108   00:10
%--------------------------------------------------------------
%LABEL:vp
    \input amstexl
    % LAMSTEX.TEX   VERSION 2.01
% COPYRIGHT (C) 1989, 1990, 1991 BY THE TEXPLORATORS CORPORATION
%  3701 W. ALABAMA, SUITE 450-273, HOUSTON, TX 77027
% ALL RIGHTS RESERVED

% ABSOLUTELY NO CHANGES SHOULD BE MADE TO THIS FILE;
% CHANGES SHOULD BE MADE ONLY IN STYLE FILES.

\catcode`\@=11
\ifx\amstexloaded@\relax\else
 \errmessage{AmS-TeX must be loaded before LamS-TeX}\fi
\ifx\laxread@\undefined\else\catcode`\@=\active \fi
\def\err@#1{\errmessage{LamS-TeX error: #1}}
\def^^L{\par}
\let\+\tabalign
\def\newcount{\alloc@0\count\countdef\insc@unt}
\def\newdimen{\alloc@1\dimen\dimendef\insc@unt}
\def\newskip{\alloc@2\skip\skipdef\insc@unt}
\def\newmuskip{\alloc@3\muskip\muskipdef\@cclvi}
\def\newbox{\alloc@4\box\chardef\insc@unt}
\let\newtoks\relax
\def\newhelp#1#2{\newtoks#1#1\expandafter{\csname#2\endcsname}}
\def\newtoks{\alloc@5\toks\toksdef\@cclvi}
\def\newread{\alloc@6\read\chardef\sixt@@n}
\def\newwrite{\alloc@7\write\chardef\sixt@@n}
\def\newfam{\alloc@8\fam\chardef\sixt@@n}
\def\newlanguage{\alloc@9\language\chardef\@cclvi}
\def\newinsert#1{\global\advance\insc@unt by\m@ne
  \ch@ck0\insc@unt\count
  \ch@ck1\insc@unt\dimen
  \ch@ck2\insc@unt\skip
  \ch@ck4\insc@unt\box
  \allocationnumber=\insc@unt
  \global\chardef#1=\allocationnumber
  \wlog{\string#1=\string\insert\the\allocationnumber}}
\def\newif#1{\count@\escapechar \escapechar\m@ne
  \expandafter\expandafter\expandafter
   \edef\@if#1{true}{\let\noexpand#1=\noexpand\iftrue}%
  \expandafter\expandafter\expandafter
   \edef\@if#1{false}{\let\noexpand#1=\noexpand\iffalse}%
  \@if#1{false}\escapechar\count@}

\def\Err@#1{\errhelp\defaulthelp@\err@{#1}}
{\catcode`\@=\active
 \edef\next{\gdef\noexpand@{\futurelet\noexpand\next
  \csname at\string@\endcsname}}
 \next
}
\def\at@{\ifcat\noexpand\next a\let\next@\at@@\else
 \ifcat\noexpand\next0\let\next@\at@@\else
 \ifcat\noexpand\next\relax\let\next@\at@@\else
 \let\next@\at@@@\fi\fi\fi\next@}
\def\at@@@{\errhelp\athelp@\err@{Invalid use of @}}
\def\at@@#1{\expandafter
 \ifx\csname\string#1@at\endcsname\relax\let\next@\at@@@\else
 \DN@{\csname\string#1@at\endcsname}\fi\next@}
\def\atdef@#1{\expandafter\def\csname\string#1@at\endcsname}
\newif\iftest@
\def\tagin@#1{\tagin@false
 \DN@##1\tag##2##3\next@{\test@true\ifx\tagin@##2\test@false\fi}%
 \next@#1\tag\tagin@\next@\tagin@false\iftest@\tagin@true\fi}
\let\lkerns@\relax
\def\nolinebreak{\RIfM@\mathmodeerr@\nolinebreak\else
 \ifhmode\saveskip@\lastskip\unskip
 \nobreak\ifdim\saveskip@>\z@\hskip\saveskip@\fi\lkerns@
 \else\vmodeerr@\nolinebreak\fi\fi}
\def\allowlinebreak{\RIfM@\mathmodeerr@\allowlinebreak\else
 \ifhmode\saveskip@\lastskip\unskip
 \allowbreak\ifdim\saveskip@>\z@\hskip\saveskip@\fi\lkerns@
 \else\vmodeerr@\allowlinebreak\fi\fi}
\def\linebreak{\RIfM@\mathmodeerr@\linebreak\else
 \ifhmode\unskip\unkern\break\lkerns@
 \else\vmodeerr@\linebreak\fi\fi}
\let\nkerns@\relax
\def\newline{\RIfM@\mathmodeerr@\newline\else
 \ifhmode\unskip\unkern\null\hfill\break\nkerns@
 \else\vmodeerr@\newline\fi\fi}%
\def\newbox@{\alloc@@4\box\chardef\insc@unt}
\def\newcount@{\alloc@@0\count\countdef\insc@unt}
\def\accentedsymbol#1#2{\expandafter\newbox@\csname\exstring@#1@box\endcsname
 \setbox\csname\exstring@#1@box\endcsname\hbox{$\m@th#2$}%
 \define#1{\copy\csname\exstring@#1@box\endcsname{}}}
\def\rightadd@#1\to#2{\toks@{\\#1}\toks@@\expandafter{#2}\xdef#2{\the\toks@@
 \the\toks@}\toks@{}\toks@@{}}
\def\fontlist@{\\\tenrm\\\sevenrm\\\fiverm\\\teni\\\seveni\\\fivei
 \\\tensy\\\sevensy\\\fivesy\\\tenex\\\tenbf\\\sevenbf\\\fivebf
 \\\tensl\\\tenit}
\def\font@#1=#2 {\rightadd@#1\to\fontlist@\font#1=#2 }
\def\ismember@#1#2{\global\let\Next@ F\let\next@= #2%
 {\def\\##1{\let\nextii@##1\ifx\nextii@\next@\global\let\Next@ T\fi}#1}%
 \test@false\ifx\Next@ T\test@true\fi\let\next@\relax}
\def\FNSS@#1{\let\FNSS@@#1\FN@\FNSS@@@}
\def\FNSS@@@{\ifx\next\space@\def\FNSS@@@@. {\FN@\FNSS@@@}\else
 \def\FNSS@@@@.{\FNSS@@}\fi\FNSS@@@@.}
\atdef@"{\unskip
 \DN@{\ifx\next`\DN@`{\FN@\nextii@}%
  \else\ifx\next\lq\DN@\lq{\FN@\nextii@}%
  \else\DN@####1{\FN@\nextiii@}\fi\fi
  \next@}%
 \DNii@{\ifx\next`\DN@`{\sldl@``}%
  \else\ifx\next\lq\DN@\lq{\sldl@``}%
  \else\DN@{\dlsl@`}\fi\fi\next@}%
 \def\nextiii@{\ifx\next'\DN@'{\srdr@''}%
  \else\ifx\next\rq\DN@\rq{\srdr@''}%
  \else\DN@{\drsr@'}\fi\fi\next@}%
 \FNSS@\next@}
\def\root{%
  \DN@{\ifx\next\uproot\let\next@\nextii@\else
   \ifx\next\leftroot\let\next@\nextiii@\else
   \let\next@\plainroot@\fi\fi\next@}%
  \DNii@\uproot##1{\uproot@##1\relax\FNSS@\nextiv@}%
  \def\nextiv@{\ifx\next\leftroot\let\next@\nextv@\else
   \let\next@\plainroot@\fi\next@}%
  \def\nextv@\leftroot##1{\leftroot@##1\relax\plainroot@}%
  \def\nextiii@\leftroot##1{\leftroot@##1\relax\FNSS@\nextvi@}%
  \def\nextvi@{\ifx\next\uproot\let\next@\nextvii@\else
   \let\next@\plainroot@\fi\next@}%
  \def\nextvii@\uproot##1{\uproot@##1\relax\plainroot@}%
  \bgroup\uproot@\z@\leftroot@\z@
 \FNSS@\next@}
\def\loop#1\repeat{\def\iterate{#1\relax\expandafter\iterate\fi}%
 \iterate\let\iterate\relax}
\def\gloop@#1\repeat{\gdef\iterate@{#1\relax\expandafter\iterate@\fi}%
 \iterate@\global\let\iterate@\relax}
\def\printoptions{\W@{Do you want S(yntax check),
  G(alleys) or P(ages)?^^JType S, G or P, follow by <return>: }\loop
 \read\m@ne to\ans@
 \edef\next@{\def\noexpand\Ans@{\ans@}}\uppercase\expandafter{\next@}%
 \ifx\Ans@\S@\test@true\syntax\else
 \ifx\Ans@\G@\test@true\galleys\else
 \ifx\Ans@\P@\test@true\else
 \test@false\fi\fi\fi
 \iftest@\else\W@{Type S, G or P, follow by <return>: }%
 \repeat}
\expandafter\let\csname A@;\endcsname;
\expandafter\let\csname A@:\endcsname:
\expandafter\let\csname A@?\endcsname?
\expandafter\let\csname A@!\endcsname!
\def\APdef#1{\def\next@{\expandafter\let\csname A@\string#1\endcsname#1}%
 \afterassignment\next@\def#1}
\let\fextra@\,
\def\tdots@{\unskip
 \DN@{$\m@th\mathinner{\ldotp\ldotp\ldotp}\,
   \ifx\next,\,$\else\ifx\next.\,$\else
   \ifx\next;\,$\else
   \expandafter\ifx\csname A@\string;\endcsname\next\fextra@$\else
   \ifx\next:\,$\else
   \expandafter\ifx\csname A@\string:\endcsname\next\fextra@$\else
   \ifx\next?\,$\else
   \expandafter\ifx\csname A@\string?\endcsname\next\fextra@$\else
   \ifx\next!\,$\else
   \expandafter\ifx\csname A@\string!\endcsname\next\fextra@$\else
   $ \fi\fi\fi\fi\fi\fi\fi\fi\fi\fi}%
 \ \FN@\next@}
\def\extrap@#1{%
 \ifx\next,\DN@{#1\,}\else
 \ifx\next;\DN@{#1\,}\else
 \expandafter\ifx\csname A@\string;\endcsname\next\DN@{#1\fextra@}\else
 \ifx\next.\DN@{#1\,}\else\extra@
 \ifextra@\DN@{#1\,}\else
 \let\next@#1\fi\fi\fi\fi\fi\next@}
\def\dotsc{\DN@{\ifx\next;\plainldots@\,\else
 \expandafter\ifx\csname A@\string;\endcsname\next\plainldots@\fextra@\else
 \ifx\next.\plainldots@\,\else\extra@\plainldots@
 \ifextra@\,\fi\fi\fi\fi}%
 \FN@\next@}
\def\keybin@{\keybin@true
 \ifx\next+\else\ifx\next=\else\ifx\next<\else\ifx\next>\else\ifx\next-\else
 \ifx\next*\else\ifx\next:\else
 \expandafter\ifx\csname A@\string;\endcsname\next\else
 \keybin@false\fi\fi\fi\fi\fi\fi\fi\fi}
\def\boldkey#1{\ifcat\noexpand#1A%
  \ifcmmibloaded@{\fam\cmmibfam#1}\else
   \Err@{First bold symbol font not loaded}\fi
 \else
 \let\next=#1%
 \ifx#1!\mathchar"5\bffam@21 \else
 \expandafter\ifx\csname A@\string!\endcsname\next\mathchar"5\bffam@21 \else
 \ifx#1(\mathchar"4\bffam@28 \else\ifx#1)\mathchar"5\bffam@29 \else
 \ifx#1+\mathchar"2\bffam@2B \else\ifx#1:\mathchar"3\bffam@3A \else
 \expandafter\ifx\csname A@\string:\endcsname\next\mathchar"3\bffam@3A \else
 \ifx#1;\mathchar"6\bffam@3B \else
 \expandafter\ifx\csname A@\string;\endcsname\next\mathchar"6\bffam@3B \else
 \ifx#1=\mathchar"3\bffam@3D \else
 \ifx#1?\mathchar"5\bffam@3F \else
 \expandafter\ifx\csname A@\string?\endcsname\next\mathchar"5\bffam@3F \else
 \ifx#1[\mathchar"4\bffam@5B \else
 \ifx#1]\mathchar"5\bffam@5D \else
 \ifx#1,\mathchari@63B \else
 \ifx#1-\mathcharii@200 \else
 \ifx#1.\mathchari@03A \else
 \ifx#1/\mathchari@03D \else
 \ifx#1<\mathchari@33C \else
 \ifx#1>\mathchari@33E \else
 \ifx#1*\mathcharii@203 \else
 \ifx#1|\mathcharii@06A \else
 \ifx#10\bold0\else\ifx#11\bold1\else\ifx#12\bold2\else\ifx#13\bold3\else
 \ifx#14\bold4\else\ifx#15\bold5\else\ifx#16\bold6\else\ifx#17\bold7\else
 \ifx#18\bold8\else\ifx#19\bold9\else
  \Err@{\noexpand\boldkey can't be used with #1}%
 \fi\fi\fi\fi\fi\fi\fi\fi\fi\fi\fi\fi\fi\fi\fi
 \fi\fi\fi\fi\fi\fi\fi\fi\fi\fi\fi\fi\fi\fi\fi\fi\fi\fi}
\def\arabic#1{#1}
\def\alph#1{\count@#1\relax\advance\count@96 \ifnum\count@>122
 \Err@{\noexpand\alph invalid for numbers > 26}\else\char\count@\fi}
\def\Alph#1{\count@#1\relax\advance\count@64 \ifnum\count@>90
 \Err@{\noexpand\Alph invalid for numbers > 26}\else\char\count@\fi}

\def\Roman#1{\uppercase\expandafter{\romannumeral#1}}
\def\fnsymbol#1{\count@#1\relax
 \count@@\count@
 \advance\count@\m@ne\divide\count@7
 \count@@@\count@\advance\count@@@\@ne
 \multiply\count@7 \advance\count@@-\count@
 \count@\count@@@
 {\loop
  \ifcase\count@@\or*\or\dag\or\ddag\or\P\or\S\or\text{$\|$}\or\#\fi
  \advance\count@\m@ne\ifnum\count@>\z@\repeat}}
\def\cardnine@#1{\ifcase#1\or one\or two\or three\or four\or five\or
 six\or seven\or eight\or nine\fi}
\let\alloc@\alloc@@
\newcount\ten@
\ten@10
\def\cardinal#1{\count@#1\relax
 \ifnum\count@>99 \number\count@
 \else
  \ifnum\count@=\z@ zero%
  \else
   \ifnum\count@<\ten@\cardnine@\count@
   \else
    \ifnum\count@<20
     \advance\count@-\ten@
     \ifcase\count@ ten\or eleven\or twelve\or thirteen\or fourteen\or
      fifteen\or sixteen\or seventeen\or eighteen\or nineteen\fi
    \else
     \count@@\count@\count@@@\count@@
     \divide\count@\ten@\multiply\count@\ten@
     \advance\count@@@-\count@\divide\count@\ten@
     \ifcase\count@\or\or twenty\or thirty\or forty\or fifty\or sixty\or
      seventy\or eighty\or ninety\fi
     \ifnum\count@@@=\z@\else-\cardnine@\count@@@\fi
    \fi
   \fi
  \fi
 \fi}
\def\ordnine@#1{\ifcase#1\or first\or second\or third\or fourth\or fifth\or
 sixth\or seventh\or eighth\or ninth\fi}
\newcount\count@@@@
\def\ordsuffix@{\count@@@@\count@
 \divide\count@\ten@
 \count@@@\count@\count@@\count@
 \divide\count@@\ten@\multiply\count@@\ten@
 \advance\count@@@-\count@@
 \ifnum\count@@@=\@ne th%
 \else
  \count@@@\count@@@@
  \count@@\count@@@@
  \divide\count@@\ten@\multiply\count@@\ten@
  \advance\count@@@-\count@@
  \ifcase\count@@@ th\or st\or nd\or rd\else th\fi
 \fi}
\def\nordinal#1{\count@#1\relax\number\count@\ordsuffix@}
\def\spordinal#1{\count@#1\relax\number\count@$^{\text{\ordsuffix@}}$}
\def\ordinal#1{\count@#1\relax
 \ifnum\count@>99 \number\count@\ordsuffix@
 \else
   \ifnum\count@=\z@ zeroth%
  \else
    \ifnum\count@<\ten@\ordnine@\count@
    \else
     \ifnum\count@<20 \advance\count@-\ten@
      \ifcase\count@ tenth\or eleventh\or twelfth\or thirteenth\or
       fourteenth\or fifteenth\or sixteenth\or seventeenth\or eighteenth\or
       nineteenth\fi
     \else
      \count@@\count@
      \divide\count@\ten@\multiply\count@\ten@
      \count@@@\count@@\advance\count@@@-\count@
      \divide\count@\ten@
      \ifcase\count@\or\or twent\or thirt\or fort\or fift\or sixt\or sevent\or
       eight\or ninet\fi
      \ifnum\count@@@=\z@ ieth\else y-\ordnine@\count@@@\fi
     \fi
    \fi
  \fi
 \fi}
\font@\tensmc=cmcsc10
\textonlyfont@\smc\tensmc
\newtoks\noexpandtoks@
\noexpandtoks@{\let\arabic\relax\let\alph\relax\let\Alph\relax
 \let\Roman\relax\let\fnsymbol\relax\let\rm\relax
 \let\it\relax\let\bf\relax\let\sl\relax\let\smc\relax
 \let\/\relax\let\null\relax}
\def\noexpands@{\the\noexpandtoks@}
\def\Nonexpanding#1{\global\noexpandtoks@
 \expandafter{\the\noexpandtoks@\let#1\relax}}
\def\prevanish@{\saveskip@\z@\ifhmode\saveskip@\lastskip\unskip\fi}
\def\postvanish@{\ifdim\saveskip@>\z@\hskip\saveskip@\fi\FN@\postvanish@@}
\def\postvanish@@{\DN@.{}%
 \ifx\next\space@\ifdim\saveskip@>\z@\DN@. {}\fi\fi\next@.}
\def\invisible#1{\prevanish@\ignorespaces#1\unskip\postvanish@}
\def\vanishlist@{\\\invisible}
\let\noindent@\noindent
\def\noindent{\par\noindent@\FN@\pretendspace@}
\def\pretendspace@{\ismember@\vanishlist@\next
 \iftest@\nobreak\hskip-\p@\hskip\p@\fi}
\let\flushpar\noindent
\newtoks\everypartoks@
\def\noindent@@{\par\everypartoks@\expandafter{\the\everypar}\everypar{}%
 \noindent@\everypar\expandafter{\the\everypartoks@}}
\def\page{\Err@{\noexpand\page has no meaning by itself}}
\let\page@C\pageno
\let\page@P\empty
\let\page@Q\empty
\def\page@S#1{#1\/}
\def\page@F{\rm}
\def\page@N{\arabic}   % cannot be \let
\newif\ifindexing@
\def\indexfile{\ifindexing@\else
 \alloc@@7\write\chardef\sixt@@n\ndx@
 \immediate\openout\ndx@=\jobname.ndx
 \global\indexing@true\fi}
\global\advance\insc@unt\m@ne
\ch@ck0\insc@unt\count
\ch@ck1\insc@unt\dimen
\ch@ck2\insc@unt\skip
\ch@ck4\insc@unt\box
\allocationnumber\insc@unt
\global\chardef\margin@\allocationnumber
\dimen\margin@\maxdimen
\count\margin@\z@
\skip\margin@\z@
\newif\ifindexproofing@
\def\indexproofing{\indexproofing@true}
\def\noindexproofing{\indexproofing@false}
\def\unmacro@#1:#2->#3\unmacro@{\def\macpar@{#2}\def\macdef@{#3}}
\def\starparts@#1{\def\stari@{#1}\def\starii@{#1}\let\stariii@\empty
 \test@false
 \DN@##1*##2##3\next@{\ifx\starparts@##2\test@false\else\test@true\fi}%
 \next@#1*\starparts@\next@
 \iftest@\DN@{\starparts@@#1\starparts@@}\else\let\next@\relax\fi\next@}
\def\starparts@@#1*#2\starparts@@{\def\starii@{#1}\def\stariii@{*#2}}
\def\windex@{\ifindexing@
 \expandafter\unmacro@\meaning\stari@\unmacro@
 \edef\macdef@{\string"\macdef@\string"}%
 \edef\next@{\write\ndx@{\macdef@}}\next@
 \write\ndx@{{\number\pageno}{\page@N}{\page@P}{\page@Q}}%
 \fi
 \ifindexproofing@
  \ifx\stariii@\empty\else
   \expandafter\unmacro@\meaning\stariii@\unmacro@\fi
  \insert\margin@{\hbox{\rm\vrule\height9\p@\depth2\p@\width\z@\starii@
  \ifx\stariii@\empty\else\tt\macdef@\fi}}\fi}
\catcode`\"=\active
\def"{\FN@\quote@}
\def\quote@{\ifx\next"\expandafter\quote@@\else\expandafter\quote@@@\fi}
\def\quote@@@#1"{\starparts@{#1}\starii@\windex@}
\def\quote@@"#1"{\prevanish@\starparts@{#1}\windex@\FN@\quote@@@@}
\def\quote@@@@{\ifx\next"\DN@"{\postvanish@}\else
 \let\next@\postvanish@\fi\next@}
\rightadd@"\to\vanishlist@
\def\idefine#1{\DN@{#1}\DNii@{\noexpand#1}%
 \afterassignment\idefine@\def\nextiii@}
\def\idefine@{\ifindexing@
 \expandafter\let\next@\nextiii@
 \expandafter\unmacro@\meaning\nextiii@\unmacro@
 \immediate\write\ndx@{\noexpand\define\nextii@\macpar@{\macdef@}}\fi}
\def\iabbrev*#1#2{\ifindexing@\toks@{#2}%
 \immediate\write\ndx@{\noexpand\abbrev*\noexpand#1{\the\toks@}}\fi}
\newread\laxread@
\newwrite\laxwrite@
\let\fnpages@\empty
\def\Finit@#1#2\Finit@{\let\nextii@#1\def\nextiii@{#2}}
\catcode`\~=11
\def\getparts@ @#1~#2~#3~#4~#5~#6{\def\nextiv@{#1}%
 \def\nextiii@{#2~#3~#4~#5~}\count@#6\relax}
\newif\ifdocument@
\def\document{\ifdocument@\else\global\document@true
 \let\fontlist@\empty
 \immediate\openin\laxread@=\jobname.lax\relax
 {\endlinechar\m@ne\noexpands@\catcode`\@=11 \catcode`\~=11
  \loop\ifeof\laxread@\else
   \read\laxread@ to\next@
   \ifx\next@\empty
   \else
    \expandafter\Finit@\next@\Finit@
    \if\nextii@ F%
     \expandafter\rightadd@\nextiii@\to\fnpages@
    \else
     \expandafter\getparts@\next@
     \edef\next@{\gdef\csname\nextiv@ @L\endcsname{\nextiii@\number\count@}}%
     \next@
    \fi
   \fi
  \repeat}%
 \immediate\closein\laxread@
 \immediate\openout\laxwrite@=\jobname.lax\relax\fi}
\let\thelabel@\relax
\def\thelabels@{\thelabel@ ~\thelabel@@ ~\thelabel@@@ ~\thelabel@@@@ ~}
\def\label#1{\prevanish@
 \ifx\thelabel@\relax
  \Err@{There's nothing here to be labelled}%
 \else
  {\noexpands@
  \expandafter\ifx\csname#1@L\endcsname\relax
   \expandafter\xdef\csname#1@L\endcsname{\thelabels@0}%
   \immediate\write\laxwrite@{@#1~\thelabels@1}%
  \else
   \edef\next@{@~\csname#1@L\endcsname}%
    \expandafter\getparts@\next@
    \ifodd\count@
    \expandafter\xdef\csname#1@L\endcsname{\thelabels@0}%
    \immediate\write\laxwrite@{@#1~\thelabels@1}%
   \else
    \Err@{Label #1 already used}%
   \fi
  \fi
  }%
 \fi
 \postvanish@}
\rightadd@\label\to\vanishlist@
\def\thepages@{\page@N{\number\page@C}~%
 \page@S{\page@P\page@N{\number\page@C}\page@Q}~%
 \number\page@C ~\page@P\page@N{\number\page@C}\page@Q ~}
\def\pagelabel#1{\prevanish@
 \expandafter\ifx\csname#1@L\endcsname\relax
  {\noexpands@
  \expandafter\xdef\csname#1@L\endcsname{\thepages@2}}%
  \write\laxwrite@{@#1~\thepages@3}%
 \else
  {\noexpands@
  \edef\next@{@~\csname#1@L\endcsname}%
  \expandafter\getparts@\next@
  \ifodd\count@
   \ifnum\count@=\@ne
    \expandafter\xdef\csname#1@L\endcsname{\thelabels@2}%
   \fi
   \write\laxwrite@{@#1~\thepages@3}%
  \else
   \Err@{Label #1 already used}%
  \fi
  }%
 \fi
 \postvanish@}
\rightadd@\pagelabel\to\vanishlist@
\newif\ifreferr@
\referr@true
\def\RefErrors{\global\referr@true}
\def\RefWarnings{\global\referr@false}
\setbox\z@\hbox{\global\count@=`^^30}
\ifnum\count@=48 \let\versionthree@\relax\fi
\def\nolabel@#1#2#3{\expandafter\ifx\csname#2@L\endcsname\relax
 \ifreferr@\Err@{No \noexpand\label found for #2}\else
 \W@{Warning: No \noexpand\label found for #2.}%
 \ifx\versionthree@\relax\W@{l.\number\inputlineno\space ... \string#1{#2}}\fi
 \fi#3\else}
\def\csL@#1{{\noexpands@\xdef\Next@{\csname#1@L\endcsname}}}
\def\ref#1{\nolabel@\ref{#1}\relax
 \DNii@##1~##2\nextii@{##1}%
 \csL@{#1}\expandafter\nextii@\Next@\nextii@\fi}
\def\Ref#1{\nolabel@\Ref{#1}\relax
 \DNii@##1~##2~##3\nextii@{##2}%
 \csL@{#1}\expandafter\nextii@\Next@\nextii@\fi}
\def\nref#1{\nolabel@\nref{#1}\relax
 \DNii@##1~##2~##3~##4\nextii@{##3}%
 \csL@{#1}\expandafter\nextii@\Next@\nextii@\fi}
\def\pref#1{\nolabel@\pref{#1}\relax
 \DNii@##1~##2~##3~##4~##5\nextii@{##4}%
 \csL@{#1}\expandafter\nextii@\Next@\nextii@\fi}
\let\pref@\pref
\def\Evaluatenref#1{\nolabel@\Evaluatenref{#1}{\gdef\Nref{-10000 }}%
 \DNii@##1~##2~##3~##4\nextii@{\DNii@{##3}}%
 \csL@{#1}\expandafter\nextii@\Next@\nextii@
 \xdef\Nref{\nextii@}\fi}
\def\Evaluatepref#1{\nolabel@\Evaluatepref{#1}{\global\let\Pref\empty}%
 \DNii@##1~##2~##3~##4~##5\nextii@{\DNii@{##4}}%
 \csL@{#1}\expandafter\nextii@\Next@\nextii@
 \xdef\Pref{\nextii@}\fi}
\def\readlax#1{\immediate\openin\laxread@=#1.lax\relax
 \ifeof\laxread@\W@{}\W@{File #1.lax not found.}\W@{}\fi
 {\endlinechar\m@ne\noexpands@\catcode`\@=11 \catcode`\~=11
  \loop\ifeof\laxread@\else
   \read\laxread@ to\nextv@
   \ifx\nextv@\empty
   \else
    \expandafter\Finit@\nextv@\Finit@
    \ifx\nextii@ F%
    \else
     \expandafter\getparts@\nextv@
     \expandafter\ifx\csname\nextiv@ @L\endcsname\relax
      \edef\next@{\gdef\csname\nextiv@ @L\endcsname
       {\nextiii@\ifnum\count@=\@ne0\else2\fi}}%
      \next@
     \else
      \Err@{Label \nextiv@\space in #1.lax already used}%
     \fi
    \fi
   \fi
  \repeat}%
 \immediate\closein\laxread@}
\catcode`\~=\active

\def\FNSSP@{\FNSS@\pretendspace@}
\everydisplay{\csname displaymath \endcsname}
\expandafter\def\csname displaymath \endcsname#1$${#1$$\FNSSP@}
\def\locallabel@{\let\thelabel@\Thelabel@\let\thelabel@@\Thelabel@@
 \let\thelabel@@@\Thelabel@@@\let\thelabel@@@@\Thelabel@@@@}
\newcount\tag@C
\tag@C\z@
\let\tag@P\empty
\let\tag@Q\empty
\def\tag@S#1{{\rm(}{#1\/}{\rm)}}
\let\tag@N\arabic
\def\tag@F{\rm}
\def\maketag@{\FN@\maketag@@}
\def\maketag@@{\ifx\next\relax\DN@\relax{\FN@\maketag@@}\else
 \ifx\next"\let\next@\maketag@@@\else
 \let\next@\maketag@@@@\fi\fi\next@}
\def\xdefThelabel@#1{\xdef\Thelabel@{#1{\Thelabel@@@}}}
\def\xdefThelabel@@#1{\xdef\Thelabel@@{#1{\Thelabel@@@@}}}
\def\maketag@@@@#1\maketag@{\global\advance\tag@C\@ne
 {\noexpands@
  \xdef\Thelabel@@@{\number\tag@C}%
  \xdefThelabel@\tag@N
  \xdef\Thelabel@@@@{\ifmathtags@$\tag@P\Thelabel@\tag@Q$\else
   \tag@P\Thelabel@\tag@Q\fi}%
  \xdefThelabel@@\tag@S
  }%
 \locallabel@
 \hbox{\tag@F\thelabel@@}%
 #1}
\def\Qlabel@#1{{\noexpands@\xdef\Thelabel@@{#1}%
 \let\style\empty\xdef\Thelabel@@@@{#1}%
 \let\pre\empty\let\post\empty\xdef\Thelabel@{#1}%
 \let\numstyle\empty\xdef\Thelabel@@@{#1}}}
\def\maketag@@@"#1"#2\maketag@{%
 {\let\pre\tag@P\let\post\tag@Q\let\style\tag@S\let\numstyle\tag@N
  \hbox{\tag@F#1}%
  \noexpands@
  \Qlabel@{#1}%
  }%
 \locallabel@
 #2}
\def\align@{\inalign@true\inany@true
 \vspace@\allowdisplaybreak@\displaybreak@\intertext@
 \def\tag{\global\tag@true\ifnum\and@=\z@
  \DN@{&\omit\global\rwidth@\z@&\relax}\else
  \DN@{&\relax}\fi\next@}%
 \iftagsleft@\DN@{\csname align \endcsname}\else
  \DN@{\csname align \space\endcsname}\fi\next@}
\def\noset@{\def\Offset##1##2{\prevanish@\postvanish@}%
 \def\Reset##1##2{\prevanish@\postvanish@}}
\def\measure@#1\endalign{\global\lwidth@\z@\global\rwidth@\z@
 \global\maxlwidth@\z@\global\maxrwidth@\z@
 \global\and@\z@
 \setbox\z@\vbox
  {\noset@\everycr{\noalign{\global\tag@false\global\and@\z@}}\Let@
  \halign{\setboxz@h{$\m@th\displaystyle{\@lign##}$}%
   \global\lwidth@\wdz@
   \ifdim\lwidth@>\maxlwidth@\global\maxlwidth@\lwidth@\fi
   \global\advance\and@\@ne
   &\setboxz@h{$\m@th\displaystyle{{}\@lign##}$}\global\rwidth@\wdz@
   \ifdim\rwidth@>\maxrwidth@\global\maxrwidth@\rwidth@\fi
   \global\advance\and@\@ne
   &\Tag@\eat@{##}\crcr#1\crcr}}%
 \totwidth@\maxlwidth@\advance\totwidth@\maxrwidth@}
\def\prepost@{\global\let\tag@P@\tag@P\global\let\tag@Q@\tag@Q}
\def\reprepost@{\let\tag@P\tag@P@\let\tag@Q\tag@Q@}
\expandafter\def\csname align \space\endcsname#1\endalign
 {\measure@#1\endalign\global\and@\z@
 \ifingather@\everycr{\noalign{\global\and@\z@}}\else\displ@y@\fi
 \Let@\tabskip\centering@
 \halign to\displaywidth
  {\hfil\strut@\setboxz@h{$\m@th\displaystyle{\@lign##\prepost@}$}%
  \boxz@\global\advance\and@\@ne
  \tabskip\z@skip
  &\setboxz@h{$\m@th\displaystyle{{}\@lign##\prepost@}$}%
  \global\rwidth@\wdz@\boxz@\hfil\global\advance\and@\@ne
  \tabskip\centering@
  &\setboxz@h{\@lign\strut@\reprepost@\maketag@##\maketag@}%
  \dimen@\displaywidth\advance\dimen@-\totwidth@
  \divide\dimen@\tw@\advance\dimen@\maxrwidth@\advance\dimen@-\rwidth@
  \ifdim\dimen@<\tw@\wdz@\llap{\vtop{\normalbaselines\null\boxz@}}%
  \else\llap{\boxz@}\fi
  \tabskip\z@skip
  \crcr#1\crcr
  \black@\totwidth@}}
\expandafter\def\csname align \endcsname#1\endalign{\measure@#1\endalign
 \global\and@\z@
 \ifdim\totwidth@>\displaywidth\let\displaywidth@\totwidth@\else
  \let\displaywidth@\displaywidth\fi
 \ifingather@\everycr{\noalign{\global\and@\z@}}\else\displ@y@\fi
 \Let@\tabskip\centering@\halign to\displaywidth
  {\hfil\strut@\setboxz@h{$\m@th\displaystyle{\@lign##\prepost@}$}%
  \global\lwidth@\wdz@\global\lineht@\ht\z@
  \boxz@\global\advance\and@\@ne
  \tabskip\z@skip&\setboxz@h{$\m@th\displaystyle{{}\@lign##\prepost@}$}%
  \ifdim\ht\z@>\lineht@\global\lineht@\ht\z@\fi
  \boxz@\hfil\global\advance\and@\@ne
  \tabskip\centering@&\kern-\displaywidth@
  \setboxz@h{\@lign\strut@\reprepost@\maketag@##\maketag@}%
  \dimen@\displaywidth\advance\dimen@-\totwidth@
  \divide\dimen@\tw@\advance\dimen@\maxlwidth@\advance\dimen@-\lwidth@
  \ifdim\dimen@<\tw@\wdz@
   \rlap{\vbox{\normalbaselines\boxz@\vbox to\lineht@{}}}\else
   \rlap{\boxz@}\fi
  \tabskip\displaywidth@\crcr#1\crcr\black@\totwidth@}}
\def\attag@#1{\let\Maketag@\maketag@\let\TAG@\Tag@
 \let\Prepost@\prepost@\let\Reprepost@\reprepost@
 \let\Tag@\relax\let\maketag@\relax
 \let\prepost@\relax\let\reprepost@\relax
 \ifmeasuring@
  \def\llap@##1{\setboxz@h{##1}\hbox to\tw@\wdz@{}}%
  \def\rlap@##1{\setboxz@h{##1}\hbox to\tw@\wdz@{}}%
 \else\let\llap@\llap\let\rlap@\rlap\fi
 \toks@{\hfil\strut@
  $\m@th\displaystyle{\@lign\the\hashtoks@\prepost@}$%
  \tabskip\z@skip\global\advance\and@\@ne&
  $\m@th\displaystyle{{}\@lign\the\hashtoks@\prepost@}$\hfil
  \ifxat@\tabskip\centering@\fi\global\advance\and@\@ne}%
 \iftagsleft@
  \toks@@{\tabskip\centering@&\Tag@\kern-\displaywidth
   \rlap@{\@lign\reprepost@\maketag@\the\hashtoks@\maketag@}%
   \global\advance\and@\@ne\tabskip\displaywidth}\else
  \toks@@{\tabskip\centering@&\Tag@\llap@{\@lign\reprepost@\maketag@
   \the\hashtoks@\maketag@}\global\advance\and@\@ne\tabskip\z@skip}\fi
 \atcount@#1\relax\advance\atcount@\m@ne
 \loop\ifnum\atcount@>\z@
  \toks@\expandafter{\the\toks@&\hfil$\m@th\displaystyle{\@lign
  \the\hashtoks@\prepost@}$\global\advance\and@\@ne
  \tabskip\z@skip
  &$\m@th\displaystyle{{}\@lign\the\hashtoks@\prepost@}$\hfil\ifxat@
  \tabskip\centering@\fi\global\advance\and@\@ne}\advance\atcount@\m@ne
 \repeat
 \edef\preamble@{\the\toks@\the\toks@@}%
 \edef\preamble@@{\preamble@}%
 \let\maketag@\Maketag@\let\Tag@\TAG@
 \let\prepost@\Prepost@\let\reprepost@\Reprepost@}
\def\unlabel@{\def\label##1{\prevanish@\postvanish@}%
 \def\pagelabel##1{\prevanish@\postvanish@}}
\newcount\tag@CC
\expandafter\def\csname alignat \endcsname#1#2\endalignat
 {\inany@true\xat@false
 \def\tag{\global\tag@true
  \count@#1\relax\multiply\count@\tw@\advance\count@\m@ne
  \gdef\tag@{&}%
  \loop\ifnum\count@>\and@\xdef\tag@{&\omit\tag@}%
  \advance\count@\m@ne\repeat
  \tag@\relax}%
 \vspace@\allowdisplaybreak@\displaybreak@\intertext@
 \displ@y@\measuring@true\tag@CC\tag@C
 \setbox\savealignat@\hbox{\noset@\unlabel@$\m@th\displaystyle\Let@
  \attag@{#1}\vbox{\halign{\span\preamble@@\crcr#2\crcr}}$}%
 \measuring@false
 \Let@\attag@{#1}\tag@C\tag@CC
 \tabskip\centering@\halign to\displaywidth
  {\span\preamble@@\crcr#2\crcr\black@{\wd\savealignat@}}}
\expandafter\def\csname xalignat \endcsname#1#2\endxalignat
 {\inany@true\xat@true
 \def\tag{\global\tag@true
  \count@#1\relax\multiply\count@\tw@\advance\count@\m@ne
  \gdef\tag@{&}%
  \loop\ifnum\count@>\and@\xdef\tag@{&\omit\tag@}%
  \advance\count@\m@ne\repeat
  \tag@\relax}%
 \vspace@\allowdisplaybreak@\displaybreak@\intertext@
 \displ@y@\measuring@true\tag@CC\tag@C
 \setbox\savealignat@\hbox{\noset@\unlabel@$\m@th\displaystyle\Let@
  \attag@{#1}\vbox{\halign{\span\preamble@@\crcr#2\crcr}}$}%
 \measuring@false\Let@\attag@{#1}\tag@C\tag@CC
 \tabskip\centering@\halign to\displaywidth
 {\span\preamble@@\crcr#2\crcr\black@{\wd\savealignat@}}}
\def\gather{\RIfMIfI@\DN@{\onlydmatherr@\gather}\else
 \ingather@true\inany@true\def\tag{&\relax}%
 \vspace@\allowdisplaybreak@\displaybreak@\intertext@
 \displ@y\Let@
 \iftagsleft@\DN@{\csname gather \endcsname}\else
  \DN@{\csname gather \space\endcsname}\fi\fi
 \else\DN@{\onlydmatherr@\gather}\fi\next@}
\def\exstring@{\expandafter\eat@\string}
\def\newcounter#1{\define#1{}%
 \edef\next@{\def\noexpand#1{\futurelet\noexpand\next
  \csname\exstring@#1@Z\endcsname}}\next@
 \edef\next@{\def\csname\exstring@#1@Z\endcsname
  {\global\advance\csname\exstring@#1@C\endcsname\@ne
  {\csname\exstring@#1@F\endcsname\csname\exstring@#1@S\endcsname
   {\csname\exstring@#1@P\endcsname\csname\exstring@#1@N\endcsname
   {\noexpand\number\csname\exstring@#1@C\endcsname}%
   \csname\exstring@#1@Q\endcsname}}%
  \noexpand\ifx\noexpand\next\noexpand\label
   \def\noexpand\next@\noexpand\label########1{{\noexpand\noexpands@
    \xdef\noexpand\Thelabel@{\csname\exstring@#1@N\endcsname
     {\noexpand\number\csname\exstring@#1@C\endcsname}}%
    \xdef\noexpand\Thelabel@@@{\noexpand\number
     \csname\exstring@#1@C\endcsname}%
    \xdef\noexpand\Thelabel@@{\csname\exstring@#1@S\endcsname
     {\csname\exstring@#1@P\endcsname
     \csname\exstring@#1@N\endcsname
     {\noexpand\number\csname\exstring@#1@C\endcsname}%
     \csname\exstring@#1@Q\endcsname}}%
    \xdef\noexpand\Thelabel@@@@{\csname\exstring@#1@P\endcsname
     \csname\exstring@#1@N\endcsname
     {\noexpand\number\csname\exstring@#1@C\endcsname}%
     \csname\exstring@#1@Q\endcsname}}%
    {\noexpand\locallabel@\noexpand\label{########1}}}%
   \noexpand\else\let\noexpand\next@\relax\noexpand\fi\noexpand\next@}}\next@
 \expandafter\newcount@\csname\exstring@#1@C\endcsname
 \expandafter\let\csname\exstring@#1@N\endcsname\arabic
 \expandafter\def\csname\exstring@#1@S\endcsname##1{##1\/}%
 \expandafter\let\csname\exstring@#1@P\endcsname\empty
 \expandafter\let\csname\exstring@#1@Q\endcsname\empty
 \expandafter\def\csname\exstring@#1@F\endcsname{\rm}%
 }
\def\HASH@#1#2{\ifnum#2=\z@\else
 \edef\next@{\toks@{\the\toks@\the\hashtoks@#2}%
 \toks@@{\the\toks@@{\the\hashtoks@#2}}}\next@\expandafter\HASH@\fi}
\def\HASH@@{\toks@{}\toks@@{}\expandafter\HASH@\macpar@00}
\def\usecounter#1#2{\expandafter\ifx\csname\exstring@#1@Z\endcsname
 \relax\Err@{\noexpand#1not created with \string\newcounter}\fi
 \expandafter\let\csname\exstring@#1@@Z\endcsname\relax
 \expandafter\let\csname\exstring@#1@@Z@\endcsname\relax
 \expandafter\let\csname\exstring@#1@@Z@@\endcsname\relax
 \edef\next@{\def\noexpand#2{\futurelet\noexpand\next
  \csname\exstring@#1@@Z\endcsname}}\next@
 \edef\next@{\def\csname\exstring@#1@@Z\endcsname{\noexpand\ifx
  \noexpand\next\noexpand\label\def\noexpand\next@\noexpand\label
   ########1{\csname\exstring@#1@@Z@\endcsname
   {\noexpand#1\noexpand\label{########1}}}%
   \noexpand\else\noexpand\ifx\noexpand\next
   \noexpand"\def\noexpand\next@\noexpand"########1\noexpand"%
   {\csname\exstring@#1@@Z@\endcsname{{\expandafter\noexpand
   \csname\exstring@#1@F\endcsname
   \let\noexpand\pre\expandafter\noexpand\csname\exstring@#1@P\endcsname
   \let\noexpand\post\expandafter\noexpand\csname\exstring@#1@Q\endcsname
   \let\noexpand\style\expandafter\noexpand\csname\exstring@#1@S\endcsname
   \let\noexpand\numstyle\expandafter\noexpand\csname\exstring@#1@N\endcsname
   ########1}}}\noexpand\else
   \def\noexpand\next@{\csname\exstring@#1@@Z@\endcsname{\noexpand#1}}%
   \noexpand\fi\noexpand\fi\noexpand\next@}}\next@
 \def\next@{\expandafter\expandafter\expandafter\unmacro@\expandafter
  \meaning\csname\exstring@#1@@Z@@\endcsname\unmacro@
  \HASH@@
  \edef\next@{\def\csname\exstring@#1@@Z@\endcsname\the\toks@{%
   \expandafter\noexpand\csname\exstring@#1@@Z@@\endcsname\the\toks@@
   \noexpand\FNSSP@}}\next@}%
 \afterassignment\next@
 \expandafter\def\csname\exstring@#1@@Z@@\endcsname}
\def\listbi@{\penalty50 \medskip}
\def\listbii@{\penalty100 \smallskip}
\let\listbiii@\relax
\let\listbiv@\relax
\let\listbv@\relax
\def\listmi@{\advance\leftskip30\p@\relax}
\let\listmii@\listmi@
\let\listmiii@\listmi@
\let\listmiv@\listmi@
\let\listmv@\listmi@
\def\itemi@#1{\noindent@@\llap{#1\hskip5\p@}}
\let\itemii@\itemi@
\let\itemiii@\itemi@
\let\itemiv@\itemi@
\let\itemv@\itemi@
\def\liste@{\penalty-50 \medskip}
\def\listei@{\penalty-100 \smallskip}
\let\listeii@\relax
\let\listeiii@\relax
\let\listeiv@\relax
\expandafter\newcount\csname list@C1\endcsname
\csname list@C1\endcsname\z@
\expandafter\newcount\csname list@C2\endcsname
\csname list@C2\endcsname\z@
\expandafter\newcount\csname list@C3\endcsname
\csname list@C3\endcsname\z@
\expandafter\newcount\csname list@C4\endcsname
\csname list@C4\endcsname\z@
\expandafter\newcount\csname list@C5\endcsname
\csname list@C5\endcsname\z@
\expandafter\let\csname list@P1\endcsname\empty
\expandafter\let\csname list@P2\endcsname\empty
\expandafter\let\csname list@P3\endcsname\empty
\expandafter\let\csname list@P4\endcsname\empty
\expandafter\let\csname list@P5\endcsname\empty
\expandafter\let\csname list@Q1\endcsname\empty
\expandafter\let\csname list@Q2\endcsname\empty
\expandafter\let\csname list@Q3\endcsname\empty
\expandafter\let\csname list@Q4\endcsname\empty
\expandafter\let\csname list@Q5\endcsname\empty
\expandafter\def\csname list@S1\endcsname#1{{\rm(}{#1\/}{\rm)}}
\expandafter\def\csname list@S2\endcsname#1{{\rm(}{#1\/}{\rm)}}
\expandafter\def\csname list@S3\endcsname#1{{\rm(}{#1\/}{\rm)}}
\expandafter\def\csname list@S4\endcsname#1{{\rm(}{#1\/}{\rm)}}
\expandafter\def\csname list@S5\endcsname#1{{\rm(}{#1\/}{\rm)}}
\expandafter\let\csname list@N1\endcsname\arabic
\expandafter\let\csname list@N2\endcsname\arabic
\expandafter\let\csname list@N3\endcsname\arabic
\expandafter\let\csname list@N4\endcsname\arabic
\expandafter\let\csname list@N5\endcsname\arabic
\expandafter\def\csname list@F1\endcsname{\rm}
\expandafter\def\csname list@F2\endcsname{\rm}
\expandafter\def\csname list@F3\endcsname{\rm}
\expandafter\def\csname list@F4\endcsname{\rm}
\expandafter\def\csname list@F5\endcsname{\rm}
\newcount\listlevel@
\listlevel@\z@
\def\list@@C{\csname list@C\number\listlevel@\endcsname}
\def\list@@P{\csname list@P\number\listlevel@\endcsname}
\def\list@@Q{\csname list@Q\number\listlevel@\endcsname}
\def\list@@S{\csname list@S\number\listlevel@\endcsname}
\def\list@@N{\csname list@N\number\listlevel@\endcsname}
\def\list@@F{\csname list@F\number\listlevel@\endcsname}
\newif\iffirstitemi@
\newif\iffirstitemii@
\newif\iffirstitemiii@
\newif\iffirstitemiv@
\newif\iffirstitemv@
\def\Firstitem@true{\csname firstitem\romannumeral\listlevel@
 @true\endcsname}
\def\Firstitem@false{\csname firstitem\romannumeral\listlevel@
 @false\endcsname}
\def\Listm@{\csname listm\romannumeral\listlevel@ @\endcsname}
\def\Item@{\csname item\romannumeral\listlevel@ @\endcsname}
\def\Liste@{\csname liste\romannumeral\listlevel@ @\endcsname}
\newif\iflistcontinue@
\def\keepitem{\listcontinue@true}
\newcount\list@C@
\def\list{%
 \iflistcontinue@\csname list@C1\endcsname\csname list@C@\endcsname\fi
 \global\csname list@C2\endcsname\z@
 \global\csname list@C3\endcsname\z@
 \global\csname list@C4\endcsname\z@
 \global\csname list@C5\endcsname\z@
 \begingroup
 \firstitemi@true
 \listlevel@\@ne
 \def\item{\FN@\item@}%
 \FN@\list@}
\Invalid@\runinitem
\def\list@{\ifx\next\par
 \DN@\par{\FN@\list@}\else
 \ifx\next\runinitem
  \DN@\runinitem{\FN@\runinitem@}\else
  \DN@{\par\dimen@\parskip\parskip\dimen@}\fi\fi\next@}
\newif\ifoutlevel@
\newif\ifrunin@
\def\item@{%
 \ifoutlevel@\Liste@\outlevel@false\fi
 \ifrunin@\runin@false\par
  \dimen@\parskip\parskip\dimen@
  \Listm@\fi
 \iffirstitemi@\listbi@\listmi@\firstitemi@false\else\par\fi
 \iffirstitemii@\listbii@\listmii@\firstitemii@false\else\par\fi
 \iffirstitemiii@\listbiii@\listmiii@\firstitemiii@false\else\par\fi
 \iffirstitemiv@\listbiv@\listmiv@\firstitemiv@false\else\par\fi
 \iffirstitemv@\listbv@\listmv@\firstitemv@false\else\par\fi
 \DN@"##1"{{\let\pre\list@@P\let\post\list@@Q
  \let\style\list@@S\let\numstyle\list@@N
  \vskip-\parskip
  \Item@{\list@@F##1}%
  \noexpands@
  \Qlabel@{##1}}%
  \locallabel@
  \FNSSP@}%
 \DNii@{\global\advance\list@@C\@ne
  {\noexpands@
   \xdef\Thelabel@@@{\number\list@@C}%
   \xdefThelabel@\list@@N
   \xdef\Thelabel@@@@{\list@@P\Thelabel@\list@@Q}%
   \xdefThelabel@@\list@@S
  }%
  \locallabel@
  \vskip-\parskip
  \Item@{\list@@F\thelabel@@}%
  \FN@\pretendspace@}%
 \ifx\next"\expandafter\next@\else\expandafter\nextii@\fi}
\def\runinitem@{%
  \runin@true
  \Firstitem@false
  \DN@"##1"{{\let\pre\list@@P\let\post\list@@Q
   \let\style\list@@S\let\numstyle\list@@N
   \unskip\space{\list@@F##1} %
   \noexpands@
   \Qlabel@{##1}}%
   \locallabel@
   \ignorespaces}%
  \DNii@{\global\advance\list@@C\@ne
   {\noexpands@
    \xdef\Thelabel@@@{\number\list@@C}%
    \xdefThelabel@\list@@N
    \xdef\Thelabel@@@@{\list@@P\Thelabel@\list@@Q}%
    \xdefThelabel@@\list@@S
   }%
   \locallabel@
   \unskip\space{\list@@F\thelabel@@} }%
  \ifx\next"\expandafter\next@\else\expandafter\nextii@\fi}
\def\inlevel{\ifnum\listlevel@=5
 \DN@{\Err@{Already 5 levels down}}\else
 \DN@{\begingroup\advance\listlevel@\@ne
 \Firstitem@true\FN@\inlevel@}\fi\next@}
\def\inlevel@{\ifx\next\par
 \DN@\par{\FN@\inlevel@}\else
 \ifx\next\runinitem
  \DN@\runinitem{\FN@\runinitem@}\else
  \let\next@\relax\fi\fi\next@}
\def\outlevel{\ifnum\listlevel@=\@ne
 \Err@{At top level}\else
 \par\global\list@@C\z@\endgroup\outlevel@true\fi}
\def\endlist{%
 \expandafter\global\csname list@C@\endcsname\csname list@C1\endcsname
 \par
 \global\toks\@ne{}\count@\listlevel@
 {\loop
  \ifnum\count@>\z@\global\toks\@ne\expandafter{\the\toks\@ne\endgroup}%
  \advance\count@\m@ne
  \repeat}%
 \the\toks\@ne
 \liste@
 \listcontinue@false\global\csname list@C1\endcsname\z@
 \vskip-\parskip
 \noindent@@
 \FN@\pretendspace@}
\newif\iffirstdescribe@
\def\describe{\par
 \begingroup\firstdescribe@true
 \def\item##1{%
  \iffirstdescribe@\penalty50 \medskip\vskip-\parskip
  \firstdescribe@false\else\par\fi
  \noindent@@\hangindent2pc\hangafter\@ne
  {\bf##1}\hskip.5em}}

\Invalid@\pullin
\Invalid@\pullinmore
\newif\iffirstpull@
\def\margins{\par\begingroup\firstpull@true
 \def\pullin##1##2{\par
  \iffirstpull@\firstpull@false\else\endgroup\fi
  \begingroup\DN@{##1}%
  \ifx\next@\empty\leftskip\z@\else\ifx\next@\space\leftskip\z@
  \else\leftskip##1\fi\fi
  \DN@{##2}\ifx\next@\empty\rightskip\z@\else\ifx\next@\space
  \rightskip\z@\else\rightskip##2\fi\fi\ignorespaces}%
 \def\pullinmore##1##2{\par
  \xdef\Next@{\leftskip\the\leftskip\relax\rightskip\the\rightskip\relax}%
  \iffirstpull@\firstpull@false\else\endgroup\fi
  \begingroup\Next@
  \DN@{##1}%
  \ifx\next@\empty\else\ifx\next@\space\else\advance\leftskip##1\fi\fi
  \DN@{##2}\ifx\next@\empty\else\ifx\next@\space\else
  \advance\rightskip##2\fi\fi\ignorespaces}}

\newif\ifnopunct@
\newif\ifnospace@
\newif\ifoverlong@
\let\nofrillslist@\empty
\let\overlonglist@\empty
\def\nopunct{\nopunct@true\FN@\nopunct@}
\def\nospace{\nospace@true\FN@\nospace@}
\def\overlong{\overlong@true\FN@\overlong@}
\def\nopunct@{\ifx\next\nospace
 \DN@\nospace{\nospace@true\FN@\nopnos@}\else\ifx\next\overlong
 \DN@\overlong{\overlong@true\FN@\nopol@}\else
 \let\next@\nopunct@@\fi\fi\next@}
\def\nopunct@@#1{\ismember@\nofrillslist@#1%
 \iftest@\let\next@#1\else
 \DN@{\nopunct@false\Err@{\noexpand\nopunct can't be used with
 \string#1}#1}\fi\next@}
\def\nospace@{\ifx\next\nopunct
 \DN@\nopunct{\nopunct@true\FN@\nopnos@}\else\ifx\next\overlong
 \DN@\overlong{\overlong@true\FN@\nosol@}\else
 \let\next@\nospace@@\fi\fi\next@}
\def\nospace@@#1{\ismember@\nofrillslist@#1%
 \iftest@\let\next@#1\else
 \DN@{\nospace@false\Err@{\noexpand\nospace can't be used with
 \string#1}#1}\fi\next@}
\def\overlong@{\ifx\next\nopunct
 \DN@\nopunct{\nopunct@true\FN@\nopol@}\else\ifx\next\nospace
 \DN@\nospace{\nospace@true\FN@\nosol@}\else
 \let\next@\overlong@@\fi\fi\next@}
\def\overlong@@#1{\ismember@\overlonglist@#1%
 \iftest@\let\next@#1\else
 \DN@{\overlong@false\Err@{\noexpand\overlong can't be used with
 \string#1}#1}\fi\next@}
\def\nopnos@{\ifx\next\overlong
 \DN@\overlong{\overlong@true\nopnosol@}\else
 \let\next@\nopnos@@\fi\next@}
\def\nopol@{\ifx\next\nospace
 \DN@\nospace{\nospace@true\nopnosol@}\else
 \let\next@\nopol@@\fi\next@}
\def\nosol@{\ifx\next\nopunct
 \DN@\nopunct{\nopunct@true\nopnosol@}\else
 \let\next@\nosol@@\fi\next@}
\def\nopnos@@#1{\ismember@\nofrillslist@#1%
 \iftest@\let\next@#1\else
 \DN@{\nopunct@false\nospace@false
  \Err@{\noexpand\nopunct\noexpand\nospace
   can't be used with \string#1}#1}\fi\next@}
\def\testii@#1{\ismember@\nofrillslist@#1%
 \iftest@\let\nextiii@ T\else\let\nextiii@ F\fi
 \ismember@\overlonglist@#1%
 \iftest@\let\nextiv@ T\else\let\nextiv@ F\fi
 \test@false\if\nextiii@ T\if\nextiv@ T\test@true\fi\fi}
\def\nopol@@#1{\testii@{#1}%
 \iftest@\let\next@#1%
 \else\DN@{\if\nextiii@ T\else\nopunct@false\fi
  \if\nextiv@ T\else\overlong@false\fi
  \Err@{\if\nextiii@ T\else\noexpand\nopunct\fi
  \if\nextiv@ T\else\noexpand\overlong\fi can't be used
  with \string#1}#1}\fi\next@}
\def\nosol@@#1{\testii@{#1}%
 \iftest@\let\next@#1%
 \else\DN@{\if\nextiii@ T\else\nospace@false\fi
  \if\nextiv@ T\else\overlong@false\fi
  \Err@{\if\nextiii@ T\else\noexpand\nospace\fi
  \if\nextiv@ T\else\noexpand\overlong\fi can't be used
  with \string#1}#1}\fi\next@}
\def\nopnosol@#1{\testii@{#1}%
 \iftest@\let\next@#1%
 \else\DN@{\if\nextiii@ T\else\nopunct@false\nospace@false\fi
  \if\nextiv@ T\else\overlong@false\fi
  \Err@{\if\nextiii@ T\else\noexpand\nopunct\noexpand\nospace\fi
  \if\nextiv@ T\else\noexpand\overlong\fi can't be used
  with \string#1}#1}\fi\next@}
\def\punct@#1{\ifnopunct@\else#1\fi}
\def\addspace@#1{\ifnospace@\else#1\fi}
\def\hss@{\ifoverlong@\z@ plus\@m\p@ minus\@m\p@
 \else \z@ plus\@m\p@\fi}
\rightadd@\demo\to\nofrillslist@
\newif\ifclaim@
\def\exxx@{\expandafter\expandafter\expandafter\eat@\expandafter\string}
\let\colon@:
\def\demo#1{\ifclaim@
 \Err@{Previous \expandafter\noexpand\claimtype@ has
  no matching \string\end\exxx@\claimtype@}%
 \let\next@\relax
 \else
  \par
  \ifdim\lastskip<\smallskipamount\removelastskip\smallskip\fi
  \begingroup
  \noindent@@{\smc\ignorespaces#1\unskip
   \punct@{\null\colon@}\addspace@\enspace}%
  \nopunct@false\nospace@false
  \rm
  \DN@{\FNSSP@}%
 \fi
 \next@}
\def\enddemo{\par\endgroup\nopunct@false\nospace@false\smallskip}
\rightadd@\claim\to\nofrillslist@
\def\claim@F{\smc}
\def\claim@@@F{\csname\exxx@\claimtype@ @F\endcsname}
\def\claimformat@#1#2#3{%
 \medbreak\noindent@@{\smc#1 {\claim@@@F#2} #3%
 \punct@{\null.}\addspace@\enspace}\sl}
\def\claimformat@@#1#2{\claimformat@{\ignorespaces#1\unskip}%
 {\ifx\thelabel@@\empty\unskip\else\thelabel@@\fi}%
 {\ignorespaces#2\unskip}%
 \let\Claimformat@@\claimformat@@\FNSSP@}
\let\Claimformat@@\claimformat@@
\def\claim@@@P{\csname\exxx@\claimtype@ @P\endcsname}
\def\claim@@@Q{\csname\exxx@\claimtype@ @Q\endcsname}
\def\claim@@@S{\csname\exxx@\claimtype@ @S\endcsname}
\def\claim@@@N{\csname\exxx@\claimtype@ @N\endcsname}
\def\claim@@@C{\csname claim@C\claimclass@\endcsname}
\newcount\claim@C
\claim@C\z@
\let\claim@P\empty
\let\claim@Q\empty
\def\claim@S#1{#1\/}
\let\claim@N\arabic
\def\claim{\claim@true\let\claimclass@\empty
 \def\claimtype@{\claim}\FN@\claim@}
\def\claim@{%
 \ifx\next\c
  \let\next@\claim@c
 \else
  \ifx\next"%
   \let\next@\claim@q
  \else
   \begingroup\global\advance\claim@C\@ne
   {\noexpands@
    \xdef\Thelabel@@@{\number\claim@C}%
    \xdefThelabel@\claim@N
    \xdef\Thelabel@@@@{\claim@P\Thelabel@\claim@Q}%
    \xdefThelabel@@\claim@S
   }%
   \locallabel@
   \let\next@\Claimformat@@
  \fi
 \fi
 \next@}
\def\claim@c\c#1{\claim@true\begingroup
 \expandafter
 \ifx\csname claim@C#1\endcsname\relax
  \expandafter\newcount@\csname claim@C#1\endcsname
  \global\csname claim@C#1\endcsname\@ne
 \else
  \global\advance\csname claim@C#1\endcsname\@ne
 \fi
 \def\claimclass@{#1}%
 {\noexpands@
  \xdef\Thelabel@@@{\number\claim@@@C}%
  \xdefThelabel@\claim@@@N
  \xdef\Thelabel@@@@{\claim@@@P\Thelabel@\claim@@@Q}%
  \xdefThelabel@@\claim@@@S
 }%
 \locallabel@
 \FNSS@\claim@c@}
\def\claim@q"#1"{\begingroup
 {\let\pre\claim@@@P\let\post\claim@@@Q
  \let\style\claim@@@S\let\numstyle\claim@@@N
  \noexpands@
  \Qlabel@{#1}}%
 \locallabel@
 \FNSS@\claim@q@}
\def\claim@c@{\ifx\next"%
 \global\advance\claim@@@C\m@ne\let\next@\claim@cq
 \else\let\next@\Claimformat@@\fi\next@}
\def\claim@cq"#1"{{\let\pre\claim@@@P\let\post\claim@@@Q
 \let\style\claim@@@S\let\numstyle\claim@@@N
 \noexpands@
 \Qlabel@{#1}}%
 \locallabel@
 \FNSS@\Claimformat@@}
\def\claim@q@{\ifx\next\c\expandafter\claim@qc
 \else\expandafter\Claimformat@@\fi}
\def\claim@qc\c#1{\expandafter\ifx\csname claim@C#1\endcsname\relax
 \expandafter\newcount@\csname claim@C#1\endcsname
 \global\csname claim@C#1\endcsname\z@\fi
 \FNSS@\Claimformat@@}
\def\endclaim{\endgroup\claim@false\nopunct@false\nospace@false
 \let\Claimformat@@\claimformat@@\medbreak}
\Invalid@\claimclause
\def\newclaim{\FN@\newclaim@}
\def\newclaim@{\ifx\next\claimclause
 \DN@\claimclause##1{\newclaim@@{##1}}\else
 \DN@{\newclaim@@\relax}\fi\next@}
\def\claimlist@{\\\claim}
\newtoks\claim@i
\newtoks\claim@v
\let\noclaimclause@=F
\def\newclaim@@#1#2#3\c#4#5{\define#2{}%
 \rightadd@#2\to\claimlist@\rightadd@#2\to\nofrillslist@%
 \expandafter\def\csname\exstring@#2@P\endcsname{\claim@P}%
 \expandafter\def\csname\exstring@#2@Q\endcsname{\claim@Q}%
 \expandafter\def\csname\exstring@#2@S\endcsname{\claim@S}%
 \expandafter\def\csname\exstring@#2@N\endcsname{\claim@N}%
 \expandafter\def\csname\exstring@#2@F\endcsname{\claim@F}%
 \expandafter\def\csname end\exstring@#2\endcsname{\endclaim}%
 \expandafter\ifx\csname claim@C#4\endcsname\relax
  \expandafter\newcount@\csname claim@C#4\endcsname
  \global\csname claim@C#4\endcsname\z@\fi
 \edef\next@{\let\csname\exstring@#2@C\endcsname
   \csname claim@C#4\endcsname}\next@
 \def#2{\ifx\noclaimclause@ T\else#1\fi
  \global\claim@i{#1}\gdef\claim@iv{#4}\global\claim@v{#5}%
  \def\claimtype@{#2}\def\Claimformat@@{\claimformat@@{#5}}\claim@c\c{#4}}}
\def\shortenclaim#1#2{\define#2{}%
 \ismember@\claimlist@#1%
 \iftest@
  \rightadd@#2\to\nofrillslist@%
  \expandafter\def\csname\exstring@#2@P\endcsname
   {\csname\exstring@#1@P\endcsname}%
  \expandafter\def\csname\exstring@#2@Q\endcsname
   {\csname\exstring@#1@Q\endcsname}%
  \expandafter\def\csname\exstring@#2@S\endcsname
   {\csname\exstring@#1@S\endcsname}%
  \expandafter\def\csname\exstring@#2@N\endcsname
   {\csname\exstring@#1@N\endcsname}%
  \expandafter\def\csname\exstring@#2@F\endcsname
   {\csname\exstring@#1@F\endcsname}%
  \expandafter\def\csname end\exstring@#2\endcsname{\endclaim}%
  \edef\next@{\let\csname\exstring@#2@C\endcsname
    \csname claim\exstring@#1C\endcsname}\next@
  \setbox\z@\vbox{\let\noclaimclause@ T#1""\relax\endgroup}%
  \edef#2{\the\claim@i
   \def\noexpand\claimtype@{\noexpand#2}%
   \def\noexpand\Claimformat@@{\noexpand\claimformat@@{\the\claim@v}\relax}%
   \noexpand\claim@c\noexpand\c{\claim@iv}}%
 \else
  \Err@{\noexpand#1not yet created by \string\newclaim}%
 \fi}
\def\classtest@#1{\DN@{#1}\ifx\next@\claimclass@
 \test@true\else\test@false\fi}
\def\typetest@#1{\DN@{#1}\ifx\next@\claimtype@\test@true\else
  \test@false\fi}
\newif\iftoc@
\def\tocfile{\iftoc@\else\alloc@@7\write\chardef\sixt@@n\toc@
 \immediate\openout\toc@=\jobname.toc
 \alloc@@7\write\chardef\sixt@@n\tic@
 \immediate\openout\tic@=\jobname.tic
 \global\toc@true\fi}
\rightadd@\hl\to\nofrillslist@
\rightadd@\HL\to\overlonglist@
\def\HL@@C{\csname HL@C\HLlevel@\endcsname}
\def\HL@@P{\csname HL@P\HLlevel@\endcsname}
\def\HL@@Q{\csname HL@Q\HLlevel@\endcsname}
\def\HL@@S{\csname HL@S\HLlevel@\endcsname}
\def\HL@@N{\csname HL@N\HLlevel@\endcsname}
\def\HL@@F{\csname HL@F\HLlevel@\endcsname}
\def\HL@@@C{\csname\exxx@\HLtype@ @C\endcsname}
\def\HL@@@P{\csname\exxx@\HLtype@ @P\endcsname}
\def\HL@@@Q{\csname\exxx@\HLtype@ @Q\endcsname}
\def\HL@@@S{\csname\exxx@\HLtype@ @S\endcsname}
\def\HL@@@N{\csname\exxx@\HLtype@ @N\endcsname}
\def\HL#1{\expandafter
 \ifx\csname HL@C#1\endcsname\relax
  \DN@{\Err@{\string\HL#1 not defined in this style}}%
 \else
  \DN@{\gdef\HLlevel@{#1}\def\HLname@{\HL{#1}}\let\HLtype@\relax\FNSS@\HL@}%
 \fi
 \next@}%
\newif\ifquoted@
\let\aftertoc@\relax
\def\HL@{%
 \DN@"##1"##2\endHL{\def\entry@{##2}\quoted@true
  {\noexpands@
  \ifx\HLtype@\relax
   \let\pre\HL@@P\let\post\HL@@Q\let\style\HL@@S\let\numstyle\HL@@N
  \else
   \let\pre\HL@@@P\let\post\HL@@@Q\let\style\HL@@@S\let\numstyle\HL@@@N
  \fi
  \Qlabel@{##1}\let\style\relax\xdef\Qlabel@@@@{##1}%
  \xdef\Thepref@{\Thelabel@@@@}}%
  \csname HL@\HLlevel@\endcsname##2\endHL
  \let\pref\Thepref@
  \csname HL@I\HLlevel@\endcsname
  \csname HL@J\HLlevel@\endcsname
  \let\pref\pref@
  \HLtoc@	
  \aftertoc@
  \let\aftertoc@\relax\overlong@false}%
 \DNii@##1\endHL{\def\entry@{##1}\quoted@false
  {\noexpands@
  \ifx\HLtype@\relax
   \global\advance\HL@@C\@ne
   \xdef\Thelabel@@@{\number\HL@@C}%
   \xdefThelabel@{\HL@@N}%
   \xdef\Thelabel@@@@{\HL@@P\Thelabel@\HL@@Q}%
   \xdefThelabel@@{\HL@@S}%
  \else
   \global\advance\HL@@@C\@ne
   \xdef\Thelabel@@@{\number\HL@@@C}%
   \xdefThelabel@{\HL@@@N}%
   \xdef\Thelabel@@@@{\HL@@@P\Thelabel@\HL@@@Q}%
   \xdefThelabel@@{\HL@@@S}%
  \fi
  \xdef\Thepref@{\Thelabel@@@@}}%
  \csname HL@\HLlevel@\endcsname##1\endHL
  \let\pref\Thepref@
  \csname HL@I\HLlevel@\endcsname
  \csname HL@J\HLlevel@\endcsname
  \let\pref\pref@
  \HLtoc@
  \aftertoc@
  \let\aftertoc@\relax\overlong@false}%
 \ifx\next"\expandafter\next@\else\expandafter\nextii@\fi}%
\Invalid@\endHL
\def\hl@@C{\csname hl@C\hllevel@\endcsname}
\def\hl@@P{\csname hl@P\hllevel@\endcsname}
\def\hl@@Q{\csname hl@Q\hllevel@\endcsname}
\def\hl@@S{\csname hl@S\hllevel@\endcsname}
\def\hl@@N{\csname hl@N\hllevel@\endcsname}
\def\hl@@F{\csname hl@F\hllevel@\endcsname}
\def\hl@@@C{\csname\exxx@\hltype@ @C\endcsname}
\def\hl@@@P{\csname\exxx@\hltype@ @P\endcsname}
\def\hl@@@Q{\csname\exxx@\hltype@ @Q\endcsname}
\def\hl@@@S{\csname\exxx@\hltype@ @S\endcsname}
\def\hl@@@N{\csname\exxx@\hltype@ @N\endcsname}
\def\hl#1{\expandafter
 \ifx\csname hl@C#1\endcsname\relax
  \DN@{\Err@{\string\hl#1 not defined in this style}}%
 \else
  \DN@{\gdef\hllevel@{#1}\def\hlname@{\hl{#1}}\let\hltype@\relax\FNSS@\hl@}%
 \fi
 \next@}
\def\hl@{%
 \DN@"##1"##2{\def\entry@{##2}\quoted@true
  {\noexpands@
  \ifx\hltype@\relax
   \let\pre\hl@@P\let\post\hl@@Q\let\style\hl@@S\let\numstyle\hl@@N
  \else
   \let\pre\hl@@@P\let\post\hl@@@Q\let\style\hl@@@S\let\numstyle\hl@@@N
  \fi
  \Qlabel@{##1}\let\style\relax\xdef\Qlabel@@@@{##1}%
  \xdef\Thepref@{\Thelabel@@@@}}%
  \csname hl@\hllevel@\endcsname{##2}%
  \let\pref\Thepref@
  \csname hl@I\hllevel@\endcsname
  \csname hl@J\hllevel@\endcsname
  \let\pref\pref@
  \hltoc@
  \aftertoc@
  \let\aftertoc@\relax\nopunct@false\nospace@false\FNSSP@}%
 \DNii@##1{\def\entry@{##1}\quoted@false
  {\noexpands@
  \ifx\hltype@\relax
   \global\advance\hl@@C\@ne
   \xdef\Thelabel@@@{\number\hl@@C}%
   \xdefThelabel@{\hl@@N}%
   \xdef\Thelabel@@@@{\hl@@P\Thelabel@\hl@@Q}%
   \xdefThelabel@@{\hl@@S}%
  \else
   \global\advance\hl@@@C\@ne
   \xdef\Thelabel@@@{\number\hl@@@C}%
   \xdefThelabel@{\hl@@@N}%
   \xdef\Thelabel@@@@{\hl@@@P\Thelabel@\hl@@@Q}%
   \xdefThelabel@@{\hl@@@S}%
  \fi
  \xdef\Thepref@{\Thelabel@@@@}}%
  \csname hl@\hllevel@\endcsname{##1}%
  \let\pref\Thepref@
  \csname hl@I\hllevel@\endcsname
  \csname hl@J\hllevel@\endcsname
  \let\pref\pref@
  \hltoc@
  \aftertoc@
  \let\aftertoc@\relax\nopunct@false\nospace@false\FNSSP@}%
 \ifx\next"\expandafter\next@\else\expandafter\nextii@\fi}%
\def\six@#1#2 #3 #4 #5 #6 #7 {\DN@{#2}\ifx\next@\empty
 \DN@##1\six@{}\else
 \write#1{ #2 #3 #4 #5 #6 #7}\DN@{\six@#1}\fi
 \next@}
\def\Sixtoc@{\ifx\macdef@\empty\else
 \DN@##1##2\next@{\def\macdef@{##1##2}}%
 \expandafter\next@\macdef@\next@
 \edef\next@
  {\noexpand\six@\toc@\macdef@
  \space\space\space\space\space\space\space\space\space\space\space\space
  \noexpand\six@}%
 \next@\let\macdef@\relax\fi}
\def\QorThelabel@@@@{\ifquoted@
 \noexpand\noexpand\noexpand"\Qlabel@@@@\noexpand\noexpand\noexpand"\else
 \Thelabel@@@@\fi}
\def\HLtoc@{%
 \iftoc@
 \expandafter\expandafter\expandafter\unmacro@
  \expandafter\meaning\csname HL@W\HLlevel@\endcsname\unmacro@
  {\noexpands@\let\style\relax
   \edef\next@{\write\toc@{\noexpand\noexpand\expandafter\noexpand\HLname@
   {\macdef@}{\QorThelabel@@@@}}}%
  \next@}%
  \expandafter\unmacro@\meaning\entry@\unmacro@
  \Sixtoc@
  \write\toc@{\noexpand\Page{\number\pageno}{\page@N}%
   {\page@P}{\page@Q}^^J}%
 \fi}
\def\hltoc@{%
 \iftoc@
 \expandafter\expandafter\expandafter\unmacro@
  \expandafter\meaning\csname hl@W\hllevel@\endcsname\unmacro@
  {\noexpands@\let\style\relax
  \edef\next@{\write\toc@{%
   \ifnopunct@\noexpand\noexpand\noexpand\nopunct\fi
   \ifnospace@\noexpand\noexpand\noexpand\nospace\fi
   \noexpand\noexpand\expandafter\noexpand\hlname@
   {\macdef@}{\QorThelabel@@@@}}}%
  \next@}%
  \expandafter\unmacro@\meaning\entry@\unmacro@
  \Sixtoc@
  \write\toc@{\noexpand\Page{\number\pageno}{\page@N}%
   {\page@P}{\page@Q}^^J}%
 \fi}
\def\mainfile#1{\def\mainfile@{#1}}
\def\checkmainfile@{\ifx\mainfile@\undefined
 \Err@{No \noexpand\mainfile specified}\fi}
\expandafter\newcount@\csname HL@C1\endcsname
\csname HL@C1\endcsname\z@
\expandafter\def\csname HL@S1\endcsname#1{#1\null.}
\expandafter\let\csname HL@N1\endcsname\arabic
\expandafter\let\csname HL@P1\endcsname\empty
\expandafter\let\csname HL@Q1\endcsname\empty
\expandafter\def\csname HL@F1\endcsname{\bf}
\expandafter\let\csname HL@W1\endcsname\empty
\expandafter\newcount@\csname hl@C1\endcsname
\csname hl@C1\endcsname\z@
\expandafter\def\csname hl@S1\endcsname#1{#1\/}
\expandafter\let\csname hl@N1\endcsname\arabic
\expandafter\let\csname hl@P1\endcsname\empty
\expandafter\let\csname hl@Q1\endcsname\empty
\expandafter\def\csname hl@F1\endcsname{\bf}
\expandafter\let\csname hl@W1\endcsname\empty
\expandafter\def\csname HL@1\endcsname#1\endHL{\bigbreak
 {\locallabel@
  \global\setbox\@ne\vbox{\Let@\tabskip\hss@
  \halign to\hsize{\bf\hfil\ignorespaces##\unskip\hfil\cr
  \expandafter\ifx\csname HL@W1\endcsname\empty\else
   \csname HL@W1\endcsname\space\fi
  {\HL@@F\ifx\thelabel@@\empty\else\thelabel@@\space\fi}%
  \ignorespaces#1\crcr}}%
  }%
 \unvbox\@ne\nobreak\medskip}
\expandafter\def\csname hl@1\endcsname#1{\medbreak\noindent@@
 {\locallabel@
 \bf{\hl@@F\ifx\thelabel@@\empty\else\thelabel@@\space\fi}%
 \ignorespaces#1\unskip\punct@{\null.}\addspace@\enspace}}
\expandafter\def\csname HL@I1\endcsname{\Reset\hl1{1}%
 \ifx\pref\empty\newpre\hl1{}\else\newpre\hl1{\pref.}\fi}
\def\NameHL#1#2{\define#2{}%
 \expandafter\ifx\csname HL@R#1\endcsname\relax
 \else
  \def\nextiv@{\let\nextiii@}%
  \expandafter\nextiv@\csname HL@R#1\endcsname
  \expandafter\let\nextiii@\undefined
  \expandafter\let\csname\exxx@\nextiii@ @C\endcsname\relax
  \expandafter\let\csname\exxx@\nextiii@ @P\endcsname\relax
  \expandafter\let\csname\exxx@\nextiii@ @Q\endcsname\relax
  \expandafter\let\csname\exxx@\nextiii@ @S\endcsname\relax
  \expandafter\let\csname\exxx@\nextiii@ @N\endcsname\relax
  \expandafter\let\csname\exxx@\nextiii@ @F\endcsname\relax
  \expandafter\let\csname\exxx@\nextiii@ @W\endcsname\relax
  \expandafter\let\csname end\exxx@\nextiii@\endcsname\undefined
 \fi
 \expandafter\gdef\csname HL@R#1\endcsname{#2}%
 \expandafter\gdef\csname\exstring@#2@R\endcsname{{HL}{#1}}%
 \iftoc@\write\toc@{\noexpand\NameHL#1\noexpand#2^^J}\fi
 \rightadd@#2\to\overlonglist@
 \edef\next@{\let\csname\exstring@#2@C\endcsname\expandafter\noexpand
  \csname HL@C#1\endcsname}\next@
 \edef\next@{\let\csname\exstring@#2@P\endcsname\expandafter\noexpand
  \csname HL@P#1\endcsname}\next@
 \edef\next@{\let\csname\exstring@#2@Q\endcsname\expandafter\noexpand
  \csname HL@Q#1\endcsname}\next@
 \edef\next@{\let\csname\exstring@#2@S\endcsname\expandafter\noexpand
  \csname HL@S#1\endcsname}\next@
 \edef\next@{\let\csname\exstring@#2@N\endcsname\expandafter\noexpand
  \csname HL@N#1\endcsname}\next@
 \edef\next@{\let\csname\exstring@#2@F\endcsname\expandafter\noexpand
  \csname HL@F#1\endcsname}\next@
 \edef\next@{\let\csname\exstring@#2@W\endcsname\expandafter\noexpand
  \csname HL@W#1\endcsname}\next@
 \edef\next@{\def\noexpand#2####1\expandafter\noexpand
  \csname end\exstring@#2\endcsname
  {\def\noexpand\HLtype@{\noexpand#2}%
   \def\noexpand\HLname@{\noexpand#2}%
   \gdef\noexpand\HLlevel@{#1}%
   \noexpand\FNSS@\noexpand\HL@####1\noexpand\endHL}}%
  \next@
 \edef\next@{\noexpand\Invalid@\expandafter\noexpand
  \csname end\exstring@#2\endcsname}%
 \next@}
\def\Namehl#1#2{\define#2{}%
 \expandafter\ifx\csname hl@R#1\endcsname\relax
 \else
  \def\nextiv@{\let\nextiii@}%
  \expandafter\nextiv@\csname hl@R#1\endcsname
  \expandafter\let\nextiii@\undefined
  \expandafter\let\csname\exxx@\nextiii@ @C\endcsname\relax
  \expandafter\let\csname\exxx@\nextiii@ @P\endcsname\relax
  \expandafter\let\csname\exxx@\nextiii@ @Q\endcsname\relax
  \expandafter\let\csname\exxx@\nextiii@ @S\endcsname\relax
  \expandafter\let\csname\exxx@\nextiii@ @N\endcsname\relax
  \expandafter\let\csname\exxx@\nextiii@ @F\endcsname\relax
  \expandafter\let\csname\exxx@\nextiii@ @W\endcsname\relax
 \fi
 \expandafter\gdef\csname hl@R#1\endcsname{#2}%
 \expandafter\gdef\csname\exstring@#2@R\endcsname{{hl}{#1}}%
 \iftoc@\write\toc@{\noexpand\Namehl#1\noexpand#2^^J}\fi
 \rightadd@#2\to\nofrillslist@%
 \edef\next@{\let\csname\exstring@#2@C\endcsname\expandafter\noexpand
  \csname hl@C#1\endcsname}\next@
 \edef\next@{\let\csname\exstring@#2@P\endcsname\expandafter\noexpand
  \csname hl@P#1\endcsname}\next@
 \edef\next@{\let\csname\exstring@#2@Q\endcsname\expandafter\noexpand
  \csname hl@Q#1\endcsname}\next@
 \edef\next@{\let\csname\exstring@#2@S\endcsname\expandafter\noexpand
  \csname hl@S#1\endcsname}\next@
 \edef\next@{\let\csname\exstring@#2@N\endcsname\expandafter\noexpand
  \csname hl@N#1\endcsname}\next@
 \edef\next@{\let\csname\exstring@#2@F\endcsname\expandafter\noexpand
  \csname hl@F#1\endcsname}\next@
 \edef\next@{\let\csname\exstring@#2@W\endcsname\expandafter\noexpand
  \csname hl@W#1\endcsname}\next@
 \edef\next@{\def\noexpand#2{%
  \def\noexpand\hltype@{\noexpand#2}%
  \def\noexpand\hlname@{\noexpand#2}%
  \gdef\noexpand\hllevel@{#1}%
  \noexpand\FNSS@\noexpand\hl@}}%
 \next@}%
\def\Initialize{\FN@\Init@}
\def\Init@{\ifx\next\HL\let\next@\InitH@\else\ifx\next\hl\let\next@\InitH@
  \else\let\next@\InitS@\fi\fi\next@}
\def\InitH@#1#2{\expandafter\ifx\csname\exstring@#1@C#2\endcsname\relax
 \DN@{\Err@{\noexpand#1level #2 not defined in this style}}\else
 \DN@{\expandafter\gdef\csname\exstring@#1@J#2\endcsname}\fi\next@}
\def\InitC@#1#2{\edef\nextii@{\expandafter\noexpand\csname#1\endcsname{#2}}}
\def\InitS@#1{\expandafter\ifx\csname\exstring@#1@R\endcsname\relax
 \Err@{\noexpand#1not defined in this style}\let\next@\relax\else
 \DN@{\let\next@}\expandafter\next@\csname\exstring@#1@R\endcsname
 \expandafter\InitC@\next@
 \DN@{\expandafter\InitH@\nextii@}\fi\next@}
\def\value#1{\expandafter
 \ifx\csname\exstring@#1@C\endcsname\relax
  \expandafter\ifx\csname\exstring@#1@C1\endcsname\relax
   \DN@{\Err@{\noexpand\value can't be used with \string#1}}%
  \else
   \DN@{\value@#1}%
  \fi
 \else
  \DN@{\number\csname\exstring@#1@C\endcsname\relax}%
 \fi
 \next@}
\def\value@#1#2{\expandafter
 \ifx\csname\exstring@#1@C#2\endcsname\relax
  \DN@{\Err@{\string\value\string#1 can't be followed by \string#2}}%
 \else
  \DN@{\number\csname\exstring@#1@C#2\endcsname\relax}%
 \fi
 \next@}
\newcount\Value
\def\Evaluate#1{\expandafter
 \ifx\csname\exstring@#1@C\endcsname\relax
  \expandafter\ifx\csname\exstring@#1@C1\endcsname\relax
   \DN@{\Err@{\noexpand\Evaluate can't be used with \string#1}}%
  \else
   \DN@{\Evaluate@#1}%
  \fi
 \else
  \DN@{\global\Value\csname\exstring@#1@C\endcsname}%
 \fi
 \next@}
\def\Evaluate@#1#2{\expandafter
 \ifx\csname\exstring@#1@C#2\endcsname\relax
  \DN@{\Err@{\string\Evaluate\string#1 can't be followed by \string#2}}%
 \else
  \DN@{\global\Value\csname\exstring@#1@C#2\endcsname}%
 \fi\next@}
\def\pre#1{\expandafter
 \ifx\csname\exstring@#1@P\endcsname\relax
  \expandafter\ifx\csname\exstring@#1@P1\endcsname\relax
   \DN@{\Err@{\noexpand\pre can't be used with \string#1}}%
  \else
   \DN@{\pre@#1}%
  \fi
 \else
  \DN@{{\csname\exstring@#1@P\endcsname}}%
 \fi
 \next@}
\def\pre@#1#2{\expandafter
 \ifx\csname\exstring@#1@P#2\endcsname\relax
  \DN@{\Err@{\string\pre\string#1 can't be followed by \string#2}}%
 \else
  \DN@{{\csname\exstring@#1@P#2\endcsname}}%
 \fi
 \next@}
\def\post#1{\expandafter
 \ifx\csname\exstring@#1@Q\endcsname\relax
  \expandafter\ifx\csname\exstring@#1@Q1\endcsname\relax
   \DN@{\Err@{\noexpand\post can't be used with \string#1}}%
  \else
   \DN@{\post@#1}%
  \fi
 \else
  \DN@{{\csname\exstring@#1@Q\endcsname}}%
 \fi
 \next@}
\def\post@#1#2{\expandafter
 \ifx\csname\exstring@#1@Q#2\endcsname\relax
  \DN@{\Err@{\string\post\string#1 can't be followed by \string#2}}%
 \else
  \DN@{{\csname\exstring@#1@Q#2\endcsname}}%
 \fi
 \next@}
\def\style#1{\expandafter
 \ifx\csname\exstring@#1@S\endcsname\relax
  \expandafter\ifx\csname\exstring@#1@S1\endcsname\relax
   \DN@{\Err@{\noexpand\style can't be used with \string#1}}%
  \else
   \DN@{\style@#1}%
  \fi
 \else
  \DN@{\csname\exstring@#1@S\endcsname}%
 \fi
 \next@}
\def\style@#1#2{\expandafter
 \ifx\csname\exstring@#1@S#2\endcsname\relax
  \DN@{\Err@{\string\style\string#1 can't be followed by \string#2}}%
 \else
  \DN@{\csname\exstring@#1@S#2\endcsname}%
 \fi
 \next@}
\def\fontstyle#1{\expandafter
 \ifx\csname\exstring@#1@F\endcsname\relax
  \expandafter\ifx\csname\exstring@#1@F1\endcsname\relax
   \DN@{\Err@{\noexpand\fontstyle can't be used with \string#1}}%
  \else
   \DN@{\fontstyle@#1}%
  \fi
 \else
  \DN@##1{{\csname\exstring@#1@F\endcsname##1}}%
 \fi
 \next@}
\def\fontstyle@#1#2{\expandafter
 \ifx\csname\exstring@#1@F#2\endcsname\relax
  \DN@{\Err@{\string\fontstyle\string#1 can't be followed by \string#2}}%
 \else
  \DN@##1{{\csname\exstring@#1@F#2\endcsname##1}}%
 \fi
 \next@}
\def\Reset#1{\expandafter
 \ifx\csname\exstring@#1@C\endcsname\relax
  \expandafter\ifx\csname\exstring@#1@C1\endcsname\relax
   \DN@{\Err@{\noexpand\Reset can't be used with \string#1}}%
  \else
   \DN@{\Reset@#1}%
  \fi
 \else
  \DN@##1{\count@##1\relax\ifx#1\page\else\advance\count@\m@ne\fi
   \global\csname\exstring@#1@C\endcsname\count@}%
 \fi
 \next@}
\def\Reset@#1#2{\expandafter
 \ifx\csname\exstring@#1@C#2\endcsname\relax
  \DN@{\Err@{\string\Reset\string#1 can't be followed by \string#2}}%
 \else
  \DN@##1{\count@##1\relax\advance\count@\m@ne
   \global\csname\exstring@#1@C#2\endcsname\count@}%
 \fi
 \next@}
\def\Offset#1{\expandafter
 \ifx\csname\exstring@#1@C\endcsname\relax
  \expandafter\ifx\csname\exstring@#1@C1\endcsname\relax
   \DN@{\Err@{\noexpand\Offset can't be used with \string#1}}%
  \else
   \DN@{\Offset@#1}%
  \fi
 \else
  \DN@##1{\count@##1\relax\advance\count@\m@ne\global\advance
   \csname\exstring@#1@C\endcsname\count@}%
 \fi
 \next@}
\def\Offset@#1#2{\expandafter
 \ifx\csname\exstring@#1@C#2\endcsname\relax
  \DN@{\Err@{\string\Offset\string#1 can't be followed by \string#2}}%
 \else
  \DN@##1{\count@##1\relax\advance\count@\m@ne
   \global\advance\csname\exstring@#1@C#2\endcsname\count@}%
 \fi
 \next@}
\def\getR@#1#2{\def\nextiv@{\let\nextiii@}\expandafter\nextiv@
 \csname\exstring@#1@R#2\endcsname}
\def\letR@#1#2#3{\expandafter\let\csname#1@#3#2\endcsname\Next@}
\def\letR@@#1#2{\expandafter\let\csname\exstring@#1@#2\endcsname\Next@}
\def\newpre#1{\expandafter
 \ifx\csname\exstring@#1@P\endcsname\relax
  \expandafter\ifx\csname\exstring@#1@P1\endcsname\relax
   \DN@{\Err@{\noexpand\newpre can't be used with \string#1}}%
  \else
   \DN@{\newpre@#1}%
  \fi
 \else
  \DN@{%
   \DNii@{%
    \endgroup
    \expandafter\let\csname\exstring@#1@P\endcsname\Next@
    \expandafter\ifx\csname\exstring@#1@R\endcsname\relax\else
    \getR@#1{}\expandafter\letR@\nextiii@ P\fi
    }%
   \begingroup\noexpands@\afterassignment\nextii@\xdef\Next@}%
 \fi
 \next@}
\def\newpre@#1#2{\expandafter
 \ifx\csname\exstring@#1@P#2\endcsname\relax
  \DN@{\Err@{\string\newpre\string#1 can't be followed by \string#2}}%
 \else
  \DN@{%
   \DNii@{%
    \endgroup
    \expandafter\let\csname\exstring@#1@P#2\endcsname\Next@
    \expandafter\ifx\csname\exstring@#1@R#2\endcsname\relax\else
    \getR@#1{#2}\expandafter\letR@@\nextiii@ P\fi
    }%
   \begingroup\noexpands@\afterassignment\nextii@\xdef\Next@}%
 \fi
 \next@}
\def\newpost#1{\expandafter
 \ifx\csname\exstring@#1@Q\endcsname\relax
  \expandafter\ifx\csname\exstring@#1@Q1\endcsname\relax
   \DN@{\Err@{\noexpand\newpost can't be used with \string#1}}%
  \else
   \DN@{\newpost@#1}%
  \fi
 \else
  \DN@{%
   \DNii@{%
    \endgroup
    \expandafter\let\csname\exstring@#1@Q\endcsname\Next@
    \expandafter\ifx\csname\exstring@#1@R\endcsname\relax\else
    \getR@#1{}\expandafter\letR@\nextiii@ Q\fi
    }%
   \begingroup\noexpands@\afterassignment\nextii@\xdef\Next@}%
 \fi
 \next@}
\def\newpost@#1#2{\expandafter
 \ifx\csname\exstring@#1@Q#2\endcsname\relax
  \DN@{\Err@{\string\newpost\string#1 can't be followed by \string#2}}%
 \else
  \DN@{%
   \DNii@{%
    \endgroup
    \expandafter\let\csname\exstring@#1@Q#2\endcsname\Next@
    \expandafter\ifx\csname\exstring@#1@R#2\endcsname\relax\else
    \getR@#1{#2}\expandafter\letR@@\nextiii@ Q\fi
    }%
   \begingroup\noexpands@\afterassignment\nextii@\xdef\Next@}%
 \fi
 \next@}
\def\newstyle#1{\expandafter
 \ifx\csname\exstring@#1@S\endcsname\relax
  \expandafter\ifx\csname\exstring@#1@S1\endcsname\relax
   \DN@{\Err@{\noexpand\newstyle can't be used
    with \string#1}}%
  \else
   \DN@{\newstyle@#1}%
  \fi
 \else
  \DN@{%
   \DNii@{%
    \expandafter\let\csname\exstring@#1@S\endcsname\Next@
    \expandafter\ifx\csname\exstring@#1@R\endcsname\relax\else
    \getR@#1{}\expandafter\letR@\nextiii@ S\fi
    }%
   \afterassignment\nextii@\gdef\Next@}%
 \fi
 \next@}
\def\newstyle@#1#2{\expandafter
 \ifx\csname\exstring@#1@S#2\endcsname\relax
  \DN@{\Err@{\string\newstyle\string#1 can't be followed by
   \string#2}}%
 \else
  \DN@{%
   \DNii@{%
    \expandafter\let\csname\exstring@#1@S#2\endcsname\Next@
    \expandafter\ifx\csname\exstring@#1@R#2\endcsname\relax\else
    \getR@#1{#2}\expandafter\letR@@\nextiii@ S\fi
    }%
   \afterassignment\nextii@\gdef\Next@}%
 \fi
 \next@}
\def\newnumstyle#1{\expandafter
 \ifx\csname\exstring@#1@N\endcsname\relax
  \expandafter\ifx\csname\exstring@#1@N1\endcsname\relax
   \DN@{\Err@{\noexpand\newnumstyle can't be used with
    \string#1}}%
  \else
   \DN@{\newnumstyle@#1}%
  \fi
 \else
  \DN@##1{%
   \gdef\Next@{##1}%
    \expandafter\let\csname\exstring@#1@N\endcsname\Next@
    \expandafter\ifx\csname\exstring@#1@R\endcsname\relax\else
    \getR@#1{}\expandafter\letR@\nextiii@ N\fi
    }%
 \fi
 \next@}
\def\newnumstyle@#1#2{\expandafter
 \ifx\csname\exstring@#1@N#2\endcsname\relax
  \DN@{\Err@{\string\newnumstyle\string#1 can't be followed by
   \string#2}}%
 \else
  \DN@##1{%
   \gdef\Next@{##1}%
    \expandafter\let\csname\exstring@#1@N#2\endcsname\Next@
    \expandafter\ifx\csname\exstring@#1@R#2\endcsname\relax\else
    \getR@#1{#2}\expandafter\letR@@\nextiii@ N\fi
    }%
  \fi
 \next@}
\def\newfontstyle#1{\expandafter
 \ifx\csname\exstring@#1@F\endcsname\relax
  \expandafter\ifx\csname\exstring@#1@F1\endcsname\relax
   \DN@{\Err@{\noexpand\newfontstyle can't be used with
    \string#1}}%
  \else
   \DN@{\newfontstyle@#1}%
  \fi
 \else
  \DN@##1{%
   \gdef\Next@{##1}%
    \expandafter\let\csname\exstring@#1@F\endcsname\Next@
    \expandafter\ifx\csname\exstring@#1@R\endcsname\relax\else
    \getR@#1{}\expandafter\letR@\nextiii@ F\fi
    }%
 \fi
 \next@}
\def\newfontstyle@#1#2{\expandafter
 \ifx\csname\exstring@#1@F#2\endcsname\relax
  \DN@{\Err@{\string\newfontstyle\string#1 can't be followed by
   \string#2}}%
 \else
  \DN@##1{%
   \gdef\Next@{##1}%
    \expandafter\let\csname\exstring@#1@F#2\endcsname\Next@
    \expandafter\ifx\csname\exstring@#1@R#2\endcsname\relax\else
    \getR@#1{#2}\expandafter\letR@@\nextiii@ F\fi
    }%
 \fi
 \next@}
\def\word#1{\expandafter
 \ifx\csname\exstring@#1@W\endcsname\relax
  \expandafter\ifx\csname\exstring@#1@W1\endcsname\relax
   \DN@{\Err@{\noexpand\word can't be used with \string#1}}%
  \else
   \DN@{\word@#1}%
  \fi
 \else
  \DN@{{\csname\exstring@#1@W\endcsname}}%
 \fi
 \next@}
\def\word@#1#2{\expandafter
 \ifx\csname\exstring@#1@W#2\endcsname\relax
  \DN@{\Err@{\string\word\noexpand#1can't be followed by \string#2}}%
 \else
  \DN@{{\csname\exstring@#1@W#2\endcsname}}%
 \fi
 \next@}
\def\newword#1{\expandafter
 \ifx\csname\exstring@#1@W\endcsname\relax
  \expandafter\ifx\csname\exstring@#1@W1\endcsname\relax
   \DN@{\Err@{\noexpand\newword can't be used  with \string#1}}%
  \else
   \DN@{\newword@#1}%
  \fi
 \else
  \DN@{%
   \DNii@{%
    \expandafter\let\csname\exstring@#1@W\endcsname\Next@
    \expandafter\ifx\csname\exstring@#1@R\endcsname\relax\else
     \getR@#1{}\expandafter\letR@\nextiii@ W\fi
    }%
   \afterassignment\nextii@\gdef\Next@}%
 \fi
 \next@}
\def\newword@#1#2{\expandafter
 \ifx\csname\exstring@#1@W#2\endcsname\relax
  \DN@{\Err@{\string\newword\noexpand#1can't be followed by \string#2}}%
 \else
  \DN@{%
   \DNii@{%
    \expandafter\let\csname\exstring@#1@W#2\endcsname\Next@
    \expandafter\ifx\csname\exstring@#1@R#2\endcsname\relax\else
     \getR@#1{#2}\expandafter\letR@@\nextiii@ W\fi
    }%
   \afterassignment\nextii@\gdef\Next@}%
 \fi
 \next@}
\newif\iffn@
\newcount\footmark@C
\footmark@C\z@
\def\footmark@S#1{$^{#1}$}
\let\footmark@N\arabic
\def\footmark@F{\rm}
\def\foottext@S#1{$^{#1}$}
\def\foottext@F{\rm}
\let\modifyfootnote@\relax
\def\modifyfootnote#1{\def\modifyfootnote@{#1}}
\def\vfootnote@#1{\insert\footins
 \bgroup
 \floatingpenalty\@MM\interlinepenalty\interfootnotelinepenalty
 \leftskip\z@\rightskip\z@\spaceskip\z@\xspaceskip\z@
 \rm\splittopskip\ht\strutbox\splitmaxdepth\dp\strutbox
 \locallabel@\noindent@@{\foottext@F#1}\modifyfootnote@
 \footstrut\FN@\fo@t}
\def\fo@t{\ifcat\bgroup\noexpand\next\expandafter\f@@t\else
 \expandafter\f@t\fi}
\def\f@t#1{#1\@foot}
\def\f@@t{\bgroup\aftergroup\@foot\afterassignment\FNSSP@\let\next@}
\def\@foot{\unskip\lower\dp\strutbox\vbox to\dp\strutbox{}\egroup
 \iffn@\expandafter\fn@false\else
 \expandafter\postvanish@\fi}
\newif\ifplainfn@
\plainfn@true
\def\fancyfootnotes{\plainfn@false}
\newcount\fancyfootmarkcount@
\fancyfootmarkcount@\z@
\newcount\lastfnpage@
\lastfnpage@-\@M
\let\justfootmarklist@\empty
\def\footmark{\let\@sf\empty
 \ifhmode\edef\@sf{\spacefactor\the\spacefactor}\/\fi
 \DN@{\ifx"\next\expandafter\nextii@\else\expandafter\footmark@\fi}%
 \DNii@"##1"{%
  \iffirstchoice@
   {\let\style\footmark@S\let\numstyle\footmark@N
   \footmark@F##1%
   \noexpands@
   \let\style\foottext@S
   \Qlabel@{##1}%
   }%
   \iffn@\else
    {\noexpands@
    \xdef\Next@{{\Thelabel@}{\Thelabel@@}{\Thelabel@@@}{\Thelabel@@@@}}%
    }%
    \expandafter\rightappend@\Next@\to\justfootmarklist@
   \fi
  \fi
  \@sf\relax}%
 \FN@\next@}
\def\footmark@{%
 \iffirstchoice@
  \global\advance\footmark@C\@ne
  \ifplainfn@
   \xdef\adjustedfootmark@{\number\footmark@C}%
  \else
   {\let\\\or\xdef\Next@{\ifcase\number\footmark@C\fnpages@\else
     -\@M\fi}}%
   \ifnum\Next@=-\@M
    \xdef\adjustedfootmark@{\number\footmark@C}%
   \else
    \ifnum\Next@=\lastfnpage@
     \global\advance\fancyfootmarkcount@\@ne
    \else
     \global\fancyfootmarkcount@\@ne
     \global\lastfnpage@\Next@
    \fi
    \xdef\adjustedfootmark@{\number\fancyfootmarkcount@}%
   \fi
  \fi
  {\noexpands@
  \xdef\Thelabel@@@{\adjustedfootmark@}%
  \xdefThelabel@\footmark@N
  \xdef\Thelabel@@@@{\Thelabel@}%
  \xdefThelabel@@\foottext@S
  }%
  \iffn@\else
   {\noexpands@
   \xdef\Next@{{\Thelabel@}{\Thelabel@@}{\Thelabel@@@}{\Thelabel@@@@}}%
   }%
   \expandafter\rightappend@\Next@\to\justfootmarklist@
  \fi
  \ifplainfn@
  \else
   \edef\next@{\write\laxwrite@{F\noexpand\the\pageno}}\next@
  \fi
 \fi
 \footmark@S{\footmark@N{\adjustedfootmark@}}%
 \@sf\relax}
\def\foottext{\prevanish@
 \ifx\justfootmarklist@\empty
  \Err@{There is no \noexpand\footmark for this \string\foottext}\fi
 \DN@\\##1##2\next@{\DN@{##1}\gdef\justfootmarklist@{##2}}%
 \expandafter\next@\justfootmarklist@\next@
 \expandafter\foottext@\next@}
\def\foottext@#1#2#3#4{{\noexpands@
  \xdef\Thelabel@{#1}\xdef\Thelabel@@{#2}%
  \xdef\Thelabel@@@{#3}\xdef\Thelabel@@@@{#4}}%
  \vfootnote@{\thelabel@@}}
\rightadd@\foottext\to\vanishlist@
\def\footnote{\fn@true
 \let\@sf\empty
 \ifhmode\edef\@sf{\spacefactor\the\spacefactor}\/\fi
 \DN@{\ifx"\next\expandafter\nextii@\else\expandafter\nextiii@\fi}%
 \DNii@"##1"{\footmark"##1"\vfootnote@{\let\style\foottext@S
  \let\numstyle\footmark@N##1}}%
 \def\nextiii@{\footmark\vfootnote@{\foottext@S{\footmark@N
  {\adjustedfootmark@}}}}%
 \FN@\next@}
\newdimen\litindent
\litindent20\p@
\newbox\litbox@
\newbox\Litbox@
\newcount\interlitpenalty@
\interlitpenalty@\@M
\newcount\litlines@
{\obeyspaces\gdef\defspace@{\def {\allowbreak\hskip.5emminus.15em}}}
{\obeylines\gdef\letM@{\let^^M\CtrlM@}}
\def\CtrlM@{\egroup
 \ifcase\litlines@\advance\litlines@\@ne\or
 \box\litbox@\advance\litlines@\@ne\else
 \penalty\interlitpenalty@\box\litbox@\fi
 \Lit@}
\def\Lit@{\setbox\litbox@\hbox\bgroup\litdefs@\hskip\litindent}
\newcount\littab@
\littab@8
\def\littab#1{\littab@#1\relax}
{\catcode`\^^I=\active\gdef\letTAB@{\let^^I\TAB@}}
\def\TAB@{\egroup
 \dimen@\wd\litbox@
 \advance\dimen@-\litindent
 \setboxz@h{\tt0}%
 \dimen@ii\littab@\wdz@
 \divide\dimen@\dimen@ii
 \multiply\dimen@\dimen@ii
 \advance\dimen@\littab@\wdz@
 \advance\dimen@\litindent
 \setbox\litbox@\hbox\bgroup\litdefs@\hbox to\dimen@{\unhbox\litbox@\hfil}}
{\catcode`\`=\active\gdef`{\relax\lq}}
\let\litbs@\relax
\let\litbs@@\relax
\def\litbackslash#1{%
 \edef\litbs@{\catcode`\string#1=\z@
 \def\noexpand\litbs@@{\def\expandafter\noexpand\csname\string#1\endcsname
  {\char`\string#1}}}}
\def\litcodes@{\catcode`\\=12
 \catcode`\{=12 \catcode`\}=12
 \catcode`\$=12 \catcode`\&=12
 \catcode`\#=12
 \catcode`\^=12 \catcode`\_=12
 \catcode`\@=12 \catcode`\~=12 \catcode`\"=12
 \catcode`\;=12 \catcode`\:=12 \catcode`\!=12 \catcode`\?=12
 \catcode`\%=12 \litbs@\catcode`\`=\active\obeyspaces\defspace@}
\def\activate@#1#2{{\lccode`\~=`#2%
 \lowercase{%
  \if0#1%
  \gdef\Next@{\def~{\egroup\endgroup\bigskip\vskip-\parskip
   \def\next@{\noindent@@\FN@\pretendspace@}\FNSS@\next@}}\else
  \gdef\Next@{\def~{\egroup\egroup\endgroup}}\fi
  }%
 }}
\def\litdefs@{\let\0\empty\let\1\litdelim@\def\ {\char32 }\litbs@@}%
\def\litdelimiter#1{%
 \edef\litdelim@{\char`#1}%
 \def\lit#1{\leavevmode\begingroup\litcodes@\litdefs@
  \tt\hyphenchar\tentt\m@ne\lit@}%
 \def\lit@##1#1{##1\endgroup\null}%
 \def\Lit#1{\ifhmode$$\abovedisplayskip\bigskipamount
  \abovedisplayshortskip\bigskipamount
  \belowdisplayskip\z@\belowdisplayshortskip\z@
  \postdisplaypenalty\@M
  $$\vskip-\baselineskip\else\bigskip\fi
  \begingroup\litlines@\z@
  \catcode`#1=\active\activate@0#1\Next@
  \def\displaybreak{\egroup\break\litlines@\z@\Lit@}%
  \def\allowdisplaybreak{\egroup\allowbreak\litlines@\z@\Lit@}%
  \def\allowdisplaybreaks{\egroup\allowbreak\interlitpenalty@\z@
   \litlines@\z@\Lit@}%
  \litcodes@\tt\catcode`\^^I=\active\letTAB@
  \obeylines\letM@\Lit@}%
 \def\Litbox##1=#1{\begingroup\ifodd##1\relax\aftergroup\global\fi
  \aftergroup\setbox\aftergroup##1\aftergroup\box\aftergroup\Litbox@
  \def\allowdisplaybreak{\egroup\allowbreak\litlines@\z@\Lit@}%
  \def\allowdisplaybreaks{\egroup\allowbreak\interlitpenalty@\z@
   \litlines@\z@\Lit@}%
  \catcode`#1=\active\activate@1#1\Next@
  \litcodes@\tt\catcode`\^^I=\active\letTAB@
  \obeylines\letM@\global\setbox\Litbox@\vbox\bgroup\litindent\z@%
  \litlines@\z@\Lit@}%
}
\newbox\titlebox@
\setbox\titlebox@\vbox{}
\rightadd@\title\to\overlonglist@
\def\title{\begingroup\Let@
 \global\setbox\titlebox@\vbox\bgroup\tabskip\hss@
 \halign to\hsize\bgroup\bf\hfil\ignorespaces##\unskip\hfil\cr}
\def\endtitle{\crcr\egroup\egroup\endgroup\overlong@false}
\newbox\authorbox@
\rightadd@\author\to\overlonglist@
\def\author{\begingroup\Let@
 \global\setbox\authorbox@\vbox\bgroup\tabskip\hss@
 \halign to\hsize\bgroup\rm\hfil\ignorespaces##\unskip\hfil\cr}
\def\endauthor{\crcr\egroup\egroup\endgroup\overlong@false}
\newbox\affilbox@
\def\affil{\begingroup\Let@
 \global\setbox\affilbox@\vbox\bgroup\tabskip\hss@
 \halign to\hsize\bgroup\rm\hfil\ignorespaces##\unskip\hfil\cr}%
\def\endaffil{\crcr\egroup\egroup\endgroup\overlong@false}
\let\date@\relax
\def\date#1{\gdef\date@{\ignorespaces#1\unskip}}
\def\today{\ifcase\month\or January\or February\or March\or April\or May\or
 June\or July\or August\or September\or October\or November\or December\fi
 \space\number\day, \number\year}
\def\maketitle{\hrule\height\z@\vskip-\topskip
 \vskip24\p@ plus12\p@ minus12\p@
 \unvbox\titlebox@
 \ifvoid\authorbox@\else\vskip12\p@ plus6\p@ minus3\p@\unvbox\authorbox@\fi
 \ifvoid\affilbox@\else\vskip10\p@ plus5\p@ minus2\p@\unvbox\affilbox@\fi
 \ifx\date@\relax\else\vskip6\p@ plus2\p@ minus\p@\centerline{\rm\date@}\fi
 \vskip18\p@ plus12\p@ minus6\p@}
\def\cite{%
 \DNii@(##1)##2{{\rm[}{##2}, {##1\/}{\rm]}}%
 \def\nextiii@##1{{\rm[}{##1\/}{\rm]}}%
 \DN@{\ifx\next(\expandafter\nextii@\else\expandafter\nextiii@\fi}%
 \FN@\next@}
\def\makebib@W{Bibliography}
\def\makebib{\begingroup\rm\bigbreak\centerline{\smc\makebib@W}%
 \nobreak\medskip
 \sfcode`\.=\@m\everypar{}\parindent\z@
 \def\nopunct{\nopunct@true}\def\nospace{\nospace@true}%
 \nopunct@false\nospace@false
 \def\lkerns@{\null\kern\m@ne sp\kern\@ne sp}%
 \def\nkerns@{\null\kern-\tw@ sp\kern\tw@ sp}%
}

\newif\ifnoprepunct@
\newif\ifnoprespace@
\newif\ifnoquotes@
\def\noprepunct{\noprepunct@true}
\def\noprespace{\noprespace@true}
\def\noquotes{\noquotes@true}
\newbox\nobox@
\newbox\keybox@
\newbox\bybox@
\newbox\paperbox@
\newbox\paperinfobox@
\newbox\jourbox@
\newbox\volbox@
\newbox\issuebox@
\newbox\yrbox@
\newbox\pgbox@
\newbox\ppbox@
\newbox\bookbox@
\newbox\inbookbox@
\newbox\bookinfobox@
\newbox\publbox@
\newbox\publaddrbox@
\newbox\edbox@
\newbox\edsbox@
\newbox\langbox@
\newbox\translbox@
\newbox\finalinfobox@
\def\setbibinfo@#1{\edef\next@{\ifnopunct@1\else0\fi
 \ifnospace@1\else0\fi\ifnoprepunct@1\else0\fi\ifnoprespace@1\else0\fi
 \ifnoquotes@1\else0\fi}%
 \DNii@{00000}%
 \ifx\next@\nextii@\else\xdef\bibinfo@{\bibinfo@\the#1,\next@}%
 \fi}
\def\getbibinfo@#1{\ifx\bibinfo@\empty
 \let\next@0\let\nextii@0\let\nextiii@0\let\nextiv@0\let\nextv@0\else
 \edef\next@{\def
  \noexpand\next@####1\the#1,####2####3####4####5####6####7\noexpand\next@
  {\let\noexpand\next@####2\let\noexpand\nextii@####3%
  \let\noexpand\nextiii@####4\let\noexpand\nextiv@####5%
  \let\noexpand\nextv@####6}%
  \noexpand\next@\bibinfo@\the#1,00000\noexpand\next@}\next@
 \fi}
\newif\ifbookinquotes@
\def\bookinquotes{\bookinquotes@true}
\newif\ifpaperinquotes@
\def\paperinquotes{\paperinquotes@true}
\newif\ifininbook@
\def\ininbook{\ininbook@true}
\newif\ifopenquotes@
\def\closequotes@{\ifopenquotes@''\openquotes@false\fi}
\newif\ifbeginbib@
\newif\ifendbib@
\newif\ifprevjour@
\newif\ifprevbook@
\newdimen\bibindent@
\bibindent@20\p@
\def\bib{\global\let\bibinfo@\empty\global\let\translinfo@\relax\beginbib@true
 \begingroup\noindent@
 \hangindent\bibindent@\hangafter\@ne\bib@}
\def\v@id#1{\setbox#1\box\voidb@x}
\def\bib@{\v@id\nobox@\v@id\keybox@\v@id\bybox@\v@id\paperbox@
 \v@id\paperinfobox@\v@id\jourbox@\v@id\volbox@\v@id\issuebox@
 \v@id\yrbox@\v@id\pgbox@\v@id\ppbox@\v@id\bookbox@\v@id\inbookbox@
 \v@id\bookinfobox@\v@id\publbox@\v@id\publaddrbox@\v@id\edbox@
 \v@id\edsbox@\v@id\langbox@\v@id\translbox@\v@id\finalinfobox@
 \bgroup}
\def\Setnonemptybox@#1#2{\unskip\setbibinfo@#1\egroup#2%
 \def\aftergroup@{\ifdim\wd#1=\z@\setbox#1\box\voidb@x\fi}%
 \setbox#1\vbox\bgroup\aftergroup\aftergroup@\hsize\maxdimen\leftskip\z@
 \rightskip\z@\hbadness\@M\hfuzz\maxdimen\noindent}
\def\setnonemptybox@#1{\Setnonemptybox@#1\relax}
\def\no{\setnonemptybox@\nobox@}
\def\key{\setnonemptybox@\keybox@\bf}
\def\by{\setnonemptybox@\bybox@}
\def\bysame{\setnonemptybox@\bybox@\leaders\hrule\hskip3em\null}
\def\paper{\setnonemptybox@\paperbox@
 \ifpaperinquotes@\getbibinfo@\paperbox@
 \if\nextv@1\else``\fi\else\it\fi}
\def\paperinfo{\setnonemptybox@\paperinfobox@}
\def\jour{\Setnonemptybox@\jourbox@\prevjour@true}
\def\vol{\setnonemptybox@\volbox@\bf}
\def\issue{\setnonemptybox@\issuebox@}
\def\yr{\setnonemptybox@\yrbox@}

\def\pg{\setnonemptybox@\pgbox@}
\def\pp{\setnonemptybox@\ppbox@}
\def\book{\Setnonemptybox@\bookbox@\prevbook@true
 \ifbookinquotes@\getbibinfo@\bookbox@
 \if\nextv@1\else``\fi\else\it\fi}
\def\inbook{\Setnonemptybox@\inbookbox@\prevbook@true
 \ifininbook@ in \fi\ifbookinquotes@\getbibinfo@\inbookbox@
 \if\nextv@1\else``\fi\fi}
\def\bookinfo{\setnonemptybox@\bookinfobox@}
\def\publ{\setnonemptybox@\publbox@}
\def\publaddr{\setnonemptybox@\publaddrbox@}
\def\ed{\setnonemptybox@\edbox@}
\def\eds{\setnonemptybox@\edsbox@}
\def\lang{\setnonemptybox@\langbox@}
\def\finalinfo{\setnonemptybox@\finalinfobox@}
\def\setboxzl@{\setbox\z@\lastbox}
\def\getbox@#1{\setbox\z@\vbox{\vskip-\@M\p@
 \unvbox#1%
 \setboxzl@
 \global\setbox\@ne\hbox{\unhbox\z@\unskip\unskip\unpenalty}%
 \ifdim\lastskip=-\@M\p@\else
 \loop\ifdim\lastskip=-\@M\p@
 \else\unskip\unpenalty\setboxzl@
 \global\setbox\@ne\hbox{\unhbox\z@\unhbox\@ne}%
 \repeat\fi}%
 \unhbox\@ne}
\def\adjustpunct@#1{\count@\lastkern
 \ifnum\count@=\z@#1\closequotes@\else
 \ifnum\count@>\tw@#1\closequotes@\else
 \ifnum\count@<-\tw@#1\closequotes@\else
  \unkern\unkern\setboxzl@
  \skip@\lastskip\unskip
  \count@@\lastpenalty\unpenalty
  \ifnum\count@=\tw@\unskip\setboxzl@\fi
  \ifdim\skip@=\z@\else\hskip\skip@\fi
  #1\closequotes@
  \ifnum\count@=\tw@\null\hfill\fi
  \penalty\count@@
 \fi\fi\fi}
\def\prepunct@#1#2{\getbibinfo@#2%
 \ifnopunct@
 \else
  \if\nextiii@0\adjustpunct@#1\fi
 \fi
 \closequotes@
 \ifnospace@
 \else
  \if\nextiv@0\space\else\fi
 \fi
 \nopunct@false\nospace@false
 \if\next@1\nopunct@true\fi
 \if\nextii@1\nospace@true\fi}
\def\ppunbox@#1#2{\prepunct@{#1}#2%
 \getbox@#2}
\let\semicolon@;
\def\endbib@{%
 \ifbeginbib@
  \ifvoid\nobox@
   \ifvoid\keybox@\else\hbox to\bibindent@{[\getbox@\keybox@]\hss}\fi
  \else\hbox to\bibindent@{\hss\getbox@\nobox@. }\fi
  \ifvoid\bybox@\else\getbox@\bybox@\fi
 \else
  \nopunct@true
  \ifvoid\bybox@\else\ppunbox@\relax\bybox@\fi
 \fi
 \ifvoid\translbox@\else\ppunbox@,\translbox@\fi
 \ifvoid\paperbox@\else\ppunbox@,\paperbox@\ifpaperinquotes@
  \if\nextv@1\else\openquotes@true\fi\fi
 \fi
 \ifvoid\paperinfobox@\else\ppunbox@,\paperinfobox@\fi
 \test@false
 \ifvoid\jourbox@\else\test@true\ppunbox@,\jourbox@\fi
 \ifprevjour@\test@true\fi
 \iftest@
  \ifvoid\volbox@\else\ppunbox@\relax\volbox@\fi
  \ifvoid\issuebox@
   \else\prepunct@\relax\issuebox@ no.~\getbox@\issuebox@\fi
  \ifvoid\yrbox@\else\prepunct@\relax\yrbox@(\getbox@\yrbox@)\fi
  \ifvoid\ppbox@\else\ppunbox@,\ppbox@\fi
  \ifvoid\pgbox@\else\prepunct@,\pgbox@ p.~\getbox@\pgbox@\fi
 \fi
 \test@false
 \ifvoid\bookbox@\else\test@true\ppunbox@,\bookbox@\ifbookinquotes@
  \if\nextv@1\else\openquotes@true\fi\fi\fi
 \ifvoid\inbookbox@\else\test@true\ppunbox@,\inbookbox@\ifbookinquotes@
  \if\nextv@1\else\openquotes@true\fi\fi\fi
 \ifprevbook@\test@true\fi
 \iftest@
  \ifvoid\edbox@\else\prepunct@\relax\edbox@(\getbox@\edbox@, ed.)\fi
  \ifvoid\edsbox@\else\prepunct@\relax\edsbox@(\getbox@\edsbox@, eds.)\fi
  \ifvoid\bookinfobox@\else\ppunbox@,\bookinfobox@\fi
  \ifvoid\publbox@\else\ppunbox@,\publbox@\fi
  \ifvoid\publaddrbox@\else\ppunbox@,\publaddrbox@\fi
  \ifvoid\yrbox@\else\ppunbox@,\yrbox@\fi
  \ifvoid\ppbox@\else\prepunct@,\ppbox@ pp.~\getbox@\ppbox@\fi
  \ifvoid\pgbox@\else\prepunct@,\pgbox@ p.~\getbox@\pgbox@\fi
 \fi
 \ifvoid\finalinfobox@
  \ifendbib@
   \ifnopunct@\else.\closequotes@\fi
  \else
  \ifvoid\langbox@\else\space(\getbox@\langbox@)\fi
   \/\semicolon@\closequotes@
  \fi
 \else
  \ifendbib@
   \ppunbox@{.\spacefactor3000\relax}\finalinfobox@
    \ifnopunct@\else.\fi
  \else
   \ppunbox@,\finalinfobox@\/\semicolon@\fi
 \fi
 \ifvoid\langbox@\else\space(\getbox@\langbox@)\fi
}
\def\endbib{\unskip\egroup\endbib@true\endbib@\par\endgroup}
\def\morebib{\unskip\egroup
 \endbib@false\endbib@
 \global\let\bibinfo@\empty\beginbib@false
 \bib@}
\def\anotherbib{\unskip\egroup
 \endbib@false\endbib@
 \global\let\bibinfo@\empty\beginbib@false
 \prevjour@false\prevbook@false\bib@}
\def\transl{\unskip
 \xdef\translinfo@{\the\translbox@,\ifnopunct@1\else0\fi
 \ifnospace@1\else0\fi\ifnoprepunct@1\else0\fi\ifnoprespace@1\else0\fi0}%
 \egroup\endbib@false\endbib@
 \global\let\bibinfo@\translinfo@\beginbib@false
 \bib@
 \egroup
 \def\aftergroup@{\ifdim\wd\translbox@=\z@\setbox\translbox@\box\voidb@x\fi}%
 \setbox\translbox@\vbox\bgroup\aftergroup\aftergroup@
 \hsize\maxdimen\leftskip\z@\rightskip\z@\hbadness\@M\hfuzz\maxdimen
 \noindent}
\newwrite\auxwrite@
\newread\bbl@
\def\UseBibTeX{\immediate\openout\auxwrite@=\jobname.aux
 \let\cite\BTcite@
 \def\nocite##1{\immediate\write\auxwrite@{\string\citation{##1}}}%
 \def\bibliographystyle##1{\immediate\write\auxwrite@{\string
  \bibstyle{##1}}}%
 \def\bibliography@W{Bibliography}%
 \def\bibliography##1{\immediate\write\auxwrite@{\string\bibdata{##1}}%
  \immediate\openin\bbl@=\jobname.bbl
  \ifeof\bbl@
   \W@{No .bbl file}%
  \else
   \immediate\closein\bbl@
   \begingroup\input bibtex \input\jobname.bbl \endgroup
  \fi}%
 }
\def\BTcite@{%
 \DNii@(##1)##2{{\rm[}\BTcite@@##2,\BTcite@@{\rm, }{##1\/}{\rm]}%
  \immediate\write\auxwrite@{\string\citation{##2}}}%
 \def\nextiii@##1{{\rm[}\BTcite@@##1,\BTcite@@\/{\rm]}%
  \immediate\write\auxwrite@{\string\citation{##1}}}%
 \DN@{\ifx\next(\expandafter\nextii@\else\expandafter\nextiii@\fi}%
 \FN@\next@}%
\def\BTcite@@#1,{\BTcite@@@{#1}\FN@\BTcite@@@@}
\def\BTcite@@@@{\ifx\next\BTcite@@
 \expandafter\eat@\else{\rm, }\expandafter\BTcite@@\fi}
\catcode`\~=11
\def\BTcite@@@#1{\nolabel@\cite{#1}\relax
 \DNii@##1~##2\nextii@{##1}%
 \csL@{#1}\expandafter\nextii@\Next@\nextii@\fi}
\catcode`\~=\active

\def\beginthebibliography@#1{\rm\setboxz@h{#1\ }\bibindent@\wdz@
 \bigbreak\centerline{\smc\bibliography@W}\nobreak\medskip
 \sfcode`\.=\@m\everypar{}\parindent\z@}

\newif\iffigproofing@
\def\Figureproofing{\figproofing@true}
\def\noFigureproofing{\figproofing@false}
\newif\ifHby@
\def\Hbyw#1{\global\Hby@true\hbyw\vsize{#1}}
\def\hbyw#1#2{%
 \hbox{%
  \ifHby@
  \else
   \iffigproofing@
    \setbox\z@\vbox{\hrule\width5\p@}\ht\z@\z@
    \vbox to#1{\hrule\height5\p@\width.4\p@\vfil\hrule\height5\p@\width.4\p@}%
    \kern-.4\p@\rlap{\copy\z@}\raise#1\hbox{\rlap{\copy\z@}}%
   \fi
  \fi
  \vbox to#1{\hbox to#2{}\vfil}%
  \ifHby@
  \else
   \iffigproofing@
    \vbox to#1{\hrule\height5\p@\width.4\p@\vfil\hrule\height5\p@\width.4\p@}%
    \kern-.4\p@\llap{\copy\z@}\raise#1\hbox{\llap{\boxz@}}%
   \fi
  \fi}}
\newcount\island@C
\let\island@P\empty
\let\island@Q\empty
\def\island@S#1{#1\null.}
\let\island@N\arabic
\def\island@F{\rm}
\def\island@@@P{\csname\exxx@\islandtype@ @P\endcsname}
\def\island@@@Q{\csname\exxx@\islandtype@ @Q\endcsname}
\def\island@@@S{\csname\exxx@\islandtype@ @S\endcsname}
\def\island@@@N{\csname\exxx@\islandtype@ @N\endcsname}
\def\island@@@F{\csname\exxx@\islandtype@ @F\endcsname}
\def\island@@@C{\csname island@C\islandclass@\endcsname}
\newif\ifplace@
\newif\ifisland@
\def\island{%
 \ifplace@
  \DN@{\let\islandclass@\empty\def\islandtype@{\island}\FN@\island@}%
 \else
  \long\DN@##1\endisland{\Err@{\noexpand\island must be used after some
   type of \string\...place}}%
 \fi
 \next@}
\def\island@{\ifx\next\c\let\next@\island@c\else
 \DN@{\FN@\island@@}\fi\next@}
\def\island@@{\ifcat\bgroup\noexpand\next\let\next@\island@@@\else
 \DN@{\Err@{\noexpand\island must be followed by a {prefix} for
 \string\caption's}}\fi\next@}
\newbox\islandbox@
\newcount\captioncount@
\def\island@@@#1{\def\captionprefix@{#1}\captioncount@\z@
 \global\setbox\islandbox@\vbox\bgroup}
\def\island@c\c#1{%
 \ifplace@
 \DN@{\def\islandclass@{#1}%
  \expandafter\ifx\csname island@C#1\endcsname\relax
  \expandafter\newcount@\csname island@C#1\endcsname
   \global\csname island@C#1\endcsname\z@\fi
  \FNSS@\island@c@}%
 \else
 \DN@{\edef\next@{\long\def\noexpand\next@########1\expandafter\noexpand
  \csname end\exxx@\islandtype@\endcsname{\noexpand\Err@{\noexpand\noexpand
  \expandafter\noexpand
  \islandtype@ must be used after some type of \noexpand\string
   \noexpand\...place}}}\next@\next@}%
 \fi
 \next@}
\def\island@c@{%
 \ifcat\bgroup\noexpand\next
  \let\next@\island@c@@
 \else
  \DN@{\Err@{\noexpand\island\string\c{\expandafter\string\islandclass@} must
   be followed by a {prefix} for \string\caption's}}%
 \fi\next@}
\def\island@c@@#1{\def\captionprefix@{#1}%
 \captioncount@\z@\global\setbox\islandbox@\vbox\bgroup}
\rightadd@\caption\to\nofrillslist@
\newbox\captionbox@
\newbox\Captionbox@
\def\caption{%
 \ifnum\captioncount@=\z@
  \ifnopunct@
   \DN@{\egroup\nopunct@true}%
  \else
   \let\next@\egroup
  \fi
 \else
  \let\next@\relax
 \fi
 \next@
 \advance\captioncount@\@ne
 \FN@\caption@}
\def\caption@{\ifx\next"\expandafter\caption@q\else\expandafter\caption@@\fi}
\def\caption@q"#1"{\quoted@true
 {\noexpands@
 \let\pre\island@@@P\let\post\island@@@Q
 \let\style\island@@@S\let\numstyle\island@@@N
 \Qlabel@{#1}\let\style\relax\xdef\Qlabel@@@@{#1}}%
 \finishcaption@}
\def\caption@@{\quoted@false
 \global\advance\island@@@C\@ne
 {\noexpands@
 \xdef\Thelabel@@@{\number\island@@@C}%
 \xdefThelabel@\island@@@N
 \xdef\Thelabel@@@@{\island@@@P\Thelabel@\island@@@Q}%
 \xdefThelabel@@\island@@@S
 \xdef\Thepref@{\Thelabel@@@@}}%
 \finishcaption@}
\long\def\captionformat@#1#2#3{\rm\strut#1 {\island@@@F#2} #3%
 \punct@.\strut}
\long\def\widerthanisland@#1#2#3{\test@true\setbox\z@\vbox{\hsize\maxdimen
 \noindent@@\captionformat@{#1}{#2}{#3}\par\setboxzl@}%
 \ifdim\wdz@=\z@
  \global\setbox\captionbox@\hbox{\noset@\unlabel@
   \captionformat@{#1}{#2}{#3}}%
  \ifdim\wd\captionbox@>\wd\islandbox@\else\test@false\fi
 \fi}
\long\def\captionformat@@#1#2#3{\widerthanisland@{#1}{#2}{#3}%
 \iftest@
  \global\setbox\captionbox@\vbox{\hsize\wd\islandbox@
   \vskip-\parskip\noindent@@\noset@\unlabel@
   \captionformat@{#1}{#2}{#3}\par}%
 \else
  \global\setbox\captionbox@
   \hbox to\wd\islandbox@{\hfil\box\captionbox@\hfil}%
 \fi}
\long\def\finishcaption@#1{\def\entry@{#1}%
 {\locallabel@
 \captionformat@@
  {\expandafter\ignorespaces\captionprefix@\unskip}%
  {\ifx\thelabel@@\empty\unskip\else\thelabel@@\fi}%
  {\ignorespaces#1\unskip}%
 \ifnum\captioncount@=\@ne
  \global\setbox\islandbox@\vbox{\ticwrite@\vbox{\box\islandbox@}}%
  \global\setbox\Captionbox@\vbox{\box\captionbox@}%
 \else
  \global\setbox\islandbox@\vbox{\unvbox\islandbox@\setboxzl@
   \ticwrite@\boxz@}%
  \global\setbox\Captionbox@\vbox{\unvbox\Captionbox@
   \smallskip\box\captionbox@}%
 \fi}%
 \nopunct@false\nospace@false\ignorespaces}
\def\Sixtic@{\ifx\macdef@\empty\else
 \DN@##1##2\next@{\def\macdef@{##1##2}}%
 \expandafter\next@\macdef@\next@
 \edef\next@
  {\noexpand\six@\tic@\macdef@
  \space\space\space\space\space\space\space\space\space\space\space\space
  \noexpand\six@}%
 \next@\let\macdef@\relax\fi}
\def\ticwrite@{%
 \iftoc@
  {\noexpands@\let\style\relax
  \DN@{\island}%
  \edef\next@{\write\tic@{%
   \ifnopunct@\noexpand\noexpand\noexpand\nopunct\fi
   \ifx\islandtype@\next@\noexpand\noexpand\noexpand\island
    \noexpand\string\noexpand\c{\islandclass@}{\captionprefix@}%
     {\QorThelabel@@@@}\else\noexpand\noexpand\expandafter\noexpand
     \islandtype@{\QorThelabel@@@@}}\fi}%
  \next@}%
  \expandafter\unmacro@\meaning\entry@\unmacro@
  \Sixtic@
  \write\tic@{\noexpand\Page{\number\pageno}{\page@N}{\page@P}{\page@Q}^^J}%
 \fi}
\def\Htrim@#1{%
 \ifHby@
  \dimen@\vsize
  \ifnum\captioncount@=\z@
  \else
   \advance\dimen@-\ht\Captionbox@
   \advance\dimen@-#1%
  \fi
  \global\Hby@false
  \dimen@ii\wd\islandbox@
  \global\setbox\islandbox@\vbox
   {\unvbox\islandbox@\setboxzl@
   \vbox to\z@{\vss\boxz@}\nointerlineskip\hbyw\dimen@\dimen@ii}%
  \global\Hby@true
 \fi}
\newif\ifdata@
\def\iclasstest@#1{\DN@{#1}\ifx\next@\islandclass@
 \test@true\else\test@false\fi}
\skipdef\skipi@=1
\def\endisland{\ifnum\captioncount@=\z@\expandafter\egroup\fi
 \ifdata@
 \else
  \iclasstest@{T}%
  \iftest@
   {\rm\global\skipi@-\dp\strutbox}\global\advance\skipi@\bigskipamount
   \Htrim@\skipi@
   \global\setbox\islandbox@\vbox
    {\ifnum\captioncount@=\z@\else
     \box\Captionbox@
     \nointerlineskip
     \vskip\skipi@\fi
     \box\islandbox@}%
  \else
   {\rm\global\skipi@\dp\strutbox}\global\advance\skipi@\medskipamount
   \Htrim@\skipi@
   \global\setbox\islandbox@\vbox
    {\box\islandbox@
     \ifnum\captioncount@=\z@\else
     \nointerlineskip
     \vskip\skipi@
     \box\Captionbox@
     \fi}%
  \fi
  \ifHby@
  \else
   \dimen@\ht\islandbox@\advance\dimen@\dp\islandbox@
   \ifdim\dimen@>\vsize
    \DN@{\island}%
    \Err@{%
     \ifx\islandtype@\next@\noexpand\island\else
      \expandafter\noexpand\islandtype@\fi
     \ifnum\captioncount@=\z@\else
       with \noexpand\caption\fi
      is larger than page}%
     \ht\islandbox@=\vsize
   \fi
  \fi
 \fi
 \global\Hby@false\island@true}
\def\newisland#1\c#2#3{\define#1{}%
 \iftoc@\immediate\write\tic@{\noexpand\newisland\noexpand#1%
  \string\c{#2}{#3}^^J}\fi
 \expandafter\def\csname\exstring@#1@S\endcsname{\island@S}%
 \expandafter\def\csname\exstring@#1@N\endcsname{\island@N}%
 \expandafter\def\csname\exstring@#1@P\endcsname{\island@P}%
 \expandafter\def\csname\exstring@#1@Q\endcsname{\island@Q}%
 \expandafter\def\csname\exstring@#1@F\endcsname{\island@F}%
 \expandafter\def\csname end\exstring@#1\endcsname{\endisland}%
 \expandafter
 \ifx\csname island@C#2\endcsname\relax
  \expandafter\newcount@\csname island@C#2\endcsname
  \global\csname island@C#2\endcsname\z@
 \fi
 \edef\next@{\noexpand\expandafter\noexpand\let\noexpand
  \csname\exstring@#1@C\noexpand\endcsname
  \csname island@C#2\endcsname}%
 \next@
 \def#1{\def\islandtype@{#1}\island@c\c{#2}{#3}}}
\newisland\Figure\c{F}{Figure}
\newisland\Table\c{T}{Table}
\newbox\islandboxi
\newbox\islandboxii
\newbox\islandboxiii
\newbox\captionboxi
\newbox\captionboxii
\newbox\captionboxiii
\long\def\islandpairdata#1#2{{\data@true
 \place@true
 #1%
 \global\setbox\islandboxi\box\islandbox@
 \global\setbox\captionboxi\box\Captionbox@
 #2%
 \global\setbox\islandboxii\box\islandbox@
 \global\setbox\captionboxii\box\Captionbox@
 }}
\long\def\islandpairbox#1#2{\islandpairdata{#1}{#2}%
 \dimen@\ht\captionboxi
 \ifdim\ht\captionboxii>\dimen@\dimen@\ht\captionboxii\fi
 \ifdim\dimen@>\z@
  \ifdim\ht\captionboxi<\dimen@
   \global\setbox\captionboxi\vbox to\dimen@{\unvbox\captionboxi\vfil}\fi
  \ifdim\ht\captionboxii<\dimen@
   \global\setbox\captionboxii\vbox to\dimen@{\unvbox\captionboxii\vfil}\fi
 \fi
 \global\setbox\islandbox@\vbox
 {\hbox to\hsize{\hfil\box\islandboxi\hfil\box\islandboxii\hfil}%
 \ifdim\dimen@>\z@\nointerlineskip
 {\rm\global\skipi@\dp\strutbox}\global\advance\skipi@\medskipamount
  \vskip\skipi@
  \hbox to\hsize{\hfil\box\captionboxi\hfil\box\captionboxii\hfil}\fi}}	
\long\def\islandpairboxa#1#2{\islandpairdata{#1}{#2}%
 \dimen@\ht\captionboxi
 \ifdim\ht\captionboxii>\dimen@\dimen@\ht\captionboxii\fi
 \ifdim\dimen@>\z@
  \ifdim\ht\captionboxi<\dimen@
   \global\setbox\captionboxi\vbox to\dimen@{\vfil\unvbox\captionboxi}\fi
  \ifdim\ht\captionboxii<\dimen@
   \global\setbox\captionboxii\vbox to\dimen@{\vfil\unvbox\captionboxii}\fi
 \fi
 \dimen@ii\ht\islandboxi
 \ifdim\ht\islandboxii>\dimen@ii \dimen@ii\ht\islandboxii\fi
 \ifdim\dimen@ii>\z@
  \ifdim\ht\islandboxi<\dimen@ii
   \global\setbox\islandboxi\vbox to\dimen@ii{\box\islandboxi\vfil}\fi
  \ifdim\ht\islandboxii<\dimen@ii
   \global\setbox\islandboxii\vbox to\dimen@ii{\box\islandboxii\vfil}\fi
 \fi
 \global\setbox\islandbox@\vbox{\ifdim\dimen@>\z@
  \hbox to\hsize{\hfil\box\captionboxi\hfil\box\captionboxii\hfil}%
  \nointerlineskip{\rm\global\skipi@-\dp\strutbox}%
  \global\advance\skipi@\bigskipamount\vskip\skipi@\fi
  \hbox to\hsize{\hfil\box\islandboxi\hfil\box\islandboxii\hfil}}}
\long\def\islandtripledata#1#2#3{{\data@true\place@true
 #1%
 \global\setbox\islandboxi\box\islandbox@
 \global\setbox\captionboxi\box\Captionbox@
 #2%
 \global\setbox\islandboxii\box\islandbox@
 \global\setbox\captionboxii\box\Captionbox@
 #3%
 \global\setbox\islandboxiii\box\islandbox@
 \global\setbox\captionboxiii\box\Captionbox@
 }}
\long\def\islandtriplebox#1#2#3{\islandtripledata{#1}{#2}{#3}%
 \dimen@\ht\captionboxi
 \ifdim\ht\captionboxii>\dimen@ \dimen@\ht\captionboxii\fi
 \ifdim\ht\captionboxiii>\dimen@ \dimen@\ht\captionboxiii\fi
 \ifdim\dimen@>\z@
  \ifdim\ht\captionboxi<\dimen@
   \global\setbox\captionboxi\vbox to\dimen@{\unvbox\captionboxi\vfil}\fi
  \ifdim\ht\captionboxii<\dimen@
   \global\setbox\captionboxii\vbox to\dimen@{\unvbox\captionboxii\vfil}\fi
  \ifdim\ht\captionboxiii<\dimen@
   \global\setbox\captionboxiii\vbox to\dimen@{\unvbox\captionboxiii\vfil}\fi
 \fi
 \global\setbox\islandbox@\vbox
  {\hbox to\hsize{\hfil\box\islandboxi\hfil\box\islandboxii\hfil
   \box\islandboxiii\hfil}%
 \ifdim\dimen@>\z@\nointerlineskip
  {\rm\global\skipi@\dp\strutbox}\global\advance\skipi@\medskipamount
  \vskip\skipi@
  \hbox to\hsize{\hfil\box\captionboxi\hfil\box\captionboxii\hfil
   \box\captionboxiii\hfil}\fi}}
\def\islandtripleboxa#1#2#3{\islandtripledata{#1}{#2}{#3}%
 \dimen@\ht\captionboxi
 \ifdim\ht\captionboxii>\dimen@ \dimen@\ht\captionboxii\fi
 \ifdim\ht\captionboxiii>\dimen@ \dimen@\ht\captionboxiii\fi
 \ifdim\dimen@>\z@
  \ifdim\ht\captionboxi<\dimen@
   \global\setbox\captionboxi\vbox to\dimen@{\vfil\unvbox\captionboxi}\fi
  \ifdim\ht\captionboxii<\dimen@
   \global\setbox\captionboxii\vbox to\dimen@{\vfil\unvbox\captionboxii}\fi
  \ifdim\ht\captionboxiii<\dimen@
   \global\setbox\captionboxiii\vbox to\dimen@{\vfil\unvbox\captionboxiii}\fi
 \fi
 \dimen@ii\ht\islandboxi
 \ifdim\ht\islandboxii>\dimen@ii \dimen@ii\ht\islandboxii\fi
 \ifdim\ht\islandboxiii>\dimen@ii \dimen@ii\ht\islandboxiii\fi
 \ifdim\dimen@ii>\z@
  \ifdim\ht\islandboxi<\dimen@ii
   \global\setbox\islandboxi\vbox to\dimen@ii{\box\islandboxi\vfil}\fi
  \ifdim\ht\islandboxii<\dimen@ii
   \global\setbox\islandboxii\vbox to\dimen@ii{\box\islandboxii\vfil}\fi
  \ifdim\ht\islandboxiii<\dimen@ii
   \global\setbox\islandboxiii\vbox to\dimen@ii{\box\islandboxiii\vfil}\fi
 \fi
 \global\setbox\islandbox@\vbox
  {\ifdim\dimen@>\z@
  \hbox to\hsize{\hfil\box\captionboxi\hfil\box\captionboxii\hfil
   \box\captionboxiii\hfil}%
  \nointerlineskip{\rm\global\skipi@-\dp\strutbox}%
  \global\advance\skipi@\bigskipamount\vskip\skipi@\fi
  \hbox to\hsize{\hfil\box\islandboxi\hfil\box\islandboxii\hfil
   \box\islandboxiii\hfil}}}
\def\Figurepair#1\and#2\endFigurepair{\island@true
 \islandpairbox{\Figure#1\endFigure}{\Figure#2\endFigure}}
\def\Figuretriple#1\and#2\and#3\endFiguretriple{\island@true
 \islandtriplebox{\Figure#1\endFigure}{\Figure#2\endFigure}%
  {\Figure#3\endFigure}}
\def\Tablepair#1\and#2\endTablepair{\island@true
 \islandpairboxa{\Table#1\endTable}{\Table#2\endTable}}
\def\Tabletriple#1\and#2\and#3\endTabletriple{\island@true
 \islandtripleboxa{\Table#1\endTable}{\Table#2\endTable}%
 {\Table#3\endTable}}
\def\place#1{\place@true\island@false
 #1%
 \ifisland@
  \box\islandbox@
 \else
  \Err@{Whoa ... there's no \string\Figure, \string\Table,
   etc., here}%
 \fi
 \place@false}
\newskip\belowtopfigskip
\belowtopfigskip 15\p@ plus 5\p@ minus5\p@
\newskip\abovebotfigskip
\abovebotfigskip 18\p@ plus 6\p@ minus6\p@
\newdimen\minpagesize
\minpagesize 5pc
\dimen@\belowtopfigskip
\advance\dimen@-\abovebotfigskip
\skip\topins\dimen@
\dimen\topins\z@
\newcount\topinscount@
\newbox\topinsdims@
\def\storedim@{\global\setbox\topinsdims@
 \vbox{\hbox to\dimen@{}\unvbox\topinsdims@}}
\def\advancedimtopins@{%
 \ifnum\pageno=\@ne
 \else
   \advance\dimen@\dimen\topins
   \global\dimen\topins\dimen@
 \fi}
\newcount\flipcount@
\def\fliptopins@{%
 \global\flipcount@\z@
 \ifvoid\topins\else
 \setbox\z@\vbox
  {\vskip\p@
   \unvbox\topins
   \global\setbox\topins\vbox{}%
   \loop
    \test@false
    \ifdim\lastskip=\z@\unskip
     \ifdim\lastskip=\z@
      \test@true\fi\fi
    \iftest@
    \global\advance\flipcount@\@ne
    \setboxzl@
    \global\setbox\topins\vbox{\unvbox\topins\boxz@}%
    \unpenalty
   \repeat}\fi}
\newif\ifPar@
\newcount\Parcount@
\newbox\Parbox@
\expandafter\newbox\csname Parfigbox1\endcsname
\expandafter\newbox\csname Parfigbox2\endcsname
\expandafter\newbox\csname Parfigbox3\endcsname
\expandafter\newbox\csname Parfigbox4\endcsname
\expandafter\newbox\csname Parfigbox5\endcsname
\expandafter\newdimen\csname Parprev1\endcsname
\expandafter\newdimen\csname Parprev2\endcsname
\expandafter\newdimen\csname Parprev3\endcsname
\expandafter\newdimen\csname Parprev4\endcsname
\expandafter\newdimen\csname Parprev5\endcsname
\expandafter\newdimen\csname Parprev6\endcsname
\def\Par{\par\global\csname Parprev1\endcsname\prevdepth
 \global\Parcount@\@ne
 \global\Par@true\global\let\Parlist@\empty
 \global\setbox\Parbox@\vbox\bgroup\break}
\def\place@#1#2{%
 \ifisland@
  \ifhmode
   \ifPar@
    \ifnum\Parcount@>5
     \Err@{Only 5 \string\place's allowed per
      \string\Par...\noexpand\endPar paragraph}%
    \else
     \expandafter\expandafter\expandafter
      \global\expandafter\setbox
       \csname Parfigbox\number\Parcount@\endcsname\box\islandbox@
     \global\advance\Parcount@\@ne
     \xdef\Parlist@{\Parlist@#1}%
    \fi
   \else
    \vadjust{#2}%
   \fi
  \else
   #2%
  \fi
 \else
  \Err@{Whoa ... there's no \string\Figure,
   \string\Table, etc., here}%
 \fi
 \place@false}
\long\def\Aplace#1{\prevanish@
 \place@true\island@false
 #1%
 \place@ a\Aplace@
 \postvanish@}
\long\def\AAplace#1{\prevanish@\place@true\island@false
 #1%
 \place@ A\AAplace@
 \postvanish@}
\newif\ifAA@
\def\AAplace@{\AA@true\Aplace@\AA@false}
\let\AAlist@\empty
\def\Aplace@{\allowbreak
 \dimen@=\ht\islandbox@
 \advance\dimen@\abovebotfigskip
 \ht\islandbox@\dimen@
 \advance\dimen@\dp\islandbox@
 \storedim@
 \ifAA@
  \xdef\AAlist@{\AAlist@1}%
  \advancedimtopins@
 \else
  \xdef\AAlist@{\AAlist@0}%
  \ifnum\topinscount@>\@ne\else\advancedimtopins@\fi
 \fi
 \insert\topins{\penalty\z@\splittopskip\z@\floatingpenalty\z@
  \box\islandbox@}%
 \global\advance\topinscount@\@ne}
\long\def\Bplace#1{\prevanish@\place@true\island@false
 #1%
 \place@ b\Bplace@
 \postvanish@}
\def\Bplace@{\allowbreak
 \ifnum\topinscount@=\z@
  \setbox\z@\vbox{\vbox to-\belowtopfigskip{}}%
  \dimen@-\skip\topins
  \ht\z@\dimen@
  \storedim@
  \advancedimtopins@
  \insert\topins{\boxz@}%
  \global\advance\topinscount@\@ne
  \xdef\AAlist@{\AAlist@0}%
 \fi
 \dimen@\ht\islandbox@
 \advance\dimen@\abovebotfigskip
 \ht\islandbox@\dimen@
 \advance\dimen@\dp\islandbox@
 \storedim@
 \xdef\AAlist@{\AAlist@0}%
 \ifnum\topinscount@>\@ne\else\advancedimtopins@\fi
 \insert\topins{\penalty\z@\splittopskip\z@
  \floatingpenalty\z@
  \box\islandbox@}%
 \global\advance\topinscount@\@ne}
\def\breakisland@{\global\setbox\@ne\lastbox\global\skipi@\lastskip\unskip
 \global\setbox\thr@@\lastbox}%
\def\printisland@{\centerline{\box\thr@@}\nobreak\nointerlineskip
 \vskip\skipi@
 \ifdim\ht\@ne<\z@\box\@ne\else\centerline{\box\@ne}\fi}
\def\bottomfigs@{%
 \count@\@ne
 \loop
  \ifnum\count@<\flipcount@
  \nointerlineskip
  \vskip\abovebotfigskip
  \global\setbox\topins\vbox{\unvbox\topins\setboxzl@
   \unvbox\z@
   \breakisland@}%
  \printisland@
  \advance\count@\@ne
  \repeat}
\def\resetdimtopins@{%
 \global\advance\topinscount@-\flipcount@
 \global\setbox\topinsdims@\vbox
  {\unvbox\topinsdims@
   \count@\z@
   \DN@##1##2\next@{\gdef\AAlist@{##2}}%
   \loop
    \ifnum\count@<\flipcount@\setboxzl@
    \expandafter\next@\AAlist@\next@
    \advance\count@\@ne
    \repeat
   \dimen@\z@
   \count@\z@
   \setbox\tw@\vbox{}%
   \edef\nextiii@{\AAlist@}%
   \DN@##1##2\next@{\DNii@{##1}\def\nextiii@{##2}}%
   \loop
    \test@false
    \ifnum\count@<\topinscount@
    \expandafter\next@\nextiii@\next@
     \ifnum\count@<\tw@
      \test@true
     \else
      \if\nextii@ 1\test@true\fi
     \fi
    \fi
    \iftest@
     \setboxzl@
     \advance\dimen@\wdz@
     \setbox\tw@\vbox{\boxz@\unvbox\tw@}%
     \advance\count@\@ne
    \repeat
    \unvbox\tw@
    \global\dimen\topins\dimen@}}
\def\Place@#1#2{%
 \ifisland@
  \ifhmode
   \ifPar@
    \ifnum\Parcount@>5
     \Err@{Only 5 \string\place's allowed per
       \string\Par...\noexpand\endPar paragraph}%
    \else
     \expandafter\expandafter\expandafter\global\expandafter\setbox
      \csname Parfigbox\number\Parcount@\endcsname\box\islandbox@
     \global\advance\Parcount@\@ne
     \xdef\Parlist@{\Parlist@#1}%
     \vadjust{\break}%
    \fi
   \else
    \Err@{\noexpand#2allowed only in a \string\Par...\noexpand\endPar
     paragraph}%
   \fi
  \else
   #2%
  \fi
 \else
  \Err@{Who ... there's no \string\Figure, \string\Table,
   etc., here}%
 \fi
 \place@false}
\newif\ifC@
\newdimen\Cdim@
\long\def\Cplace#1{\prevanish@\place@true\island@false
 #1%
 \Place@ c\Cplace@
 \postvanish@}
\def\Cplace@{\allowbreak
 \ifnum\topinscount@>\z@\else
  \global\C@true\global\Cdim@\pagetotal\fi
 \Aplace@}
\long\def\Mplace#1{\prevanish@\place@true\island@false
 #1%
 \Place@ m\Mplace@
 \postvanish@}
\long\def\MXplace#1{\prevanish@\place@true\island@false
 #1%
 \Place@ M\MXplace@
 \postvanish@}
\newif\ifMX@
\def\MXplace@{\MX@true\Mplace@\MX@false}
\def\Mplace@{\allowbreak
 \dimen@\ht\islandbox@\advance\dimen@\dp\islandbox@
 \ifdim\pagetotal=\z@\else
  \ifdim\lastskip<\abovebotfigskip\advance\dimen@\abovebotfigskip
  \advance\dimen@-\lastskip\fi
 \fi
 \advance\dimen@\pagetotal
 \ifdim\dimen@>\pagegoal
  \Aplace@
 \else
  \nointerlineskip
  \ifdim\lastskip<\abovebotfigskip\removelastskip\vskip\abovebotfigskip\fi
  \setbox\z@\vbox{\unvbox\islandbox@
   \breakisland@}%
  \printisland@
  \ifnum\topinscount@=\z@
   \setbox\z@\vbox{\vbox to-\belowtopfigskip{}}%
   \dimen@-\skip\topins
   \ht\z@\dimen@
   \storedim@
   \advancedimtopins@
   \insert\topins{\boxz@}%
   \global\advance\topinscount@\@ne
   \xdef\AAlist@{\AAlist@0}%
  \fi
  \ifMX@
   \ifnum\topinscount@=\@ne
    \setbox\z@\vbox{\vbox to-\abovebotfigskip{}}%
    \ht\z@\z@
    \dimen@\z@
    \storedim@
    \advancedimtopins@
    \insert\topins{\boxz@}%
    \global\advance\topinscount@\@ne
    \xdef\AAlist@{\AAlist@0}%
   \fi
  \fi
  \nointerlineskip
  \vskip\belowtopfigskip
 \fi}
\expandafter\newbox\csname Parbox1\endcsname
\expandafter\newbox\csname Parbox2\endcsname
\expandafter\newbox\csname Parbox3\endcsname
\expandafter\newbox\csname Parbox4\endcsname
\expandafter\newbox\csname Parbox5\endcsname
\def\endPar{\egroup
 \count@\@ne
 {\vbadness\@M\vfuzz\maxdimen\splitmaxdepth\maxdimen\splittopskip\ht\strutbox
 \setbox\z@\vsplit\Parbox@ to\ht\Parbox@
 \loop
  \ifnum\count@<\Parcount@
  \expandafter\expandafter\expandafter\global\expandafter\setbox
   \csname Parbox\number\count@\endcsname\vsplit\Parbox@ to\ht\Parbox@
  \count@@\count@\advance\count@@\@ne
  \global\csname Parprev\number\count@@\endcsname
   \dp\csname Parbox\number\count@\endcsname
  \advance\count@\@ne
  \repeat}%
 \vskip\parskip
 \count@\@ne
 \def\nextv@##1##2\nextv@{\DN@{##1}\gdef\Parlist@{##2}}%
 \loop
  \ifnum\count@<\Parcount@
   \dimen@\csname Parprev\number\count@\endcsname
   \advance\dimen@\ht\strutbox
   \ifdim\dimen@<\baselineskip
    \advance\dimen@-\baselineskip\vskip-\dimen@
   \else
    \vskip\lineskip
   \fi
   \unvbox\csname Parbox\number\count@\endcsname
   \global\setbox\islandbox@\box\csname Parfigbox\number\count@\endcsname
   \expandafter\nextv@\Parlist@\nextv@
   \if a\next@\Aplace@\else
   \if A\next@\AAplace@\else
   \if b\next@\Bplace@\else
   \if c\next@\Cplace@\else
   \if m\next@\Mplace@\else
   \if M\next@\MXplace@\fi\fi\fi\fi\fi\fi
  \advance\count@\@ne
  \repeat
 \global\Par@false
 \ifvoid\Parbox@
  \prevdepth\csname Parprev\number\count@\endcsname
 \else
  \dimen@\csname Parprev\number\count@\endcsname\advance\dimen@\ht\strutbox
  \ifdim\dimen@<\baselineskip
    \advance\dimen@-\baselineskip\vskip-\dimen@
  \else
    \vskip\lineskip
  \fi
  \dimen@\dp\Parbox@
  \unvbox\Parbox@
  \prevdepth\dimen@
 \fi}
\def\folio{{\page@F\page@S{\page@P\page@N{\number\page@C}\page@Q}}}
\def\advancepageno{\global\advance\pageno\@ne}
\newif\ifspecialsplit@
\newbox\outbox@
\let\shipout@\shipout
\def\plainoutput{\specialsplit@false\ifvoid\topins\else\ifdim\ht\topins=\z@
 \specialsplit@true\advance\minpagesize-\skip\topins\fi\fi
 \fliptopins@
 \setbox\outbox@\vbox{\makeheadline\pagebody\makefootline}%
 {\noexpands@\let\style\relax
 \shipout@\box\outbox@}%
 \advancepageno
 \resetdimtopins@
 \ifvoid\@cclv\else\unvbox\@cclv\penalty\outputpenalty\fi
 \ifnum\outputpenalty>-\@MM\else\dosupereject\fi}
\def\pagebody{\vbox to\vsize{\boxmaxdepth\maxdepth
 \ifvoid\margin@\else
 \rlap{\kern\hsize\vbox to\z@{\kern4\p@\box\margin@\vss}}\fi
 \pagecontents}}
\newif\ifonlytop@
\def\pagecontents{%
 \onlytop@false
 \ifdim\ht\@cclv<\minpagesize\ifnum\flipcount@<\tw@\ifvoid\footins
  \onlytop@true\fi\fi\fi
 \test@false
 \ifC@
  \ifnum\flipcount@=\@ne
   \global\multiply\Cdim@\tw@
   \ifdim\Cdim@>\ht\@cclv
    \test@true
   \fi
  \fi
 \fi
 \global\C@false
 \iftest@
  \dimen@\ht\@cclv
  \advance\dimen@\skip\topins
  {\vfuzz\maxdimen\vbadness\@M
  \splitmaxdepth\maxdepth\splittopskip\topskip
  \setbox\z@\vsplit\@cclv to\dimen@
  \unvbox\z@}%
  \global\setbox\topins\vbox{\unvbox\topins
   \global\setbox\@ne\lastbox}%
  \setbox\z@\vbox{\unvbox\@ne
   \breakisland@}%
  \nointerlineskip
  \vskip\abovebotfigskip
  \printisland@
 \else
  \ifnum\flipcount@>\z@
   \global\setbox\topins\vbox{\unvbox\topins\global\setbox\@ne\lastbox}%
   \setbox\z@\vbox{\unvbox\@ne
    \breakisland@}%
   \printisland@
   \ifonlytop@\kern-\prevdepth\vfill\else\vskip\belowtopfigskip\fi
  \fi
 \fi
 \ifdim\ht\@cclv<\minpagesize
  \ifonlytop@\else\vfill\fi
 \else
  \ifspecialsplit@
   {\vfuzz\maxdimen\vbadness\@M
   \splitmaxdepth\maxdepth\splittopskip\topskip
   \dimen@ii\ht\@cclv \advance\dimen@ii\skip\topins
   \setbox\z@\vsplit\@cclv to\dimen@ii
   \unvbox\z@}%
  \else
   \unvbox\@cclv
  \fi
 \fi
 \bottomfigs@
 \ifvoid\footins\else\vskip\skip\footins\footnoterule\unvbox\footins\fi}
\newread\readdata@
\def\readthedata@#1{\expandafter
 \ifx\csname#1@D\endcsname\relax
  \immediate\openin\readdata@=#1.dat
  \ifeof\readdata@
   \Err@{No file #1.dat}%
  \else
   {\endlinechar\m@ne\gdef\Next@{}%
   \DNii@##1 ##2 ##3pt{\global\data@ht##1\global\data@dp##2%
    \global\data@wd##3pt}%
   \loop
    \ifeof\readdata@
    \else
    \read\readdata@ to\next@
    \ifx\next@\empty\else
     \edef\next@{\expandafter\nextii@\next@}%
     \expandafter\rightadd@\next@\to\Next@
    \fi
    \repeat}%
   \immediate\closein\readdata@
   \expandafter\expandafter\expandafter\global\expandafter
    \let\csname#1@D\endcsname\Next@\global\let\Next@\relax
  \fi
 \fi}
\newdimen\data@ht
\newdimen\data@dp
\newdimen\data@wd
\newif\ifgetdata@
\def\getdata@#1#2{\global\getdata@true\count@#2\relax
 {\let\\\or\xdef\Next@{\ifcase\number\count@#1\else
 \global\noexpand\getdata@false\fi}}\Next@}
\def\paste#1#2{\readthedata@{#1}%
 \getdata@{\csname#1@D\endcsname}{#2}%
 \ifgetdata@
 \dimen@\data@ht \advance\dimen@\data@dp
  \hbox{\special{dvipaste: #1 #2}%
   \lower\data@dp\vbox to\dimen@{\hbox to\data@wd{}\vfil}}%
 \else
  {\lccode`\Z=`\#\lccode`\N=`\N\lccode`\F=`\F%
   \lowercase{\Err@{No data for File [#1], Z#2}}}%
 \fi}
\newdimen\httable
\newdimen\dptable
\newdimen\wdtable
\def\measuretable#1#2{\readthedata@{#1}%
 \getdata@{\csname#1@D\endcsname}{#2}%
 \ifgetdata@
  \httable\data@ht \dptable\data@dp \wdtable\data@wd
 \else
  {\lccode`\Z=`\#\lccode`\N=`\N\lccode`\F=`\F%
  \lowercase{\Err@{No data for File [#1], Z#2}}}%
 \fi}
\def\East#1#2{\setboxz@h{$\m@th\ssize\;{#1}\;\;$}%
 \setbox\tw@\hbox{$\m@th\ssize\;{#2}\;\;$}\setbox4=\hbox{$\m@th#2$}%
 \dimen@\minaw@
 \ifdim\wdz@>\dimen@\dimen@\wdz@\fi\ifdim\wd\tw@>\dimen@\dimen@\wd\tw@\fi
 \ifdim\wd4 >\z@
  \mathrel{\mathop{\hbox to\dimen@{\rightarrowfill}}\limits^{#1}_{#2}}%
 \else
  \mathrel{\mathop{\hbox to\dimen@{\rightarrowfill}}\limits^{#1}}%
 \fi}
\def\West#1#2{\setboxz@h{$\m@th\ssize\;\;{#1}\;$}%
 \setbox\tw@\hbox{$\m@th\ssize\;\;{#2}\;$}\setbox4=\hbox{$\m@th#2$}%
 \dimen@\minaw@
 \ifdim\wdz@>\dimen@\dimen@\wdz@\fi\ifdim\wd\tw@>\dimen@\dimen@\wd\tw@\fi
 \ifdim\wd4 >\z@
  \mathrel{\mathop{\hbox to\dimen@{\leftarrowfill}}\limits^{#1}_{#2}}%
 \else
  \mathrel{\mathop{\hbox to\dimen@{\leftarrowfill}}\limits^{#1}}%
 \fi}
\font\arrow@i=lams1
\font\arrow@ii=lams2
\font\arrow@iii=lams3
\font\arrow@iv=lams4
\font\arrow@v=lams5
\newdimen\standardcgap
\standardcgap40\p@
\newdimen\hunit
\hunit\tw@\p@
\newdimen\standardrgap
\standardrgap32\p@
\newdimen\vunit
\vunit1.6\p@
\def\Cgaps#1{\RIfM@
 \standardcgap#1\standardcgap\relax\hunit#1\hunit\relax
 \else\nonmatherr@\Cgaps\fi}
\def\Rgaps#1{\RIfM@
 \standardrgap#1\standardrgap\relax\vunit#1\vunit\relax
 \else\nonmatherr@\Rgaps\fi}
\newdimen\getdim@
\def\getcgap@#1{\ifcase#1\or\getdim@\z@\else\getdim@\standardcgap\fi}
\def\getrgap@#1{\ifcase#1\getdim@\z@\else\getdim@\standardrgap\fi}
\def\cgaps{\RIfM@\expandafter\cgaps@\else\expandafter\nonmatherr@
 \expandafter\cgaps\fi}
\def\cgaps@{\ifnum\catcode`\;=\active\expandafter\cgapsA@\else
 \expandafter\cgapsO@\fi}
\def\cgapsO@#1{\toks@{\ifcase\i@\or\getdim@=\z@}%
 \gaps@@\standardcgap#1;\gaps@@\gaps@@
 \edef\next@{\the\toks@\noexpand\else\noexpand\getdim@\noexpand\standardcgap
  \noexpand\fi}%
 \toks@=\expandafter{\next@}%
 \edef\getcgap@##1{\i@##1\relax\the\toks@}\toks@{}}
{\catcode`\;=\active
 \gdef\cgapsA@#1{\toks@{\ifcase\i@\or\getdim@=\z@}%
 \gaps@@\standardcgap#1;\gaps@@\gaps@@
 \edef\next@{\the\toks@\noexpand\else\noexpand\getdim@\noexpand\standardcgap
  \noexpand\fi}%
 \toks@=\expandafter{\next@}%
 \edef\getcgap@##1{\i@##1\relax\the\toks@}\toks@{}}
}
\def\Gaps@@{\gaps@@}
\def\gaps@@#1#2;#3{\mgaps@#1#2\mgaps@
 \edef\next@{\the\toks@\noexpand\or\noexpand\getdim@
  \noexpand#1\the\mgapstoks@@}%
 \toks@\expandafter{\next@}%
 \DN@{#3}%
 \ifx\next@\Gaps@@\def\next@##1\gaps@@{}\else
  \def\next@{\gaps@@#1#3}\fi\next@}
{\catcode`\;=\active
 \gdef\rgaps#1{\RIfM@{\ifnum\catcode`\;=\active\def;{\string;}\fi
   \xdef\Next@{\noexpand\rgaps@{#1}}}%
  \Next@\edef\getrgap@##1{\i@##1\relax\the\toks@}\toks@{}\else
  \nonmatherr@\rgaps\fi}
}
\def\rgaps@#1{\toks@{\ifcase\i@\getdim@=\z@}%
 \gaps@@\standardrgap#1;\gaps@@\gaps@@
 \edef\next@{\the\toks@\noexpand\else\noexpand\getdim@\noexpand\standardrgap
  \noexpand\fi}%
 \toks@=\expandafter{\next@}}
\newbox\ZER@
\def\mgaps@#1{\let\mgapsnext@#1\FNSS@\mgaps@@}
\def\mgaps@@{\ifx\next\w\expandafter\mgaps@@@\else
 \expandafter\mgaps@@@@\fi}
\newtoks\mgapstoks@@
\def\mgaps@@@@#1\mgaps@{\getdim@\mgapsnext@\getdim@#1\getdim@
 \edef\next@{\noexpand\getdim@\the\getdim@}%
 \mgapstoks@@\expandafter{\next@}}
\def\mgaps@@@\w#1#2\mgaps@{\mgaps@@@@#2\mgaps@
 \setbox\ZER@\hbox{$\m@th\hskip15\p@\tsize@#1$}%
 \dimen@\wd\ZER@
 \ifdim\dimen@>\getdim@\getdim@\dimen@\fi
 \edef\next@{\noexpand\getdim@\the\getdim@}%
 \mgapstoks@@\expandafter{\next@}}
\def\changewidth#1#2{\setbox\ZER@{$\m@th#2}%
 \hbox to\wd\ZER@{\hss$\m@th#1$\hss}}
\atdef@({\FN@\ARROW@}
\def\ARROW@{\ifx\next)\let\next@\OPTIONS@\else
 \DN@{\csname\string @(\endcsname}\fi\next@}
\newif\ifoptions@
\def\OPTIONS@){\ifoptions@\let\next@\relax\else
 \DN@{\global\options@true\begingroup\optioncodes@}\fi\next@}
\newif\ifN@
\newif\ifE@
\newif\ifNESW@
\newif\ifH@
\newif\ifV@
\newif\ifHshort@
\expandafter\def\csname\string @(\endcsname #1,#2){%
 \ifoptions@\expandafter\endgroup\fi
 \N@false\E@false\H@false\V@false\Hshort@false
 \ifnum#1>\z@\E@true\fi
 \ifnum#1=\z@\V@true\global\tX@false\global\tY@false\global\a@false\fi
 \ifnum#2>\z@\N@true\fi
 \ifnum#2=\z@\H@true\global\tX@false\global\tY@false\global\a@false
  \ifshort@\Hshort@true\fi\fi
 \NESW@false
 \ifN@\ifE@\NESW@true\fi\else\ifE@\else\NESW@true\fi\fi
 \arrow@{#1}{#2}%
 \global\options@false
 \global\scount@\z@\global\tcount@\z@\global\arrcount@\z@
 \global\s@false\global\sxdimen@\z@\global\sydimen@\z@
 \global\tX@false\global\tXdimen@i\z@\global\tXdimen@ii\z@
 \global\tY@false\global\tYdimen@i\z@\global\tYdimen@ii\z@
 \global\a@false\global\exacount@\z@
 \global\x@false\global\xdimen@\z@
 \global\X@false\global\Xdimen@\z@
 \global\y@false\global\ydimen@\z@
 \global\Y@false\global\Ydimen@\z@
 \global\p@false\global\pdimen@\z@
 \global\label@ifalse\global\label@iifalse
 \global\dl@ifalse\global\ldimen@i\z@
 \global\dl@iifalse\global\ldimen@ii\z@
 \global\short@false\global\unshort@false}
\newif\iflabel@i
\newif\iflabel@ii
\newcount\scount@
\newcount\tcount@
\newcount\arrcount@
\newif\ifs@
\newdimen\sxdimen@
\newdimen\sydimen@
\newif\iftX@
\newdimen\tXdimen@i
\newdimen\tXdimen@ii
\newif\iftY@
\newdimen\tYdimen@i
\newdimen\tYdimen@ii
\newif\ifa@
\newcount\exacount@
\newif\ifx@
\newdimen\xdimen@
\newif\ifX@
\newdimen\Xdimen@
\newif\ify@
\newdimen\ydimen@
\newif\ifY@
\newdimen\Ydimen@
\newif\ifp@
\newdimen\pdimen@
\newif\ifdl@i
\newif\ifdl@ii
\newdimen\ldimen@i
\newdimen\ldimen@ii
\newif\ifshort@
\newif\ifunshort@
\def\zero@#1{\ifnum\scount@=\z@
 \if#1e\global\scount@\m@ne\else
 \if#1t\global\scount@\tw@\else
 \if#1h\global\scount@\thr@@\else
 \if#1'\global\scount@6 \else
 \if#1`\global\scount@7 \else
 \if#1(\global\scount@8 \else
 \if#1)\global\scount@9 \else
 \if#1s\global\scount@12 \else
 \if#1H\global\scount@13 \else
 \Err@{\Invalid@@ option \string\0}\fi\fi\fi\fi\fi\fi\fi\fi\fi
 \fi}
\def\one@#1{\ifnum\tcount@=\z@
 \if#1e\global\tcount@\m@ne\else
 \if#1h\global\tcount@\tw@\else
 \if#1t\global\tcount@\thr@@\else
 \if#1'\global\tcount@4 \else
 \if#1`\global\tcount@5 \else
 \if#1(\global\tcount@\ten@ \else
 \if#1)\global\tcount@11 \else
 \if#1s\global\tcount@12 \else
 \if#1H\global\tcount@13 \else
 \Err@{\Invalid@@ option \string\1}\fi\fi\fi\fi\fi\fi\fi\fi\fi
 \fi}
\def\a@#1{\ifnum\arrcount@=\z@
 \if#10\global\arrcount@\m@ne\else
 \if#1+\global\arrcount@\@ne\else
 \if#1-\global\arrcount@\tw@\else
 \if#1=\global\arrcount@\thr@@\else
 \Err@{\Invalid@@ option \string\a}\fi\fi\fi\fi
 \fi}
\def\ds@{\ifnum\catcode`\;=\active\expandafter\dsA@\else
 \expandafter\dsO@\fi}
\def\dsO@(#1;#2){\ds@@{#1}{#2}}
\def\ds@@#1#2{\ifs@\else
 \global\s@true
 \global\sxdimen@\hunit\global\sxdimen@#1\sxdimen@\relax
 \global\sydimen@\vunit\global\sydimen@#2\sydimen@\relax
 \fi}
\def\dtX@{\ifnum\catcode`\;=\active\expandafter\dtXA@\else
 \expandafter\dtXO@\fi}
\def\dtXO@(#1;#2){\dtX@@{#1}{#2}}
\def\dtX@@#1#2{\iftX@\else
 \global\tX@true
 \global\tXdimen@i\hunit\global\tXdimen@i#1\tXdimen@i\relax
 \global\tXdimen@ii\vunit\global\tXdimen@ii#2\tXdimen@ii\relax
 \fi}
\def\dtY@{\ifnum\catcode`\;=\active\expandafter\dtYA@\else
 \expandafter\dtYO@\fi}
\def\dtYO@(#1;#2){\dtY@@{#1}{#2}}
\def\dtY@@#1#2{\iftY@\else
 \global\tY@true
 \global\tYdimen@i\hunit\global\tYdimen@i#1\tYdimen@i\relax
 \global\tYdimen@ii\vunit\global\tYdimen@ii#2\tYdimen@ii\relax
 \fi}
{\catcode`\;=\active
 \gdef\dsA@(#1;#2){\ds@@{#1}{#2}}
 \gdef\dtXA@(#1;#2){\dtX@@{#1}{#2}}
 \gdef\dtYA@(#1;#2){\dtY@@{#1}{#2}}
}
\def\da@#1{\ifa@\else\global\a@true\global\exacount@#1\relax\fi}
\def\dx@#1{\ifx@\else
 \global\x@true
 \global\xdimen@\hunit\global\xdimen@#1\xdimen@\relax
 \fi}
\def\dX@#1{\ifX@\else
 \global\X@true
 \global\Xdimen@\hunit\global\Xdimen@#1\Xdimen@\relax
 \fi}
\def\dy@#1{\ify@\else
 \global\y@true
 \global\ydimen@\vunit\global\ydimen@#1\ydimen@\relax
 \fi}
\def\dY@#1{\ifY@\else
 \global\Y@true
 \global\Ydimen@\vunit\global\Ydimen@#1\Ydimen@\relax
 \fi}
\def\p@@#1{\ifp@\else
 \global\p@true
 \global\pdimen@\hunit\global\divide\pdimen@\tw@
 \global\pdimen@#1\pdimen@\relax
 \fi}
\def\L@#1{\iflabel@i\else
 \global\label@itrue\gdef\label@i{#1}%
 \fi}
\def\l@#1{\iflabel@ii\else
 \global\label@iitrue\gdef\label@ii{#1}%
 \fi}
\def\dL@#1{\ifdl@i\else
 \global\dl@itrue\global\ldimen@i\hunit\global\ldimen@i#1\ldimen@i\relax
 \fi}
\def\dl@#1{\ifdl@ii\else
 \global\dl@iitrue\global\ldimen@ii\hunit\global\ldimen@ii#1\ldimen@ii\relax
 \fi}
\def\s@{\ifunshort@\else\global\short@true\fi}
\def\uns@{\ifshort@\else\global\unshort@true\global\short@false\fi}
\def\optioncodes@{\let\0\zero@\let\1\one@\let\a\a@\let\ds\ds@\let\dtX\dtX@
 \let\dtY\dtY@\let\da\da@\let\dx\dx@\let\dX\dX@\let\dY\dY@\let\dy\dy@
 \let\p\p@@\let\L\L@\let\l\l@\let\dL\dL@\let\dl\dl@\let\s\s@\let\uns\uns@}
\def\slopes@{\\161\\152\\143\\134\\255\\126\\357\\238\\349\\45{10}\\56{11}%
 \\11{12}\\65{13}\\54{14}\\43{15}\\32{16}\\53{17}\\21{18}\\52{19}\\31{20}%
 \\41{21}\\51{22}\\61{23}}
\newcount\tan@i
\newcount\tan@ip
\newcount\tan@ii
\newcount\tan@iip
\newdimen\slope@i
\newdimen\slope@ip
\newdimen\slope@ii
\newdimen\slope@iip
\newcount\angcount@
\newcount\extracount@
\def\slope@{{\slope@i\secondy@\advance\slope@i-\firsty@
 \ifN@\else\multiply\slope@i\m@ne\fi
 \slope@ii\secondx@\advance\slope@ii-\firstx@
 \ifE@\else\multiply\slope@ii\m@ne\fi
 \ifdim\slope@ii<\z@
  \global\tan@i6 \global\tan@ii\@ne\global\angcount@23
 \else
  \dimen@\slope@i\multiply\dimen@6
  \ifdim\dimen@<\slope@ii
   \global\tan@i\@ne\global\tan@ii6 \global\angcount@\@ne
  \else
   \dimen@\slope@ii\multiply\dimen@6
   \ifdim\dimen@<\slope@i
    \global\tan@i6 \global\tan@ii\@ne\global\angcount@23
   \else
    \global\tan@ip\z@\global\tan@iip\@ne
    \def\\##1##2##3{\global\angcount@##3\relax
     \slope@ip\slope@i\slope@iip\slope@ii
     \multiply\slope@iip##1\relax\multiply\slope@ip##2\relax
     \ifdim\slope@iip<\slope@ip
      \global\tan@ip##1\relax\global\tan@iip##2\relax
     \else
      \global\tan@i##1\relax\global\tan@ii##2\relax
      \def\\####1####2####3{}%
     \fi}%
    \slopes@
    \slope@i\secondy@\advance\slope@i-\firsty@
    \ifN@\else\multiply\slope@i\m@ne\fi
    \multiply\slope@i\tan@ii\multiply\slope@i\tan@iip\multiply\slope@i\tw@
    \count@\tan@i\multiply\count@\tan@iip
    \extracount@\tan@ip\multiply\extracount@\tan@ii
    \advance\count@\extracount@
    \slope@ii\secondx@\advance\slope@ii-\firstx@
    \ifE@\else\multiply\slope@ii\m@ne\fi
    \multiply\slope@ii\count@
    \ifdim\slope@i<\slope@ii
     \global\tan@i\tan@ip\global\tan@ii\tan@iip
     \global\advance\angcount@\m@ne
    \fi
   \fi
  \fi
 \fi}%
}
\def\slope@a#1{{\def\\##1##2##3{\ifnum##3=#1\global\tan@i##1\relax
 \global\tan@ii##2\relax\fi}\slopes@}}
\newcount\i@
\newcount\j@
\newcount\colcount@
\newcount\Colcount@
\newcount\tcolcount@
\newdimen\rowht@
\newdimen\rowdp@
\newcount\rowcount@
\newcount\Rowcount@
\newcount\maxcolrow@
\newtoks\colwidthtoks@
\newtoks\Rowheighttoks@
\newtoks\Rowdepthtoks@
\newtoks\widthtoks@
\newtoks\Widthtoks@
\newtoks\heighttoks@
\newtoks\Heighttoks@
\newtoks\depthtoks@
\newtoks\Depthtoks@
\newif\iffirstCDcr@
\def\dotoks@i{%
 \global\widthtoks@\expandafter{\the\widthtoks@\else\getdim@\z@\fi}%
 \global\heighttoks@\expandafter{\the\heighttoks@\else\getdim@\z@\fi}%
 \global\depthtoks@\expandafter{\the\depthtoks@\else\getdim@\z@\fi}}
\def\dotoks@ii{%
 \global\widthtoks@{\ifcase\j@}%
 \global\heighttoks@{\ifcase\j@}%
 \global\depthtoks@{\ifcase\j@}}
\def\preCD@#1\endCD{\setbox\ZER@
 \vbox{%
  \def\arrow@##1##2{{}}%
  \global\rowcount@\m@ne\global\colcount@\z@\global\Colcount@\z@
  \global\firstCDcr@true\toks@{}%
  \global\widthtoks@{\ifcase\j@}%
  \global\Widthtoks@{\ifcase\i@}%
  \global\heighttoks@{\ifcase\j@}%
  \global\Heighttoks@{\ifcase\i@}%
  \global\depthtoks@{\ifcase\j@}%
  \global\Depthtoks@{\ifcase\i@}%
  \global\Rowheighttoks@{\ifcase\i@}%
  \global\Rowdepthtoks@{\ifcase\i@}%
  \Let@
  \everycr{%
   \noalign{%
    \global\advance\rowcount@\@ne
    \ifnum\colcount@<\Colcount@
    \else
     \global\Colcount@\colcount@\global\maxcolrow@\rowcount@
    \fi
    \global\colcount@\z@
    \iffirstCDcr@
     \global\firstCDcr@false
    \else
     \edef\next@{\the\Rowheighttoks@\noexpand\or\noexpand\getdim@\the\rowht@}%
      \global\Rowheighttoks@\expandafter{\next@}%
     \edef\next@{\the\Rowdepthtoks@\noexpand\or\noexpand\getdim@\the\rowdp@}%
      \global\Rowdepthtoks@\expandafter{\next@}%
     \global\rowht@\z@\global\rowdp@\z@
     \dotoks@i
     \edef\next@{\the\Widthtoks@\noexpand\or\the\widthtoks@}%
      \global\Widthtoks@\expandafter{\next@}%
     \edef\next@{\the\Heighttoks@\noexpand\or\the\heighttoks@}%
      \global\Heighttoks@\expandafter{\next@}%
     \edef\next@{\the\Depthtoks@\noexpand\or\the\depthtoks@}%
      \global\Depthtoks@\expandafter{\next@}%
     \dotoks@ii
    \fi}}%
  \tabskip\z@
  \halign{&\setbox\ZER@\hbox{\vrule\height\ten@\p@\width\z@\depth\z@     %1
   $\m@th\displaystyle{##}$}\copy\ZER@
   \ifdim\ht\ZER@>\rowht@\global\rowht@\ht\ZER@\fi
   \ifdim\dp\ZER@>\rowdp@\global\rowdp@\dp\ZER@\fi
   \global\advance\colcount@\@ne
   \edef\next@{\the\widthtoks@\noexpand\or\noexpand\getdim@\the\wd\ZER@}%
    \global\widthtoks@\expandafter{\next@}%
   \edef\next@{\the\heighttoks@\noexpand\or\noexpand\getdim@\the\ht\ZER@}%
    \global\heighttoks@\expandafter{\next@}%
   \edef\next@{\the\depthtoks@\noexpand\or\noexpand\getdim@\the\dp\ZER@}%
    \global\depthtoks@\expandafter{\next@}%
   \cr#1\crcr}}%
 \Rowcount@\rowcount@
 \global\Widthtoks@\expandafter{\the\Widthtoks@\fi\relax}%
 \edef\Width@##1##2{\i@##1\relax\j@##2\relax\the\Widthtoks@}%
 \global\Heighttoks@\expandafter{\the\Heighttoks@\fi\relax}%
 \edef\Height@##1##2{\i@##1\relax\j@##2\relax\the\Heighttoks@}%
 \global\Depthtoks@\expandafter{\the\Depthtoks@\fi\relax}%
 \edef\Depth@##1##2{\i@##1\relax\j@##2\relax\the\Depthtoks@}%
 \edef\next@{\the\Rowheighttoks@\noexpand\fi\relax}%
 \global\Rowheighttoks@\expandafter{\next@}%
 \edef\Rowheight@##1{\i@##1\relax\the\Rowheighttoks@}%
 \edef\next@{\the\Rowdepthtoks@\noexpand\fi\relax}%
 \global\Rowdepthtoks@\expandafter{\next@}%
 \edef\Rowdepth@##1{\i@##1\relax\the\Rowdepthtoks@}%
 \global\colwidthtoks@{\fi}%
 \setbox\ZER@\vbox{%
  \unvbox\ZER@
  \count@\rowcount@
  \loop
   \unskip\unpenalty
   \setbox\ZER@\lastbox
   \ifnum\count@>\maxcolrow@\advance\count@\m@ne
   \repeat
  \hbox{%
   \unhbox\ZER@
   \count@\z@
   \loop
    \unskip
    \setbox\ZER@\lastbox
    \edef\next@{\noexpand\or\noexpand\getdim@\the\wd\ZER@\the\colwidthtoks@}%
     \global\colwidthtoks@\expandafter{\next@}%
    \advance\count@\@ne
    \ifnum\count@<\Colcount@
    \repeat}}%
 \edef\next@{\noexpand\ifcase\noexpand\i@\the\colwidthtoks@}%
  \global\colwidthtoks@\expandafter{\next@}%
 \edef\Colwidth@##1{\i@##1\relax\the\colwidthtoks@}%
 \global\colwidthtoks@{}\global\Rowheighttoks@{}\global\Rowdepthtoks@{}%
 \global\widthtoks@{}\global\Widthtoks@{}\global\heighttoks@{}%
 \global\Heighttoks@{}\global\depthtoks@{}\global\Depthtoks@{}%
}
\newcount\xoff@
\newcount\yoff@
\newcount\endcount@
\newcount\rcount@
\newdimen\firstx@
\newdimen\firsty@
\newdimen\secondx@
\newdimen\secondy@
\newdimen\tocenter@
\newdimen\charht@
\newdimen\charwd@
\def\outside@{\Err@{This arrow points outside the \string\CD}}
\newif\ifsvertex@
\newif\iftvertex@
\def\arrow@#1#2{\global\xoff@#1\relax\global\yoff@#2\relax
 \count@\rowcount@\advance\count@-\yoff@
 \ifnum\count@<\@ne\outside@\else\ifnum\count@>\Rowcount@\outside@\fi\fi
 \count@\colcount@\advance\count@\xoff@
 \ifnum\count@<\@ne\outside@\else\ifnum\count@>\Colcount@\outside@\fi\fi
 \tcolcount@\colcount@\advance\tcolcount@\xoff@
 \Width@\rowcount@\colcount@\divide\getdim@\tw@\tocenter@-\getdim@
 \ifdim\getdim@=\z@
  \firstx@\z@\firsty@\mathaxis@\svertex@true
 \else
  \svertex@false
  \ifHshort@
   \Colwidth@\colcount@\divide\getdim@\tw@
   \ifE@ \firstx@\getdim@ \else \firstx@-\getdim@ \fi
  \else
   \ifE@ \firstx@\getdim@ \else \firstx@-\getdim@ \fi
  \fi
  \ifE@
   \ifH@ \advance\firstx@\thr@@\p@ \else \advance\firstx@-\thr@@\p@ \fi  %2
  \else
   \ifH@ \advance\firstx@-\thr@@\p@ \else \advance\firstx@\thr@@\p@ \fi  %3
  \fi
  \ifN@
   \Height@\rowcount@\colcount@ \firsty@\getdim@                         %4
   \ifV@ \advance\firsty@\thr@@\p@ \fi                                   %5
  \else
   \ifV@
    \Depth@\rowcount@\colcount@ \firsty@-\getdim@                        %6
    \advance\firsty@-\thr@@\p@                                           %7
   \else
    \firsty@\z@                                                          %8
   \fi
  \fi
 \fi
 \ifV@
 \else
  \Colwidth@\colcount@\divide\getdim@\tw@
  \ifE@\secondx@\getdim@\else\secondx@-\getdim@\fi
  \ifE@\else\getcgap@\colcount@\advance\secondx@-\getdim@\fi
  \endcount@\colcount@\advance\endcount@\xoff@
  \count@\colcount@
  \ifE@
   \advance\count@\@ne
   \loop
    \ifnum\count@<\endcount@
    \Colwidth@\count@\advance\secondx@\getdim@
    \getcgap@\count@\advance\secondx@\getdim@
    \advance\count@\@ne
    \repeat
  \else
   \advance\count@\m@ne
   \loop
    \ifnum\count@>\endcount@
    \Colwidth@\count@\advance\secondx@-\getdim@
    \getcgap@\count@\advance\secondx@-\getdim@
    \advance\count@\m@ne
    \repeat
  \fi
  \Colwidth@\count@\divide\getdim@\tw@
  \ifHshort@
  \else
   \ifE@\advance\secondx@\getdim@\else\advance\secondx@-\getdim@\fi
  \fi
  \ifE@\getcgap@\count@\advance\secondx@\getdim@\fi
  \rcount@\rowcount@\advance\rcount@-\yoff@
  \Width@\rcount@\count@\divide\getdim@\tw@
  \tvertex@false
  \ifH@\ifdim\getdim@=\z@\tvertex@true\Hshort@false\fi\fi
  \ifHshort@
  \else
   \ifE@\advance\secondx@-\getdim@\else\advance\secondx@\getdim@\fi
  \fi
  \iftvertex@
   \advance\secondx@.4\p@
  \else
   \ifE@\advance\secondx@-\thr@@\p@\else\advance\secondx@\thr@@\p@\fi    %9
  \fi
 \fi
 \ifH@
 \else
  \ifN@
   \Rowheight@\rowcount@\secondy@\getdim@
  \else
   \Rowdepth@\rowcount@\secondy@-\getdim@
   \getrgap@\rowcount@\advance\secondy@-\getdim@
  \fi
  \endcount@\rowcount@\advance\endcount@-\yoff@
  \count@\rowcount@
  \ifN@
   \advance\count@\m@ne
   \loop
    \ifnum\count@>\endcount@
    \Rowheight@\count@\advance\secondy@\getdim@
    \Rowdepth@\count@\advance\secondy@\getdim@
    \getrgap@\count@\advance\secondy@\getdim@
    \advance\count@\m@ne
    \repeat
  \else
   \advance\count@\@ne
   \loop
    \ifnum\count@<\endcount@
    \Rowheight@\count@\advance\secondy@-\getdim@
    \Rowdepth@\count@\advance\secondy@-\getdim@
    \getrgap@\count@\advance\secondy@-\getdim@
    \advance\count@\@ne
    \repeat
  \fi
  \tvertex@false
  \ifV@\Width@\count@\colcount@\ifdim\getdim@=\z@\tvertex@true\fi\fi
  \ifN@
   \getrgap@\count@\advance\secondy@\getdim@
   \Rowdepth@\count@\advance\secondy@\getdim@
   \iftvertex@
    \advance\secondy@\mathaxis@
   \else
    \Depth@\count@\tcolcount@\advance\secondy@-\getdim@
    \advance\secondy@-\thr@@\p@                                          %10
   \fi
  \else
   \Rowheight@\count@\advance\secondy@-\getdim@
   \iftvertex@
    \advance\secondy@\mathaxis@
   \else
    \Height@\count@\tcolcount@\advance\secondy@\getdim@
    \advance\secondy@\thr@@\p@                                           %11
   \fi
  \fi
 \fi
 \ifV@\else\advance\firstx@\sxdimen@\fi
 \ifH@\else\advance\firsty@\sydimen@\fi
 \iftX@
  \advance\secondy@\tXdimen@ii
  \advance\secondx@\tXdimen@i
  \slope@
 \else
  \iftY@
   \advance\secondy@\tYdimen@ii
   \advance\secondx@\tYdimen@i
   \slope@
   \secondy@\secondx@\advance\secondy@-\firstx@
   \ifNESW@\else\multiply\secondy@\m@ne\fi
   \multiply\secondy@\tan@i\divide\secondy@\tan@ii\advance\secondy@\firsty@
  \else
   \ifa@
    \slope@
    \ifNESW@\global\advance\angcount@\exacount@\else
     \global\advance\angcount@-\exacount@\fi
    \ifnum\angcount@>23 \global\angcount@23 \fi
    \ifnum\angcount@<\@ne\global\angcount@\@ne\fi
    \slope@a\angcount@
    \ifY@
     \advance\secondy@\Ydimen@
    \else
     \ifX@
      \advance\secondx@\Xdimen@
      \dimen@\secondx@\advance\dimen@-\firstx@
      \ifNESW@\else\multiply\dimen@\m@ne\fi
      \multiply\dimen@\tan@i\divide\dimen@\tan@ii
      \advance\dimen@\firsty@\secondy@\dimen@
     \fi
    \fi
   \else
    \ifH@\else\ifV@\else\slope@\fi\fi
   \fi
  \fi
 \fi
 \ifH@\else\ifV@\else\ifsvertex@\else
  \dimen@6\p@\multiply\dimen@\tan@ii
  \count@\tan@i\advance\count@\tan@ii\divide\dimen@\count@
  \ifE@\advance\firstx@\dimen@\else\advance\firstx@-\dimen@\fi
  \multiply\dimen@\tan@i\divide\dimen@\tan@ii
  \ifN@\advance\firsty@\dimen@\else\advance\firsty@-\dimen@\fi
 \fi\fi\fi
 \ifp@
  \ifH@\else\ifV@\else
   \getcos@\pdimen@\advance\firsty@\dimen@\advance\secondy@\dimen@
   \ifNESW@\advance\firstx@-\dimen@ii\else\advance\firstx@\dimen@ii\fi
  \fi\fi
 \fi
 \ifH@\else\ifV@\else
  \ifnum\tan@i>\tan@ii
   \charht@\ten@\p@\charwd@\ten@\p@
   \multiply\charwd@\tan@ii\divide\charwd@\tan@i
  \else
   \charwd@\ten@\p@\charht@\ten@\p@
   \divide\charht@\tan@ii\multiply\charht@\tan@i
  \fi
  \ifnum\tcount@=\thr@@
   \ifN@\advance\secondy@-.3\charht@\else\advance\secondy@.3\charht@\fi
  \fi
  \ifnum\scount@=\tw@
   \ifE@\advance\firstx@.3\charht@\else\advance\firstx@-.3\charht@\fi
  \fi
  \ifnum\tcount@=12
   \ifN@\advance\secondy@-\charht@\else\advance\secondy@\charht@\fi
  \fi
  \iftY@
  \else
   \ifa@
    \ifX@
    \else
     \secondx@\secondy@\advance\secondx@-\firsty@
     \ifNESW@\else\multiply\secondx@\m@ne\fi
     \multiply\secondx@\tan@ii\divide\secondx@\tan@i
     \advance\secondx@\firstx@
    \fi
   \fi
  \fi
 \fi\fi
 \ifH@\harrow@\else\ifV@\varrow@\else\arrow@@\fi\fi}
\newdimen\mathaxis@
\mathaxis@90\p@\divide\mathaxis@36
\def\harrow@b{\ifE@\hskip\tocenter@\hskip\firstx@\fi}
\def\harrow@bb{\ifE@\hskip\xdimen@\else\hskip\Xdimen@\fi}
\def\harrow@e{\ifE@\else\hskip-\firstx@\hskip-\tocenter@\fi}
\def\harrow@ee{\ifE@\hskip-\Xdimen@\else\hskip-\xdimen@\fi}
\def\harrow@{\dimen@\secondx@\advance\dimen@-\firstx@
 \ifE@\let\next@\rlap\else\multiply\dimen@\m@ne\let\next@\llap\fi
 \next@{%
  \harrow@b
  \smash{\raise\pdimen@\hbox to\dimen@
   {\harrow@bb\arrow@ii
    \ifnum\arrcount@=\m@ne\else\ifnum\arrcount@=\thr@@\else
     \ifE@
      \ifnum\scount@=\m@ne
      \else
       \ifcase\scount@\or\or\char118 \or\char117 \or\or\or\char119 \or
       \char120 \or\char121 \or\char122 \or\or\or\arrow@i\char125 \or
       \char117 \hskip\thr@@\p@\char117 \hskip-\thr@@\p@\fi
      \fi
     \else
      \ifnum\tcount@=\m@ne
      \else
       \ifcase\tcount@\char117 \or\or\char117 \or\char118 \or\char119 \or
       \char120 \or\or\or\or\or\char121 \or\char122 \or\arrow@i\char125
       \or\char117 \hskip\thr@@\p@\char117 \hskip-\thr@@\p@\fi
      \fi
     \fi
    \fi\fi
    \dimen@\mathaxis@\advance\dimen@.2\p@
    \dimen@ii\mathaxis@\advance\dimen@ii-.2\p@
    \ifnum\arrcount@=\m@ne
     \let\leads@\null
    \else
     \ifcase\arrcount@
      \def\leads@{\hrule\height\dimen@\depth-\dimen@ii}\or
      \def\leads@{\hrule\height\dimen@\depth-\dimen@ii}\or
      \def\leads@{\hbox to\ten@\p@{%
       \leaders\hrule\height\dimen@\depth-\dimen@ii\hfil
       \hfil
      \leaders\hrule\height\dimen@\depth-\dimen@ii\hskip\z@ plus2fil\relax
       \hfil
       \leaders\hrule\height\dimen@\depth-\dimen@ii\hfil}}\or
     \def\leads@{\hbox{\hbox to\ten@\p@{\dimen@\mathaxis@\advance\dimen@1.2\p@
       \dimen@ii\dimen@\advance\dimen@ii-.4\p@
       \leaders\hrule\height\dimen@\depth-\dimen@ii\hfil}%
       \kern-\ten@\p@
       \hbox to\ten@\p@{\dimen@\mathaxis@\advance\dimen@-1.2\p@
       \dimen@ii\dimen@\advance\dimen@ii-.4\p@
       \leaders\hrule\height\dimen@\depth-\dimen@ii\hfil}}}\fi
    \fi
    \cleaders\leads@\hfil
    \ifnum\arrcount@=\m@ne\else\ifnum\arrcount@=\thr@@\else
     \arrow@i
     \ifE@
      \ifnum\tcount@=\m@ne
      \else
       \ifcase\tcount@\char119 \or\or\char119 \or\char120 \or\char121 \or
       \char122 \or \or\or\or\or\char123 \or\char124 \or
       \char125 \or\char119 \hskip-\thr@@\p@\char119 \hskip\thr@@\p@\fi
      \fi
     \else
      \ifcase\scount@\or\or\char120 \or\char119 \or\or\or\char121 \or\char122
      \or\char123 \or\char124 \or\or\or\char125 \or
      \char119 \hskip-\thr@@\p@\char119 \hskip\thr@@\p@\fi
     \fi
    \fi\fi
    \harrow@ee}}%
  \harrow@e}%
 \iflabel@i
  \dimen@ii\z@\setbox\ZER@\hbox{$\m@th\tsize@@\label@i$}%
  \ifnum\arrcount@=\m@ne
  \else
   \advance\dimen@ii\mathaxis@
   \advance\dimen@ii\dp\ZER@\advance\dimen@ii\tw@\p@
   \ifnum\arrcount@=\thr@@\advance\dimen@ii\tw@\p@\fi
  \fi
  \advance\dimen@ii\pdimen@
  \next@{\harrow@b\smash{\raise\dimen@ii\hbox to\dimen@
   {\harrow@bb\hskip\tw@\ldimen@i\hfil\box\ZER@\hfil\harrow@ee}}\harrow@e}%
 \fi
 \iflabel@ii
  \ifnum\arrcount@=\m@ne
  \else
   \setbox\ZER@\hbox{$\m@th\tsize@\label@ii$}%
   \dimen@ii-\ht\ZER@\advance\dimen@ii-\tw@\p@
   \ifnum\arrcount@=\thr@@\advance\dimen@ii-\tw@\p@\fi
   \advance\dimen@ii\mathaxis@\advance\dimen@ii\pdimen@
   \next@{\harrow@b\smash{\raise\dimen@ii\hbox to\dimen@
    {\harrow@bb\hskip\tw@\ldimen@ii\hfil\box\ZER@\hfil\harrow@ee}}\harrow@e}%
  \fi
 \fi}
\let\tsize@\tsize
\def\tsizeCDlabels{\let\tsize@\tsize}
\def\ssizeCDlabels{\let\tsize@\ssize}
\def\tsize@@{\ifnum\arrcount@=\m@ne\else\tsize@\fi}
\def\varrow@{\dimen@\secondy@\advance\dimen@-\firsty@
 \ifN@\else\multiply\dimen@\m@ne\fi
 \setbox\ZER@\vbox to\dimen@
  {\ifN@\vskip-\Ydimen@\else\vskip\ydimen@\fi
   \ifnum\arrcount@=\m@ne\else\ifnum\arrcount@=\thr@@\else
    \hbox{\arrow@iii
     \ifN@
      \ifnum\tcount@=\m@ne
      \else
       \ifcase\tcount@\char117 \or\or\char117 \or\char118 \or\char119 \or
       \char120 \or\or\or\or\or\char121 \or\char122 \or\char123 \or
       \vbox{\hbox{\char117}\nointerlineskip\vskip\thr@@\p@
       \hbox{\char117}\vskip-\thr@@\p@}\fi
      \fi
     \else
      \ifcase\scount@\or\or\char118 \or\char117 \or\or\or\char119 \or
      \char120 \or\char121 \or\char122 \or\or\or\char123 \or
      \vbox{\hbox{\char117}\nointerlineskip\vskip\thr@@\p@
      \hbox{\char117}\vskip-\thr@@\p@}\fi
     \fi}%
    \nointerlineskip
   \fi\fi
   \ifnum\arrcount@=\m@ne
    \let\leads@\null
   \else
    \ifcase\arrcount@\let\leads@\vrule\or\let\leads@\vrule\or
    \def\leads@{\vbox to\ten@\p@{%
     \hrule\height1.67\p@\depth\z@\width.4\p@
     \vfil
     \hrule\height3.33\p@\depth\z@\width.4\p@
     \vfil
     \hrule\height1.67\p@\depth\z@\width.4\p@}}\or
    \def\leads@{\hbox{\vrule\height\p@\hskip\tw@\p@\vrule}}\fi
   \fi
  \cleaders\leads@\vfill\nointerlineskip
   \ifnum\arrcount@=\m@ne\else\ifnum\arrcount@=\thr@@\else
    \hbox{\arrow@iv
     \ifN@
      \ifcase\scount@\or\or\char118 \or\char117 \or\or\or\char119 \or
      \char120 \or\char121 \or\char122 \or\or\or\arrow@iii\char123 \or
      \vbox{\hbox{\char117}\nointerlineskip\vskip-\thr@@\p@
      \hbox{\char117}\vskip\thr@@\p@}\fi
     \else
      \ifnum\tcount@=\m@ne
      \else
       \ifcase\tcount@\char117 \or\or\char117 \or\char118 \or\char119 \or
       \char120 \or\or\or\or\or\char121 \or\char122 \or\arrow@iii\char123 \or
       \vbox{\hbox{\char117}\nointerlineskip\vskip-\thr@@\p@
       \hbox{\char117}\vskip\thr@@\p@}\fi
      \fi
     \fi}%
   \fi\fi
   \ifN@\vskip\ydimen@\else\vskip-\Ydimen@\fi}%
 \ifN@
  \dimen@ii\firsty@
 \else
  \dimen@ii-\firsty@\advance\dimen@ii\ht\ZER@\multiply\dimen@ii\m@ne
 \fi
 \rlap{\smash{\hskip\tocenter@\hskip\pdimen@\raise\dimen@ii\box\ZER@}}%
 \iflabel@i
  \setbox\ZER@\vbox to\dimen@{\vfil
   \hbox{$\m@th\tsize@@\label@i$}\vskip\tw@\ldimen@i\vfil}%
  \rlap{\smash{\hskip\tocenter@\hskip\pdimen@
  \ifnum\arrcount@=\m@ne\let\next@\relax\else\let\next@\llap\fi
  \next@{\raise\dimen@ii\hbox{\ifnum\arrcount@=\m@ne\hskip-.5\wd\ZER@\fi
   \box\ZER@\ifnum\arrcount@=\m@ne\else\hskip\tw@\p@\fi}}}}%
 \fi
 \iflabel@ii
  \ifnum\arrcount@=\m@ne
  \else
   \setbox\ZER@\vbox to\dimen@{\vfil
    \hbox{$\m@th\tsize@\label@ii$}\vskip\tw@\ldimen@ii\vfil}%
   \rlap{\smash{\hskip\tocenter@\hskip\pdimen@
   \rlap{\raise\dimen@ii\hbox{\ifnum\arrcount@=\thr@@\hskip4.5\p@\else
    \hskip2.5\p@\fi\box\ZER@}}}}%
  \fi
 \fi
}
\newdimen\goal@
\newdimen\shifted@
\newcount\Tcount@
\newcount\Scount@
\newbox\shaft@
\newcount\slcount@
\def\getcos@#1{%
 \ifnum\tan@i<\tan@ii
  \dimen@#1%
  \ifnum\slcount@<8 \count@9 \else \ifnum\slcount@<12 \count@8 \else
   \count@7 \fi\fi
  \multiply\dimen@\count@\divide\dimen@\ten@
  \dimen@ii\dimen@\multiply\dimen@ii\tan@i\divide\dimen@ii\tan@ii
 \else
  \dimen@ii#1%
  \count@-\slcount@\advance\count@24
  \ifnum\count@<8 \count@9 \else \ifnum\count@<12 \count@8
   \else\count@7 \fi\fi
  \multiply\dimen@ii\count@\divide\dimen@ii\ten@
  \dimen@\dimen@ii\multiply\dimen@\tan@ii\divide\dimen@\tan@i
 \fi}
\newdimen\adjust@
\def\Nnext@{\ifN@\let\next@\raise\else\let\next@\lower\fi}
\def\arrow@@{\slcount@\angcount@
 \ifNESW@
  \ifnum\angcount@<\ten@
   \let\arrowfont@\arrow@i\global\advance\angcount@\m@ne
   \global\multiply\angcount@13
  \else
   \ifnum\angcount@<19
    \let\arrowfont@\arrow@ii\global\advance\angcount@-\ten@
    \global\multiply\angcount@13
   \else
    \let\arrowfont@\arrow@iii\global\advance\angcount@-19
    \global\multiply\angcount@13
  \fi\fi
  \Tcount@\angcount@
 \else
  \ifnum\angcount@<5
   \let\arrowfont@\arrow@iii\global\advance\angcount@\m@ne
   \global\multiply\angcount@13 \global\advance\angcount@65
  \else
   \ifnum\angcount@<14
    \let\arrowfont@\arrow@iv\global\advance\angcount@-5
    \global\multiply\angcount@13
   \else
    \ifnum\angcount@<23
     \let\arrowfont@\arrow@v\global\advance\angcount@-14
     \global\multiply\angcount@13
    \else
     \let\arrowfont@\arrow@i\global\angcount@117
  \fi\fi\fi
  \ifnum\angcount@=117 \Tcount@115 \else\Tcount@\angcount@\fi
 \fi
 \Scount@\Tcount@
 \ifE@
  \ifnum\tcount@=\z@\advance\Tcount@\tw@\else\ifnum\tcount@=13
   \advance\Tcount@\tw@\else\advance\Tcount@\tcount@\fi\fi
  \ifnum\scount@=\z@\else\ifnum\scount@=13 \advance\Scount@\thr@@\else
   \advance\Scount@\scount@\fi\fi
 \else
  \ifcase\tcount@\advance\Tcount@\thr@@\or\or\advance\Tcount@\thr@@\or
  \advance\Tcount@\tw@\or\advance\Tcount@6 \or\advance\Tcount@7
  \or\or\or\or\or\advance\Tcount@8 \or\advance\Tcount@9 \or
  \advance\Tcount@12 \or\advance\Tcount@\thr@@\fi
  \ifcase\scount@\or\or\advance\Scount@\thr@@\or\advance\Scount@\tw@\or
  \or\or\advance\Scount@4 \or\advance\Scount@5 \or\advance\Scount@\ten@
  \or\advance\Scount@11 \or\or\or\advance\Scount@12 \or\advance
  \Scount@\tw@\fi
 \fi
 \ifcase\arrcount@\or\or\global\advance\angcount@\@ne\else\fi
 \ifN@\shifted@\firsty@\else\shifted@-\firsty@\fi
 \ifE@\else\advance\shifted@\charht@\fi
 \goal@\secondy@\advance\goal@-\firsty@
 \ifN@\else\multiply\goal@\m@ne\fi
 \setbox\shaft@\hbox{\arrowfont@\char\angcount@}%
 \ifnum\arrcount@=\thr@@
  \getcos@{1.5\p@}%
  \setbox\shaft@\hbox to\wd\shaft@{\arrowfont@
   \rlap{\hskip\dimen@ii
    \smash{\ifNESW@\let\next@\lower\else\let\next@\raise\fi
     \next@\dimen@\hbox{\arrowfont@\char\angcount@}}}%
   \rlap{\hskip-\dimen@ii
    \smash{\ifNESW@\let\next@\raise\else\let\next@\lower\fi
      \next@\dimen@\hbox{\arrowfont@\char\angcount@}}}\hfil}%
 \fi
 \rlap{\smash{\hskip\tocenter@\hskip\firstx@
  \ifnum\arrcount@=\m@ne
  \else
   \ifnum\arrcount@=\thr@@
   \else
    \ifnum\scount@=\m@ne
    \else
     \ifnum\scount@=\z@
     \else
      \setbox\ZER@\hbox{\ifnum\angcount@=117 \arrow@v\else\arrowfont@\fi
       \char\Scount@}%
      \ifNESW@
       \ifnum\scount@=\tw@
        \dimen@\shifted@\advance\dimen@-\charht@
        \ifN@\hskip-\wd\ZER@\fi
        \Nnext@
        \next@\dimen@\copy\ZER@
        \ifN@\else\hskip-\wd\ZER@\fi
       \else
        \Nnext@
        \ifN@\else\hskip-\wd\ZER@\fi
        \next@\shifted@\copy\ZER@
        \ifN@\hskip-\wd\ZER@\fi
       \fi
       \ifnum\scount@=12
        \advance\shifted@\charht@\advance\goal@-\charht@
        \ifN@\hskip\wd\ZER@\else\hskip-\wd\ZER@\fi
       \fi
       \ifnum\scount@=13
        \getcos@{\thr@@\p@}%
        \ifN@\hskip\dimen@\else\hskip-\wd\ZER@\hskip-\dimen@\fi
        \adjust@\shifted@\advance\adjust@\dimen@ii
        \Nnext@
        \next@\adjust@\copy\ZER@
        \ifN@\hskip-\dimen@\hskip-\wd\ZER@\else\hskip\dimen@\fi
       \fi
      \else
       \ifN@\hskip-\wd\ZER@\fi
       \ifnum\scount@=\tw@
        \ifN@\hskip\wd\ZER@\else\hskip-\wd\ZER@\fi
        \dimen@\shifted@\advance\dimen@-\charht@
        \Nnext@
        \next@\dimen@\copy\ZER@
        \ifN@\hskip-\wd\ZER@\fi
       \else
        \Nnext@
        \next@\shifted@\copy\ZER@
        \ifN@\else\hskip-\wd\ZER@\fi
       \fi
       \ifnum\scount@=12
        \advance\shifted@\charht@\advance\goal@-\charht@
        \ifN@\hskip-\wd\ZER@\else\hskip\wd\ZER@\fi
       \fi
       \ifnum\scount@=13
        \getcos@{\thr@@\p@}%
        \ifN@\hskip-\wd\ZER@\hskip-\dimen@\else\hskip\dimen@\fi
        \adjust@\shifted@\advance\adjust@\dimen@ii
        \Nnext@
        \next@\adjust@\copy\ZER@
        \ifN@\hskip\dimen@\else\hskip-\dimen@\hskip-\wd\ZER@\fi
       \fi	
      \fi
  \fi\fi\fi\fi
  \ifnum\arrcount@=\m@ne
  \else
   \loop
    \ifdim\goal@>\charht@
    \ifE@\else\hskip-\charwd@\fi
    \Nnext@
    \next@\shifted@\copy\shaft@
    \ifE@\else\hskip-\charwd@\fi
    \advance\shifted@\charht@\advance\goal@-\charht@
    \repeat
   \ifdim\goal@>\z@
    \dimen@\charht@\advance\dimen@-\goal@
    \divide\dimen@\tan@i\multiply\dimen@\tan@ii
    \ifE@\hskip-\dimen@\else\hskip-\charwd@\hskip\dimen@\fi
    \adjust@\shifted@\advance\adjust@-\charht@\advance\adjust@\goal@
    \Nnext@
    \next@\adjust@\copy\shaft@
    \ifE@\else\hskip-\charwd@\fi
   \else
    \adjust@\shifted@\advance\adjust@-\charht@
   \fi
  \fi
  \ifnum\arrcount@=\m@ne
  \else
   \ifnum\arrcount@=\thr@@
   \else
    \ifnum\tcount@=\m@ne
    \else
     \setbox\ZER@
      \hbox{\ifnum\angcount@=117 \arrow@v\else\arrowfont@\fi\char\Tcount@}%
     \ifnum\tcount@=\thr@@
      \advance\adjust@\charht@
      \ifE@\else\ifN@\hskip-\charwd@\else\hskip-\wd\ZER@\fi\fi
     \else
      \ifnum\tcount@=12
       \advance\adjust@\charht@
       \ifE@\else\ifN@\hskip-\charwd@\else\hskip-\wd\ZER@\fi\fi
      \else
       \ifE@\hskip-\wd\ZER@\fi
     \fi\fi
     \Nnext@
     \next@\adjust@\copy\ZER@
     \ifnum\tcount@=13
      \hskip-\wd\ZER@
      \getcos@{\thr@@\p@}%
      \ifE@\hskip-\dimen@\else\hskip\dimen@\fi
      \advance\adjust@-\dimen@ii
      \Nnext@
      \next@\adjust@\box\ZER@
     \fi
  \fi\fi\fi}}%
 \iflabel@i
  \rlap{\hskip\tocenter@
  \dimen@\firstx@\advance\dimen@\secondx@\divide\dimen@\tw@
  \advance\dimen@\ldimen@i
  \dimen@ii\firsty@\advance\dimen@ii\secondy@\divide\dimen@ii\tw@
  \global\multiply\ldimen@i\tan@i\global\divide\ldimen@i\tan@ii
  \ifNESW@\advance\dimen@ii\ldimen@i\else\advance\dimen@ii-\ldimen@i\fi
  \setbox\ZER@\hbox{\ifNESW@\else\ifnum\arrcount@=\thr@@\hskip4\p@\else
   \hskip\tw@\p@\fi\fi
   $\m@th\tsize@@\label@i$\ifNESW@\ifnum\arrcount@=\thr@@\hskip4\p@\else
   \hskip\tw@\p@\fi\fi}%
  \ifnum\arrcount@=\m@ne
   \ifNESW@\advance\dimen@.5\wd\ZER@\advance\dimen@\p@\else
    \advance\dimen@-.5\wd\ZER@\advance\dimen@-\p@\fi
   \advance\dimen@ii-.5\ht\ZER@
  \else
   \advance\dimen@ii\dp\ZER@
   \ifnum\slcount@<6 \advance\dimen@ii\tw@\p@\fi
  \fi
  \hskip\dimen@
  \ifNESW@\let\next@\llap\else\let\next@\rlap\fi
  \next@{\smash{\raise\dimen@ii\box\ZER@}}}%
 \fi
 \iflabel@ii
  \ifnum\arrcount@=\m@ne
  \else
   \rlap{\hskip\tocenter@
   \dimen@\firstx@\advance\dimen@\secondx@\divide\dimen@\tw@
   \ifNESW@\advance\dimen@\ldimen@ii\else\advance\dimen@-\ldimen@ii\fi
   \dimen@ii\firsty@\advance\dimen@ii\secondy@\divide\dimen@ii\tw@
   \global\multiply\ldimen@ii\tan@i\global\divide\ldimen@ii\tan@ii
   \advance\dimen@ii\ldimen@ii
   \setbox\ZER@\hbox{\ifNESW@\ifnum\arrcount@=\thr@@\hskip4\p@\else
    \hskip\tw@\p@\fi\fi
    $\m@th\tsize@\label@ii$\ifNESW@\else\ifnum\arrcount@=\thr@@\hskip4\p@
    \else\hskip\tw@\p@\fi\fi}%
   \advance\dimen@ii-\ht\ZER@
   \ifnum\slcount@<9 \advance\dimen@ii-\thr@@\p@\fi
   \ifNESW@\let\next@\rlap\else\let\next@\llap\fi
   \hskip\dimen@\next@{\smash{\raise\dimen@ii\box\ZER@}}}%
  \fi
 \fi
}
\def\outCD@#1{\def#1{\Err@{\noexpand#1must not be used within \string\CD}}}
\newskip\preCDskip@
\newskip\postCDskip@
\preCDskip@\z@
\postCDskip@\z@
\def\preCDspace#1{\RIfMIfI@
 \onlydmatherr@\preCDspace\else\advance\preCDskip@#1\relax\fi\else
 \onlydmatherr@\preCDspace\fi}
\def\postCDspace#1{\RIfMIfI@
 \onlydmatherr@\postCDspace\else\advance\postCDskip@#1\relax\fi\else
 \onlydmatherr@\postCDspace\fi}
\def\predisplayspace#1{\RIfMIfI@
 \onlydmatherr@\predisplayspace\else
 \advance\abovedisplayskip#1\relax
 \advance\abovedisplayshortskip#1\relax\fi
 \else\onlydmatherr@\preCDspace\fi}
\def\postdisplayspace#1{\RIfMIfI@
 \onlydmatherr@\postdisplayspace\else
 \advance\belowdisplayskip#1\relax
 \advance\belowdisplayshortskip#1\relax\fi
 \else\onlydmatherr@\postdisplayspace\fi}
\def\PreCDSpace#1{\global\preCDskip@#1\relax}
\def\PostCDSpace#1{\global\postCDskip@#1\relax}
\def\CD#1\endCD{%
 \outCD@\cgaps\outCD@\rgaps\outCD@\Cgaps\outCD@\Rgaps
 \preCD@#1\endCD
 \advance\abovedisplayskip\preCDskip@
 \advance\abovedisplayshortskip\preCDskip@
 \advance\belowdisplayskip\postCDskip@
 \advance\belowdisplayshortskip\postCDskip@
 \vcenter{\offinterlineskip
  \vskip\preCDskip@\Let@\global\colcount@\@ne\global\rowcount@\z@
  \everycr{%
   \noalign{%
    \ifnum\rowcount@=\Rowcount@
    \else
     \getrgap@\rowcount@\vskip\getdim@
     \global\advance\rowcount@\@ne\global\colcount@\@ne
    \fi}}%
  \tabskip\z@
  \halign{&\global\xoff@\z@\global\yoff@\z@
   \getcgap@\colcount@\hskip\getdim@
   \hfil\vrule\height\ten@\p@\width\z@\depth\z@
   $\m@th\displaystyle{##}$\hfil
   \global\advance\colcount@\@ne\cr
   #1\crcr}\vskip\postCDskip@}%
 \preCDskip@\z@\postCDskip@\z@
 \def\getcgap@##1{\ifcase##1\or\getdim@\z@\else\getdim@\standardcgap\fi}%
 \def\getrgap@##1{\ifcase##1\getdim@\z@\else\getdim@\standardrgap\fi}%
 \let\Width@\relax\let\Height@\relax\let\Depth@\relax\let\Rowheight@\relax
 \let\Rowdepth@\relax\let\Colwidth@\relax
}

\def\alloc@#1#2#3#4#5{\global\advance\count1#1by\@ne
  \ch@ck#1#4#2%
  \allocationnumber=\count1#1%
  \global#3#5=\allocationnumber
  \wlog{\string#5=\string#2\the\allocationnumber}}
\catcode`\@=\active

    %File information: C:\TEXTS\MET\AAMCR.TEX M     178   23/11/102   18:49
%--------------------------------------------------------------
\catcode`\"=12
%File information: C:\TEXTS\PLAZ\PLFNT.TEX P     140    9/04/101   19:28
%--------------------------------------------------------------
\font\black=cmbx10
\font\sblack=cmbx7
\font\ssblack=cmbx5
\font\blackital=cmmib10  \skewchar\blackital='177
\font\sblackital=cmmib7  \skewchar\sblackital='177
\font\ssblackital=cmmib5  \skewchar\ssblackital='177
\font\sanss=cmss10
\font\ssanss=cmss8 scaled 900
\font\sssanss=cmss8 scaled 600
\font\blackboard=msbm10
\font\sblackboard=msbm7
\font\ssblackboard=msbm5
\font\caligr=eusm10
\font\scaligr=eusm7
\font\sscaligr=eusm5

\font\fraktur=eufm10
\font\sfraktur=eufm7
\font\ssfraktur=eufm5

\font\bsymb=cmsy10 scaled\magstep2
\def\all#1{\setbox0=\hbox{\lower1.5pt\hbox{\bsymb
       \char"38}}\setbox1=\hbox{$_{#1}$} \box0\lower2pt\box1\;}
\def\exi#1{\setbox0=\hbox{\lower1.5pt\hbox{\bsymb \char"39}}
       \setbox1=\hbox{$_{#1}$} \box0\lower2pt\box1\;}

\def\tx#1{{\fam0\relax#1}}

\newfam\bifam
\textfont\bifam=\blackital
\scriptfont\bifam=\sblackital
\scriptscriptfont\bifam=\ssblackital
\def\bi#1{{\fam\bifam\relax#1}}

\newfam\blfam
\textfont\blfam=\black
\scriptfont\blfam=\sblack
\scriptscriptfont\blfam=\ssblack

\newfam\bbfam
\textfont\bbfam=\blackboard
\scriptfont\bbfam=\sblackboard
\scriptscriptfont\bbfam=\ssblackboard
\def\bb#1{{\fam\bbfam\relax#1}}

\newfam\ssfam
\textfont\ssfam=\sanss
\scriptfont\ssfam=\ssanss
\scriptscriptfont\ssfam=\sssanss
\def\ss#1{{\fam\ssfam\relax#1}}

\newfam\clfam
\textfont\clfam=\caligr
\scriptfont\clfam=\scaligr
\scriptscriptfont\clfam=\sscaligr
\def\cl#1{{\fam\clfam\relax#1}}

\newfam\frfam
\textfont\frfam=\fraktur
\scriptfont\frfam=\sfraktur
\scriptscriptfont\frfam=\ssfraktur

\font\syy=cmsy10 scaled 1150
\font\syx=cmsy10 scaled 1300

\define\PR{{\setbox0=\hbox{\syx\char"02}\box0}}
\define\WE{{\setbox0=\hbox{\syy\char"5E}\box0}}

\def\hpb#1{\setbox0=\hbox{${#1}$}
    \copy0 \kern-\wd0 \kern.2pt \box0}
\def\vpb#1{\setbox0=\hbox{${#1}$}
    \copy0 \kern-\wd0 \raise.08pt \box0}

\def\pmb#1{\setbox0\hbox{${#1}$} \copy0 \kern-\wd0 \kern.2pt \box0}
\def\pmbb#1{\setbox0\hbox{${#1}$} \copy0 \kern-\wd0
      \kern.2pt \copy0 \kern-\wd0 \kern.2pt \box0}
\def\pmbbb#1{\setbox0\hbox{${#1}$} \copy0 \kern-\wd0
      \kern.2pt \copy0 \kern-\wd0 \kern.2pt
    \copy0 \kern-\wd0 \kern.2pt \box0}
\def\pmxb#1{\setbox0\hbox{${#1}$} \copy0 \kern-\wd0
      \kern.2pt \copy0 \kern-\wd0 \kern.2pt
      \copy0 \kern-\wd0 \kern.2pt \copy0 \kern-\wd0 \kern.2pt \box0}
\def\pmxbb#1{\setbox0\hbox{${#1}$} \copy0 \kern-\wd0 \kern.2pt
      \copy0 \kern-\wd0 \kern.2pt
      \copy0 \kern-\wd0 \kern.2pt \copy0 \kern-\wd0 \kern.2pt
      \copy0 \kern-\wd0 \kern.2pt \box0}

\mathchardef\za="710B  %\alpha
\mathchardef\zb="710C  %\beta
\mathchardef\zg="710D  %\gamma
\mathchardef\zd="710E  %\delta
\mathchardef\zve="710F %\epsilon
\mathchardef\zz="7110  %\zeta
\mathchardef\zh="7111  %\eta
\mathchardef\zvy="7112 %\theta
\mathchardef\zi="7113  %\iota
\mathchardef\zk="7114  %\kappa
\mathchardef\zl="7115  %\lambda
\mathchardef\zm="7116  %\mu
\mathchardef\zn="7117  %\nu
\mathchardef\zx="7118  %\xi
\mathchardef\zp="7119  %\pi
\mathchardef\zr="711A  %\rho
\mathchardef\zs="711B  %\sigma
\mathchardef\zt="711C  %\tau
\mathchardef\zu="711D  %\upsilon
\mathchardef\zvf="711E %\phi
\mathchardef\zq="711F  %\chi
\mathchardef\zc="7120  %\psi
\mathchardef\zw="7121  %\omega
\mathchardef\ze="7122  %\varepsilon
\mathchardef\zy="7123  %\vartheta
\mathchardef\zvp="7124 %\varpi
\mathchardef\zvr="7125 %\varrho
\mathchardef\zvs="7126 %\varsigma
\mathchardef\zf="7127  %\varphi
\mathchardef\zG="7000  %\Gamma
\mathchardef\zD="7001  %\Delta
\mathchardef\zY="7002  %\Theta
\mathchardef\zL="7003  %\Lambda
\mathchardef\zX="7004  %\Xi
\mathchardef\zP="7005  %\Pi
\mathchardef\zS="7006  %\Sigma
\mathchardef\zU="7007  %\Upsilon
\mathchardef\zF="7008  %\Phi
\mathchardef\zC="7009  %\Psi
\mathchardef\zW="700A  %\Omega

\catcode`\"=\active

%\UseAMSsymbols

\loadmsam
\loadmsbm
\newsymbol\blacksquare 1004
\newsymbol\blacklozenge 1007
\newsymbol\leqslant 1336
\newsymbol\geqslant 133E
\newsymbol\centerdot 1205
\newsymbol\shortparallel 2371

\def\leqs{\leqslant}
\def\geqs{\geqslant}

\def\fpr#1{\underset{{#1}}\to\times}

\newcounter\secno
\newcounter\secna
\newcounter\secnaa
\newcounter\ssecno
\newcounter\sssecno

\font\tfont=cmb10 %scaled 1300

\define\Title#1{\bigpagebreak\flushpar\centerline{\tfont #1}\vskip1.5mm}

\define\Asect#1{\Reset\secnaa1\bigpagebreak
    \flushpar {\bf A\secna.}\,{\bf #1}\vskip1.2mm}
\newfontstyle\secna{\bf}

\define\AAsect#1{\Offset\secna0\bigpagebreak
    \flushpar {\bf A\secna.\secnaa.}\,{\bf #1}\vskip1.2mm}
\newfontstyle\secnaa{\bf}

\define\sect#1{\Reset\ssecno1\bigpagebreak
    \flushpar {\secno.}\,{\bf #1}\vskip1.2mm}
\newfontstyle\secno{\bf}

\define\ssca#1{\Offset\secno0\Reset\sssecno1\bigpagebreak\vskip-4mm
    \flushpar {\secno.\ssecno.}\,{\bf #1}\vskip1.2mm}
\newfontstyle\ssecno{\bf}

\define\sscx#1{\Offset\secno0\Reset\sssecno1\bigpagebreak
    \flushpar {\secno.\ssecno.}\,{\bf #1}\vskip1.2mm}
\newfontstyle\ssecno{\bf}

\define\sssa#1{\Offset\secno0\Offset\ssecno0\bigpagebreak\vskip-4mm
    \flushpar {\secno.\ssecno.\sssecno.}\,{\bf #1}\vskip1.2mm}
\newfontstyle\sssecno{\bf}

\define\sssx#1{\Offset\secno0\Offset\ssecno0\bigpagebreak
    \flushpar {\secno.\ssecno.\sssecno.}\,{\bf #1}\vskip1.2mm}
\newfontstyle\sssecno{\bf}

\def\*{{\textstyle *}}

\newsymbol\blacktriangle 104E

\def\proof{\demo{Proof}}
\def\endproof{\hfill \vrule height4pt width6pt depth2pt \enddemo}

\def\N{{\bb N}}
\def\R{{\bb R}}

\def\Z{{\bb Z}}

\def\ssL{{\scriptscriptstyle {\ss L}}}

\def\*{{\textstyle *}}
\def\s*{{\scriptstyle *}}

\def\by{{\bi y}}

\def\bD{{\bi D}}

\def\bc{{\bi c}}

\def\bph{{\bi p\bi h}}
\def\bPH{{\bi P\bi h}}
\def\bD{{\bi D}}

\def\bC{{\bi C}}

\def\bP{{\bi P}}

\def\sA{{\ss A}}

\def\sC{{\ss C}}

\def\sK{{\ss K}}

\def\sN{{\ss N}}
\def\sO{{\ss O}}

\def\sV{{\ss V}}

\def\sF{{\ss F}}

\def\sQ{{\ss Q}}

\def\sq{{\ss q}}
\def\sp{{\ss p}}

\def\rd{\tx{d}}
\def\xi{\tx{i}}
\def\xD{\tx{D}}

\def\sgn{\operatorname{sgn}}

\def\det{\operatorname{det}}

\def\Sup{\operatornamewithlimits{Sup}}

%Macrodefinitions
    \def\sX{{\ss X}}

    \def\wA{{\widetilde A}}

    \def\of{{\overline f}}
    \def\blangle {{\langle}}
    \def\brangle {{\rangle}}

    \def\wA{{\widetilde A}}

    \def\-{{-}}
    \def\+{{+}}

%end Macrodefinitions

    \input paper.st\relax

    \hsize=37pc
    \hoffset=-10pt
    \vsize=52pc
    \voffset=6pt

%%%%%%%% Fonts other than those  defined in plain TeX %%%%

%   \line{VARELA.TEX \hfill \today}

%    \input xy
%    \xyoption{all}

    \TagsOnRight
    \document
%    \input pictex
%    \input dcpic.sty

%Macrodefinitions

    \def\sgn{\operatorname{sgn}}

    \def\sX{{\ss X}}

    \def\wA{{\widetilde A}}

    \def\of{{\overline f}}

% DEFINIZIONI MIE

    \def\orb{{\overline \rb}}
    \def\oorb{{\overline {\overline \rb}}}

    \def\ozl{{\overline \zl}}
    \def\oozl{{\overline {\overline \zl}}}
    \def\ozm{{\overline \zm}}
    \def\oozm{{\overline {\overline \zm}}}
    \def\ozn{{\overline \zn}}
    \def\oozn{{\overline {\overline \zn}}}

% FINE DEFINIZIONI MIE
    \def\iK{\overset \circ \to K}
    
    \def\blangle {{\pmxb\langle}}
    \def\brangle {{\pmxb\rangle}}

    \def\-{{-}}
    \def\+{{+}}
    
    \def\We{{\operatorname{We}}}
    
    \def\rb{\tx{b}}

    \def\bcD{{\pmbbb{\cl D}}}

    \define\compose#1#2#3#4#5#6{{\setbox0=\hbox{\raise#2\hbox{\kern#3\hbox{${#1}$}}}
\setbox1=\hbox{\raise#5\hbox{\kern#6\hbox{${#4}$}}}\box0\box1}}

    \define\position#1#2#3{{\setbox0=\hbox{\raise#2\hbox{\kern#3\hbox{${#1}$}}}}}

    \def\lpr{{\setbox0=\hbox{\vrule height .15pt width 3.5pt depth 0pt} \setbox1=\hbox{\vrule height 5.8pt width .3pt depth
0pt}\kern2pt\box0\box1\kern3pt}}

    \def\rpr{{\setbox0=\hbox{\vrule height .15pt width 3.5pt depth 0pt} \setbox1=\hbox{\vrule height 5.8pt width .3pt depth
0pt}\kern3pt\box0\kern-3.5pt\box1\kern6.5pt}}

%endMacrodefinitions

    \title
        A variational formulation of electrodynamics with external sources
    \endtitle

    \author
        Antonio De Nicola \\
        CMUC, Department of Mathematics,\\
        University of Coimbra, \\
        3001-454 Coimbra, Portugal\\
        {\tt antondenicola\@gmail.com} \\
        \\
        W\l odzimierz M. Tulczyjew \\
        Associated with \\
        Istituto Nazionale di Fisica Nucleare,
        Sezione di Napoli, Italy \\
        Valle San Benedetto, 2 --
        62030 Montecavallo (MC), Italy\\
        {\tt tulczy\@libero.it} \\
    \endauthor
    \maketitle

{\tenpoint \vskip6mm \line{\bf\hfil Abstract \hfil}
    We present a variational formulation of electrodynamics using de Rham even and odd differential forms. Our formulation
relies on a variational principle more complete than the Hamilton principle and thus leads to field equations with external
sources and permits the derivation of the constitutive relations. We interpret a domain in space-time as an odd de Rham
4-current. This permits a treatment of different types of boundary problems in an unified way. In particular we obtain a
smooth transition to the infinitesimal version by using a current with a one point support.

\vskip5mm

    \noindent{\bf 2000 Mathematics Subject Classification:} {49S05, 70S05}.

    \noindent{\bf Keywords:} variational principles, electrodynamics.

\vskip8mm }

        \flushpar{\bf  Introduction \hfil}\vskip1.2mm

    A general framework for variational formulations of physical theories was presented in [1].  Applications to statics
and dynamics of mechanical systems appear in [2, 3].  In this paper we present a variational formulation of electrodynamics
based on that framework.  Our work is related to the general formulation of linear field theories in a symplectic framework
contained in [4] and to the earlier formulations of electrodynamics contained in [5, 6].  Some of the results contained in
this paper were announced in [7].

    We  are presenting a variational formulation of electrodynamics in an intrinsic, frame independent fashion in the
affine Minkowski space-time using de Rham odd and even differential forms ([8, 9, 10]) which permit the rigorous
formulations of electrodynamics and the description of the transformation properties of electromagnetic fields relative to
reflections (see [5, 6]). Observable quantities like the charge contained in a compact volume or the flux of the
electromagnetic field through a surface require the integration of differential forms. The use of odd quantities is not
just a matter of elegance. Even if the space-time is assumed to be orientable, the alternative approach of using standard
differential forms and the Hodge star operator require the choice of a specific orientation, i.~e. the addition of an
extrinsic structure. Other recent treatments of classical electrodynamics using odd and even differential forms can be
found in [11, 12, 13].

    Our construction is an example of application of the general variational framework [1]. Similar constructions are
needed for the variational formulation of a general field theory. The linearity of classical electrodynamics and the choice
of formulating it on the affine Minkowski space makes our presentation simpler. Relying on a variational principle more
complete than the Hamilton principle our formulation leads to field equations with external sources. This variational
principle also permits the derivation of the constitutive relations which are usually postulated separately since the
variations normally considered are not general enough to derive them from the variational principle.

    We interpret a domain in space-time as an odd de Rham 4-current.  This permits a treatment of different types of
boundary problems in an unified way.  As an example we obtain a smooth transition to the infinitesimal version by using a
current with a one point support.  De Rham currents are essentially objects dual to differential forms.

    The present paper is organized in the following way.  In the first part we provide the geometric structures needed for
the rigorous formulation of electrodynamics. Part of this material is based on [5] and is briefly reported here for the
sake of completeness. We recall the Cartan calculus for odd and even differential forms and their integration theory over
odd and even de Rham currents.

    The second part contains the main results.  We start with the construction of a suitable space of fields for
electrodynamics (not a differential manifold) and a construction of tangent and cotangent vectors.  A convenient
representation of these objects is introduced.  The definition of the space of fields is inspired by a similar construction
suited for the statics of continuous media which is contained in the final section of [1]. In Section 3 we formulate a
variational principle for electrodynamics similar to the virtual action principle of analytical mechanics with external
forces and boundary terms and derive the field equations which include the constitutive relations in addition to Maxwell's
equations.  The boundary problem in a finite domain is treated in Section 4. Section 5 contains the Lagrangian formulation
of electrodynamics. The Legendre transformation and the Hamiltonian formulation of electrodynamics in Section 6 conclude
the paper.

\vskip5mm\penalty-400
        \line{\bf A. Preliminaries \hfil}
 Here we provide the geometric structures needed for the rigorous formulation of electrodynamics in Part B. The material in
Sections 1, 2, and 4 is based on [5] and is briefly recalled here for the sake of completeness. Nevertheless, we add a more
explicit presentation of some useful details.

        \sect{Orientations of vector spaces and vector subspaces.}
    Let $V$ be a vector space of dimension $m \neq 0$.  We denote by $\sF(V)$ the space of linear isomorphisms from $V$ to
$\R^m$ called {\it frames}.  It is known that $\sF(V)$ is a homogeneous space with respect to the natural group action of
the general linear group $GL(m,\R)$ in $\sF(V)$. Let $GL^+(m,\R)$  and $GL^-(m,\R)$ be the two connected  components of the
group $GL(m,\R)$.
    The set of {\it orientations} $\sO(V) = \sF(V) \big/ GL^+(m,\R)$
     has two elements.  This set is a homogeneous space for the quotient group $H(m,\R) = GL(m,\R) \big/ GL^+(m,\R)$.
    The sets $E = GL^+(m,\R)$ and $P = GL^-(m,\R)$ are the elements of the quotient group $H(m,\R)$ which is the group of
permutations of the two elements of $\sO(V)$.
    There is an ordered base $(e_1,e_2,\ldots ,e_m)$ of $V$ associated with each frame $\zx$ in a obvious way.

    Let $W \subset V$ be a subspace of a vector space $V$.  The subspace has the set $\sO(W)$ of orientations called {\it
inner orientations} of $W$.  Orientations of the quotient space $V\big/W$ are called {\it outer orientations} of $W$.  An
outer orientation $o''$ of $W$ can be determined by specifying an inner orientation $o$ of $W$ together with an orientation
$o'$ of $V$.  Let $(e_1,\ldots,e_n)$ be the base of $W$ associated with a frame $\zx \in o$.  This base can be completed to
a base $(e'_1,\ldots,e'_m)$ of $V$ with $(e'_1,\ldots,e'_n) = (e_1,\ldots,e_n)$.  The extended base can be chosen to be
associated with a frame $\zx' \in o'$.  Let $\zp \colon V \rightarrow V\big/W$ be the canonical projection.  The sequence
        $$(e''_1,\ldots,e''_{m-n}) = (\zp(e'_{n+1}),\ldots,\zp(e'_m))
                                                                                                        \tag \label{Fvp1}$$
    is a base of $V\big/W$.  It determines an orientation $o''$ of $V\big/W$.  Hence an outer orientation of $W$.  The
outer orientation $o''$ of $W$ constructed from $o \in \sO(W)$ and $o' \in \sO(V)$ is the same as the orientation
constructed from $Po$ and $Po'$.

    We have introduced inner orientation of subspaces of dimension different from zero and outer orientation of subspaces
of codimension different from zero.  Integration theory of differential forms requires the possibility of assigning inner
orientations to the subspace $W = \{0\} \subset V$ and outer orientations to the subspace $W = V$.  Two possible
orientations are assigned to the subspace $W = \{0\} \subset V$ one of which is distinguished. The distinguished
orientation is denoted by $(+)$ and  the other orientation is denoted by $(-)$.  In agreement with the conventions
established for orientation the outer orientations of the subspace $W = \{0\} \subset V$ are the orientations of $V$. An
outer orientation of $W = \{0\}$ can be specified in terms of an inner orientation and an orientation of $V$. If the inner
orientation is $(+)$ and the orientation of $V$ is $o$, then $o$ is the outer orientation of $W$.  The orientation $Po$ is
the outer orientation derived from $(-)$ and $o$.
    The subspace $W=V$ has a distinguished outer orientation defined as an orientation of $V\big/W$.

        \sect{Multicovectors and multivectors.}
    A {\it $q$\,-\,covector} in a vector space $V$ is a mapping
        $a\, \colon \times^q V \times \sO(V) \rightarrow \R$.
    This mapping is $q$\,-\,linear and totally antisymmetric in its vector arguments.  A $q$\,-\,covector $a$ is said to be
{\it even}, if
    $$a(v_1, v_2, \ldots ,v_q,Po) = a(v_1, v_2, \ldots ,v_q,o).
                                                                                                        \tag \label{Fvp2}$$
    It is said to be {\it odd}, if
        $$a(v_1, v_2, \ldots ,v_q,Po) = - a(v_1, v_2, \ldots ,v_q,o).
                                                                                                        \tag \label{Fvp3}$$
    The vector space of even $q$\,-\,covectors will be denoted by $\wedge^q_e V^\*$ and the space of odd $q$\,-\,covectors
will be denoted by $\wedge^q_o V^\*$.  We will use the symbol $\wedge^q_p V^\*$ to denote either of the spaces in
constructions valid for both parities.  The index $p$ with the two possible values $e$ and $o$ will be used on other
occasions.

    For the definition of {\it exterior product} of (even and odd) multicovectors we refer to [5]. Here we just recall that
    if two multicovectors $a$ and $a'$ are even or both are odd, the product $a \wedge a'$ is even.  In other cases the
product is odd.  The exterior product is commutative in the graded sense and associative.

    Let $\{e_\zk\}_{\zk = 1, \ldots ,m}$ be a base of $V$ and let $\{e^\zk\}_{\zk = 1, \ldots ,m}$ be the dual base.  Each
element $e^\zk$ defines an even covector
        $$e_e^\zk \colon V \times \sO(V) \rightarrow \R \colon (v,o) \mapsto \langle e^\zk, v\rangle.
                                                                                                        \tag \label{Fvp4}$$
    We choose an orientation $o$ of $V$ and introduce odd 0-covector $e_o$ defined by
    $e_o(o) = -e_o(Po)=1$ and the even covector $e_e$ defined by $e_e(o) = e_e(Po) =1$.

We come now to multivectors.
    We denote by $\sK(\times^q V \times \sO(V))$ the vector space of formal linear combinations of sequences $(v_1, v_2,
\ldots ,v_q,o) \in \times^q V \times \sO(V)$.
    In the space $\sK(\times^q V \times \sO(V))$ we introduce subspaces
        $$\align
        \sA^{\,p}_q(V) &= \left\{\tsize{\sum_{\,i=1}^{\,n}}\, \zl_i(v^i_1, v^i_2, \ldots ,v^i_q,o^i) \in \sK(\times^q V
\times \sO(V)) ; \right.\hskip38mm \\
            &\hskip32mm \left. \tsize{\sum_{\,i=1}^{\,n}}\, \zl_i a(v^i_1, v^i_2, \ldots ,v^i_q,o^i) = 0 \;\text{ for each
}\;\; a \in \wedge^q_p V^\* \right\}.
                                                                                                        \tag \label{Fvp5}\endalign$$
    Subsequently we define quotient spaces $\wedge^q_p V = \sK(\times^q V \times \sO(V)) \big/ \sA^{\,p}_q(V)$.
    Elements of spaces $\wedge^q_e V$ and $\wedge^q_o V$ are called {\it even $q$\,-\,vectors} and {\it odd
$q$\,-\,vectors} respectively.
    A multivector is said to be {\it simple} if it is represented by a single element of the space $V^q \times \sO(V)$
interpreted as a subspace of $\sK(V^q \times \sO(V))$.
    Evaluation of $q$\,-\,covectors on sequences $(v_1, v_2, \ldots ,v_q,o) \in \times^q V \times \sO(V)$ extends to linear
combinations and their equivalence classes.  If $w$ is a $q$\,-\,vector represented by the linear combination
        $\tsize{\sum_{\,i=1}^{\,n}}\, \zl_i(v^i_1, v^i_2, \ldots ,v^i_q,o^i)$ and $a$ is a $q$\,-\,covector of the same
parity as $w$, then
        $$\langle a, w\rangle = \tsize{\sum_{\,i=1}^{\,n}}\, \zl_i a(v^i_1, v^i_2, \ldots ,v^i_q,o^i)
                                                                                                        \tag \label{Fvp6}$$
    is the evaluation of $a$ on $w$.  We have constructed pairings
        $\langle \;,\,\rangle \colon \wedge^q_p V^\* \times \wedge^q_p V \rightarrow \R$.
    The {\it exterior product} of multivectors can be easily defined using representatives. The parity of the exterior
product is odd if the parity of one of the factors is odd.  It is even otherwise.
    The exterior product is commutative in the graded sense and associative.

    The {\it left interior multiplications} are the operations $\lpr \colon \wedge^{q}_{p} V \times \wedge^{q'}_{p'} V^\*
\rightarrow \wedge^{q'-q}_{pp'} V^\*,$
    defined for $q \leqs q'$ by $\langle w \lpr a, w' \rangle = \langle a, w \wedge w' \rangle.$
    The parity $pp'$ which appears in this definition is constructed by assigning the numerical values $+1$ and $-1$ to $e$
and $p$ respectively.  The parity of the multivector $w'$ must match the parity of the multicovector $w \lpr a$.
    The {\it right interior multiplications} are the operations
        $\rpr \colon \wedge^{q}_{p} V \times \wedge^{q'}_{p'} V^\* \rightarrow \wedge^{q-q'}_{pp'} V,$
    defined for $q \geqs q'$ by $\langle a', w \rpr a \rangle = \langle a' \wedge a, w \rangle$.
    The parity of the multicovector $a'$ in this definition must match the parity of the multivector $w \rpr a$.
\vskip7mm\penalty-400
        \sect{The Weyl isomorphism and a useful formula.}
    The space $\wedge_o^m V^\*$ is one-dimensional.  This makes it possible to define the tensor product $\wedge_e^qV
\otimes \wedge_o^m V^\*$ as the set of equivalence classes of pairs $(w,e) \in \wedge_e^qV \times \wedge_o^m V^\*$.  Pairs
$(w,e)$ and $(w',e')$ are equivalent if there is a number $\zl$ such that $w' = \zl w$ and $e = \zl e'$ or $w = \zl w'$ and
$e' = \zl e$.  The equivalence class of a pair $(w,e)$ will be denoted by $w \otimes e$.  A tensor $\overline a \in
\wedge_e^qV \otimes \wedge_o^m V^\*$ will always be presented as a product $w \otimes e$.  The set $\wedge_e^qV \otimes
\wedge_o^m V^\*$ is a vector space (see [5]).

        \claim \c{p}{Proposition}{([5])}                                                 \label{C1}
    The linear mapping
        $$\We_q \colon \wedge_e^qV \otimes \wedge_o^m V^\* \rightarrow \wedge_o^{m-q}V^\* \colon w \otimes e \mapsto w \lpr
e
                                                                                                        \tag \label{Fvp7}$$
    is an isomorphism.
        \endclaim

    The mapping $\We_q$ is called the {\it Weyl isomorphism}.

    We will show now that the values of any bilinear mapping $\rb \colon \wedge_e^q V^\* \times  \wedge_e^{q'} V^\*
\rightarrow \wedge_o^m V^\*$
    can be expressed using the exterior product.  The following three technical lemmas will be used.

    \claim \c{l}{Lemma}{}                                                       \label{L1}
    Let $w\in\wedge ^q_e V$. Then
        $$w\lpr (e_o\wedge e^1_e\wedge \ldots\wedge e^m_e)  = \sum_{\zn_1< \ldots <\zn_q}
(-1)^{\sum_{i=1}^q(\zn_i-i)}\langle   e^{\zn_1}_e\wedge \ldots\wedge e^{\zn_q}_e,w \rangle e_o\wedge e^{\zn_q+1}_e\wedge
\ldots\wedge e^{\zn_m}_e
                                                                                                        \tag \label{Fvp8}$$
    where $\zn_{q+1}< \ldots <\zn_m$ and $(\zn_{q+1}, \ldots, \zn_m)$ denotes the complementary $(m-q)$-tuple of $(\zn_1,
\ldots ,\zn_q)$.
        \endclaim
        \proof
    It is enough to prove the claim for a simple $q$-vector.  If $w = w_1 \wedge \ldots \wedge w_q$, then
        $$\align
    (& w_1 \wedge \ldots \wedge w_q) \lpr(e_o\wedge e^1_e\wedge \ldots\wedge e^m_e)\\
            &= \sum_{\zn_1=1}^m (-1)^{\zn_1-1} \langle e^{\zn_1}_e,w_1 \rangle w_q\lpr\left( \ldots \lpr \left(w_2 \lpr
\left(e_o \wedge e^1_e \wedge \ldots \wedge \widehat{e^{\zn_1}_e} \wedge \ldots \wedge e^m_e\right)\right)\right)\\
            &= \ldots = \sum_{\zn_1\neq\ldots\neq\zn_q} (-1)^{\sum_{i=1}^q (\zn_i-i)} \sgn(\zn_1,\ldots,\zn_q)\langle
e^{\zn_1}_e,w_1\rangle \cdots \langle e^{\zn_q}_e,w_q\rangle\\
    &\hskip67mm e_o \wedge e^1 \wedge \ldots \wedge \widehat{e^{\zn_1}_e} \wedge \ldots \wedge\widehat{e^{\zn_q}_e} \wedge
\ldots \wedge e^m_e\\
            &= \sum_{\zn_1<\ldots<\zn_q} \sum_{\zs\in S(q)}(-1)^{\sum_{i=1}^q (\zn_i-i)} \sgn\zs \langle
e^{\zn_{\zs(1)}}_e,w_1\rangle \cdots \langle e^{\zn_{\zs(q)}}_e,w_q\rangle\\
    &\hskip67mm e_o \wedge e^1 \wedge \ldots \wedge \widehat{e^{\zn_1}_e} \wedge \ldots \wedge\widehat{e^{\zn_q}_e} \wedge
\ldots \wedge e^m_e\\
            &= \sum_{\zn_1<\ldots<\zn_q} \sum_{\zs\in S(q)}(-1)^{\sum_{i=1}^q (\zn_i-i)} \sgn\zs \langle
e^{\zn_{\zs(1)}}_e,w_1\rangle \cdots \langle e^{\zn_{\zs(q)}}_e,w_q\rangle e_o \wedge e^{\zn_{q+1}} \wedge \ldots \wedge
e^{\zn_{m}}_e\\
            &= \sum_{\zn_1<\ldots<\zn_q} (-1)^{\sum_{i=1}^q (\zn_i-i)} \det(\langle e^{\zn_r}_e,w_s\rangle) e_o \wedge
e^{\zn_{q+1}} \wedge \ldots \wedge e^{\zn_{m}}_e.
                                                                                                        \tag \label{Fvp9}\endalign$$
    In the second line of this sequence of equalities we used the well known identity
        $$v\lpr (a^1 \wedge \ldots  \wedge a^q) = \sum_{\zn=1}^q (-1)^{\zn - 1} \langle a^{\zn},v \rangle a^1 \wedge \ldots
\wedge \widehat{a^{\zn}} \wedge \ldots  \wedge a^q
                                                                                                        \tag \label{Fvp10}$$
    where $v$ is a vector and  $a^1,\ldots,a^q$ are covectors.  In the third line we applied repeatedly this identity and
we noted that when $\zn_1<\ldots<\zn_q$ each missing covector $e_e^{\zn_i}$ in the expression $e_o \wedge e^1 \wedge \ldots
\wedge \widehat{e^{\zn_1}_e} \wedge \ldots \wedge\widehat{e^{\zn_q}_e} \wedge \ldots \wedge e^m_e$ occupied the
$(\zn_i-i+1)$-th place in the exterior product, otherwise if $\zn_{i-l-1}<\zn_i<\zn_{i-l}$, then it occupied the
$(\zn_i-i-l+1)$-th place, hence we get the factor $(-1)^{\sum_{i=1}^q(\zn_i-i)}\sgn(\zn_1,\ldots,\zn_q)$ where the symbol
$\sgn(\zn_1,\ldots,\zn_q)$ denotes the sign of the permutation $(\zn_1,\ldots,\zn_q)$.
        \endproof \vskip\baselineskip
        \claim \c{l}{Lemma}{}                                                   \label{L2} If $a\in \wedge_e^q V^\*$ and $w
\otimes e \in \wedge_e^q V \otimes \wedge_o^m V^\*$, then
        $$\langle a, w \rangle e = a \wedge \We_q(w\otimes e).
                                                                                                        \tag \label{Fvp11}$$
        \endclaim
        \proof
    It is enough to prove the claim for $e=e_o \wedge e^1_e \wedge \ldots \wedge e^m_e$. By the Lemma \Ref{L1} we get
        $$\align
        & a \wedge \We_q(w\otimes e)\\
        &= a \wedge (w \lpr (e_o \wedge e^1_e \wedge \ldots \wedge e^m_e))\\
        &= \sum_{\zn_1< \ldots <\zn_q} (-1)^{\sum_{i=1}^q(\zn_i-i)} \langle w,  e^{\zn_1}_e \wedge \ldots \wedge
e^{\zn_q}_e\rangle a \wedge e_o \wedge e^{\zn_{q+1}}_e \wedge \ldots \wedge e^{\zn_m}_e\hskip45mm\\
        &= \sum_{\zn_1< \ldots < \zn_q} (-1)^{\sum_{i=1}^q(\zn_i-i)} w^{\zn_1\ldots\zn_q} \sum_{\zm_1< \ldots <\zm_q}
a_{\zm_1\ldots\zm_q} e_e \wedge e^{\zm_1}_e \wedge \ldots \wedge e^{\zm_q}_e \wedge e_o \wedge e^{\zn_{q+1}}_e \wedge
\ldots \wedge e^{\zn_m}_e.\hskip-1mm
                                                                                                        \tag \label{Fvp12}\endalign$$
    The right-hand side of \Ref{Fvp12} reduces to
        $$\sum_{\zn_1< \ldots < \zn_q} (-1)^{\sum_{i=1}^q(\zn_i-i)} w^{\zn_1\ldots\zn_q} a_{\zn_1\ldots\zn_q} e_o \wedge
e^{\zn_1}_e \wedge \ldots \wedge e^{\zn_m}_e,
                                                                                                        \tag\label{Fvp13}$$
     since  $\zn_{q+1}< \ldots <\zn_m$ and $(\zn_{q+1}, \ldots, \zn_m)$ is the complementary $(m-q)$-tuple of $(\zn_1,
\ldots ,\zn_q)$. Finally we note that  moving each $\zn_i$ (for $i= 1,\ldots,q$) to the $\zn_i$-th place in \Ref{Fvp12}
requires $\zn_i-i$ transpositions, since $\zn_1< \ldots <\zn_q$. Then each of the remaining $\zn_i$, i.e. those with
$i=q+1,\ldots,m$, will necessarily be at the $\zn_i$-th place.  Hence we obtain
        $$a \wedge \We_q(w\otimes e) = \sum_{\zn_1 < \ldots < \zn_q} a_{\zn_1\ldots\zn_q} w^{\zn_1\ldots\zn_q}  e_o \wedge
e^1_e \wedge \ldots \wedge e^m_e.
                                                                                                        \tag \label{Fvp14}$$
        \endproof \vskip\baselineskip
    We denote by $\text{Hom}(\wedge_o^m V^\*|\wedge_e^q V^\*)$ the space of linear mappings from $\wedge_e^q V^\*$ to
$\wedge_o^m V^\*$ and by
        $$i_q\colon \text{Hom}(\wedge_o^m V^\*|\wedge_e^q V^\*)\rightarrow \wedge_e^qV \otimes \wedge_o^m V^\*
                                                                                                        \tag \label{Fvp15}$$
    the isomorphism characterized by
        $$\langle i_q(l), a'\otimes u \rangle =\langle l(a'),u\rangle
                                                                                                        \tag \label{Fvp16}$$
    for each $l\in\text{Hom}(\wedge_o^m V^\*|\wedge_e^q V^\*)$, $a'\in \wedge_e^q V^\*$ and $u\in \wedge_o^m V$.  The
pairing
        $$\langle \, , \, \rangle\colon  (\wedge_e^qV \otimes \wedge_o^m V^\*)\times (\wedge_e^qV^\* \otimes \wedge_o^m
V)\rightarrow \R\colon (w\otimes e,a\otimes u)\mapsto \langle a,w \rangle \langle e,u \rangle
                                                                                                        \tag \label{Fvp17}$$
    is used.

        \claim \c{l}{Lemma}{}                                                   \label{L3} If $l\in\text{Hom}(\wedge_o^m
V^\*|\wedge_e^q V^\*)$, then
        $$l(a) = a \wedge \We_q(i_q(l)),
                                                                                                        \tag \label{Fvp18}$$
    for each $a\in \wedge^q_e V^\*. $
        \endclaim
        \proof
    Let $i_q(l)=w_l\otimes e$ with $w_l \in \wedge ^q_e V$.  Using Lemma \Ref{L2} we obtain
        $$a \wedge \We_q(i(l)) = a \wedge \We_q(w_l\otimes e) = \langle a, w_l \rangle e.
                                                                                                        \tag \label{Fvp19}$$
    We will show now that $l(a) = \langle a, w_l \rangle e$.  Indeed if we denote by $u\in\wedge ^m_o V$ the dual basis of
$e$ then we have $l(a) = \langle l(a),u \rangle e$ and
        $$\langle l(a),u \rangle = \langle i(l),a\otimes u\rangle = \langle w_l\otimes e,a\otimes u\rangle = \langle
a,w_l\rangle\langle e, u\rangle = \langle a,w_l\rangle.
                                                                                                        \tag \label{Fvp20}$$

        \endproof

    We can now finally show that a useful expression can be obtained for the values of any bilinear mapping
        $$\rb \colon \wedge_e^q V^\* \times  \wedge_e^{q'} V^\* \rightarrow \wedge_o^m V^\*.
                                                                                                        \tag \label{Fvp21}$$
    We associate with $b$ the linear mappings
        $$\orb \colon \wedge_e^q V^\* \rightarrow \wedge_e^{q'}V \otimes \wedge_o^m V^\* \colon a \mapsto
i_{q'}(\rb(a,\cdot))
                                                                                                        \tag \label{Fvp22}$$
    and
        $$\oorb \colon \wedge_e^{q'} V^\* \rightarrow \wedge_e^q V \otimes \wedge_o^m V^\* \colon a' \mapsto
i_{q}(\rb(\cdot,a')).
                                                                                                        \tag \label{Fvp23}$$

            \claim \c{p}{Proposition}{}                                             \label{C4}
    If $\rb$ as in \Ref{Fvp21} is a bilinear mapping and $\orb ,\oorb$ are the associated linear mappings \Ref{Fvp22} and
\Ref{Fvp23}, then
        $$\rb(a,a') = a' \wedge \We_{q'}(\orb(a)) = a \wedge \We_{q}(\oorb(a')).
                                                                                                        \tag \label{Fvp24}$$
    for each $(a,a')\in \wedge_e^q V^\* \times \wedge_e^{q'} V^\*$.

        \endclaim
        \proof
    Applying Lemma \Ref{L3} to $l=b(a,\cdot)$ we get
        $$b(a,a') = b(a, \cdot)(a')= a' \wedge \We_{q'}(i_{q'}(b(a, \cdot))) =a' \wedge \We_{q'}(\orb(a)).
                                                                                                        \tag \label{Fvp25}$$
    The other equality is obtained in the same way by applying Lemma \Ref{L3} to $l=b(\cdot,a')$.
        \endproof

\vskip5mm\penalty-500
        \sect{Integration of differential forms. Chains and currents.}
    Let $M$ be an affine space modelled on a vector space $V$.  A {\it differential $q$\-form} on $M$ is a differentiable
function $A \colon M \times \PR^q V \times \sO(V) \rightarrow \R$
    depending on a point, $q$ vectors and an orientation.  It is $q$\-linear and totally antisymmetric in its vector
arguments.  A differential form $A$ is said to be {\it even} or {\it odd} if for each point in $x\in M$ it defines a even
or odd multicovector, respectively.
    We note that a zero-form on $M$ is a differentiable function $f \colon M \times \sO(V) \rightarrow \R$.
    The vector space of even differential $q$\-forms will be denoted by $\zF^q_e(M)$ and space of odd differential
$q$\-forms will be denoted by $\zF^q_o(M)$.  The symbol $\zF^q_p(M)$ will be used to denote either of the two spaces when
the distinction is of no importance.

    For the definition of the exterior product of two differential forms and for that of exterior differential of a
differential form, we refer the reader to [5]. Here we just recall that if both forms $A$ and $A'$ are even or both are
odd, the product $A \wedge A'$ is even.  In other cases the product is odd. The parity of the differential $\rd A$ of a
$q$\-form $A$ is the same as the parity of the original form $A$.
    A $q$\-form $A$ can be interpreted as a {\it $q$-covector field} $\wA \colon M \rightarrow \WE^q_p V^\*$.
    The exterior product and the exterior differential are extended to this alternative interpretation of forms.
    The left and right interior multiplications of even and odd multivector fields with even and odd multicovector fields
are defined point by point in an obvious manner.

  A {\it cell} of dimension $q$ or a $q$\,-\,{\it cell }in $M$ is a pair $(\zq,o)$, where $\zq$ is a differentiable mapping
$\zq \colon \R^q \rightarrow M$ and $o$ is an orientation of $V$.  For $q=0$, $\R^0$ is the vector space $\{0\}$ with a
single element $0$. Hence a zero-cell in $M$ is a pair of a point $x\in M$ and an orientation of $V$.  The {\it integral}
of a $q$\,-\,form $A$ on a cell $(\zq,o)$ is the Riemann integral
        $$\int_{(\zq,o)} A = \int_0^1\cdots\int_0^1
A\left(\zq(s_1,\ldots,s_q),\xD_1\zq(s_1,\ldots,s_q),\ldots,\xD_q\zq(s_1,\ldots,s_q),o\right) \rd s_1 \cdots \rd s_q.
                                                                                                        \tag \label{Fvp27}$$
  The {\it integral} of a $0$\,-\,form $f$ on a zero-cell $(x,o)$ is
        the value $\int_{(x,o)} f = f(x,o)$.
    For each $q$ we introduce the space $\sK(X_q(M))$ of formal linear combinations of $q$\,-\,cells.  The formal linear
combinations turn into real linear combinations if cells are identified with elements of $\sK(X_q(M))$.  Integration of
forms is extended to linear combinations by linearity.  Subspaces $\sN^{\,p}_q(M) \subset \sK(X_q(M))$ are defined as the
sets
        $$\sN^{\,p}_q(M) = \left\{C \in \sK(X_q(M)) ;\; \int_C A = 0\;\;\text{ for each }\;\;A \in \zF^q_p(M) \right\}.
                                                                                                        \tag \label{Fvp28}$$
    Elements of the quotient spaces $\zX^p_q(M) = \sK(X_q(M))\big/\sN^{\,p}_q(M)$ are called {\it even chains} or {\it odd
chains} of dimension $q$.  We extend the sequence of even and odd chains to negative dimension $q$ by defining the spaces
$\zX^p_q(M) =\{0\}$ for each $q < 0$.
    A chain is said to be simple if it has a single cell as one of its representatives.  Integrals of $q$\,-\,forms on
$q$\,-\,chains are well defined.  The integral of a $q$\,-\,form $A$ on the class $\bC$ of $C \in \sK(X_q(M))$ is the
integral of $A$ on $C$.

    The {\it boundary operator} $\partial$ assigns to a chain $\bC \in \zX^{\,p}_q(M)$ its {\it boundary} $\partial\bC \in
\zX^{\,p}_{q-1}(M)$.  The boundary of a simple chain represented by a $q$\,-\,cell $(\zq,o)$ is the chain represented by
the combination
        $$\tsize{\sum_{i=1}^q} (-1)^{i-1} ((\zq^{(i,1)},o) - (\zq^{(i,0)},o)),
                                                                                                        \tag \label{Fvp29}$$
    where the $(q\-1)$\,-\,cells $(\zq^{(i,1)},o)$ and $(\zq^{(i,0)},o)$ are defined by
        $$\hskip-1mm\zq^{(i,*)} \colon \R^{q-1} \rightarrow M \colon (s_1,\ldots,\widehat{s_i},\ldots,s_q) \mapsto
\zq(s_1,\ldots,s_{i-1},*,s_{i+1}\ldots,s_q)
                                                                                                        \tag \label{Fvp30}$$
    with $*$ replaced by $1$ and $0$, respectively.
    The cells introduced in \Ref{Fvp30} represent the {\it faces} of the simple chain.
    The construction of the boundary is extended to generic chains by linearity.  The boundary of a boundary is the zero
chain.
   It is known that Stokes theorem holds for chains and forms of the same parity.

    An {\it even} or {\it odd de Rham current} of dimension $q$ on a manifold $M$ is a linear function
        $\bc \colon \zF^q_p(M) \rightarrow \R$.
    We will use the symbol $\int_\bc A$ to denote the value $\bc(A)$.
    The spaces of forms are given certain topologies and the continuity of the function is required.  Chains will be
treated as currents.  They form a dense subspace in the space of currents.  We will consider only very simple examples of
currents other than chains. Topological considerations are of little importance for these examples.  The boundary of a
current is defined by assuming that Stokes theorem holds for currents.  Thus if $\bc$ is a current of dimension $q$, then
the boundary of $\bc$ is the mapping
        $$\partial\bc \colon \zF_p^{q-1}(M) \rightarrow \R \colon A \mapsto \int_{\partial\bc} A = \int_\bc \rd A.
                                                                                                        \tag \label{Fvp32}$$

    In addition to chains the odd de Rham current most frequently used is the {\it Dirac current} $w\zd(x)$ of dimension
$m$ derived from an odd $m$\,-\,vector $w$ and a point $x \in M$.  If $A$ is an odd $m$\,-\,form, then
        $$\int_{w\zd(x)}A = \langle \wA(x), w\rangle.
                                                                                                        \tag
\label{Fvp33}$$ \vskip10mm\penalty-500
            \line{ \bf B. Electrodynamics \hfil}

\Reset\secno 1
        \sect{The space of fields.}
    In this section we will construct a suitable space of fields for electrodynamics which has the same role of the space
of motions in mechanics (see [3]). The space of fields is not a differentiable manifolds. Nevertheless, we will introduce
in this space enough structure to permit the construction of tangent and cotangent vectors which are the objects needed in
in the formulation of the variational principle. A class of admissible functions defined on the space of fields has also to
be specified. This class needs to be large enough to include the action, which for electrodynamics is generated (in a sense
that will be specified) from a quadratic Lagrangian density, due to the linearity of the theory.  Thus, only functions
generated from quadratic mappings will be used in our variational formulation of electrodynamics. A larger class of
admissible functions would be needed to deal with more general, non linear field theories, and the price of increased
technical difficulties should be paid.

    Let $M$ be the affine Minkowski space-time of special relativity with the $4$-dimensional model space $V$ and the non
degenerate metric tensor $g: V \rightarrow  V^\*$ of signature $(1,3)$.  The space of odd $4$-currents with compact
supports in $M$ will be denoted by $\sC M$.  Differential forms will always be presented as covector fields.

    We consider the set $\sX(\zF^1_e(M);\sC M)$ of pairs $(A,\bc)$, where $\bc$ is an odd current of dimension $4$ in $M$
with a compact support $\Sup(\bc)$ and $A$ is a local even $1$\,-\,form
        $$A \colon U \rightarrow \wedge^1_e V^\*
                                                                                                        \tag \label{Fvp34}$$
    defined in an open set $U\subset M$ containing the support of $\bc$.  The $1$\,-\,form $A$ will represent the
electromagnetic potential.

    A mapping
        $$\zk \colon M\times \wedge^1_e V^\* \times  \wedge^2_e V^\* \rightarrow  \wedge^4_o V^\*
                                                                                                        \tag \label{Fvp35}$$
     is said to be {\it quadratic} if for each $x\in M$ there exists a symmetric bilinear mapping
        $$\zd^2\zk_x\colon \left(\wedge^1_e V^\* \times  \wedge^2_e V^\*    \right) \times \left(\wedge^1_e V^\* \times
\wedge^2_e V^\*  \right) \rightarrow  \wedge^4_o V^\*
                                                                                                        \tag \label{Fvp36}$$
     such that the mappings $\zk_x=\zk(x,\cdot,\cdot)$ and $\zd^2\zk_x$ satisfy the equality
        $$\zk_x(a,f)=\frac{1}{2} \zd^2\zk_x((a,f)(a,f)),
                                                                                                        \tag \label{Fvp37}$$
    for each $(a,f)\in \wedge^1_e V^\* \times \wedge^2_e V^\*$.  We are using the standard definition of quadratic mappings
in terms of polarizations.  The polarization of the mapping $\zk_x$ is the mapping $\zd^2\zk_x$ and the equation
\Ref{Fvp37} is the standard relation between a quadratic mapping defined on a vector space and its polarization. We will
use the set of quadratic mappings \Ref{Fvp35} to introduce an equivalence relation in the set $\sX(\zF^1_e(M);\sC M)$. In
the next section they will be also used to define the admissible functions defined on the space of fields.

    Pairs $(A,\bc)$ and $(A',\bc')$ are equivalent if
        $$\int_{\textstyle{\bc'}} \zk \circ (x,A',\rd A') = \int_{\textstyle{\bc}} \zk \circ (x,A,\rd A)
                                                                                                        \tag \label{Fvp38}$$
    for each quadratic mapping $\zk \colon  M \times \wedge^1_e V^\* \times  \wedge^2_e V^\* \rightarrow  \wedge^4_o V^\*$.
    The symbol $x$ is used to indicate the identity mapping of $M$ and also a point of $M$.

    If $\zm$ is an arbitrary odd 4-form on $M$ and $\zk$ is set to be the mapping
        $$\zk \colon M\times \wedge^1_e V^\* \times  \wedge^2_e V^\* \rightarrow  \wedge^4_o V^\* \colon (x,a,f)\mapsto
\zm(x),
                                                                                                        \tag \label{Fvp39}$$
    then the equivalence condition \Ref{Fvp38} reduces to
        $$\int_{\textstyle{\bc'}} \zm = \int_{\textstyle{\bc}} \zm
                                                                                                        \tag \label{Fvp40}$$
    and implies that $\bc'=\bc$.

    Equivalence classes of elements of $\sX(\zF^1_e(M);\sC M)$ will be called {\it fields}. Our fields are similar to those
used by Freed in [14]. The space of fields will be denoted by $\sQ(\zF^1_e(M);\sC M)$ or simply $Q$.  The equivalence class
of $(A,\bc)$ will be denoted by $\sq(A,\bc)$ or simply $q$. There is a natural projection
        $$\ze \colon \sQ(\zF^1_e(M);\sC M) \rightarrow \sC M \colon \sq(A,\bc) \mapsto \bc
                                                                                                        \tag \label{Fvp41}$$
    from the space of fields to the space $\sC M$ of currents in $M$.  It is not difficult to check that each fibre
$\ze^{-1}(\bc)$ of the projection $\ze$ is a vector space which will be denoted by the symbol $\sQ(\zF^1_e(M);\bc)$ or
$Q_\bc$. Indeed, the sum of two pairs $(A,\bc)$ and $(A',\bc)$ with $A \colon U \rightarrow \wedge^1_e V^\*$ and $A' \colon
U' \rightarrow \wedge^1_e V^\*$ is the pair $(A+A',\bc)$ defined in $U\cap U'\supset \Sup(\bc)$. Descending to the quotient
with respect to the equivalence relation defined by \Ref{Fvp38} easily gives the sum in $Q_\bc$.  Therefore, the space of
fields is the disjoint union of vector spaces $Q = \bigcup_{\bc \in \sC M} Q_\bc$.

\vskip10mm
        \sect{Functions, vertical tangent vectors and covectors in the space of fields.}
    With each quadratic mapping $\zk \colon  M \times \wedge^1_e V^\* \times  \wedge^2_e V^\* \rightarrow  \wedge^4_o V^\*$
we associate the function
        $$k \colon \sQ(\zF^1_e(M);\sC M) \rightarrow \R \colon \sq(A,\bc) \mapsto \int_{\textstyle{\bc}} \zk \circ (x,A,\rd
A).
                                                                                                        \tag \label{Fvp42}$$

    The above mentioned admissible functions defined on the space of fields are those constructed in this way. They will be
called {\it differentiable}.  The space of such functions will be denoted by $\sK(\zF^1_e(M);\sC M)$.

    We note that from the definition of the space of fields $Q = \sQ(\zF^1_e(M);\sC M)$ (as a quotient space with respect
to the equivalence condition \Ref{Fvp38}) follows easily that
 the differentiable functions separate points of $Q$, i.e. if $k(q') = k(q)$ for each $k\in \sK(\zF^1_e(M);\sC M)$, then
$q'=q$.   Indeed if $k(\sq(A',\bc')) = k(\sq(A,\bc))$ for each $k\in \sK(\zF^1_e(M);\sC M)$, then equation \Ref{Fvp38}
holds for each quadratic mapping $\zk \colon  M \times \wedge^1_e V^\* \times  \wedge^2_e V^\* \rightarrow  \wedge^4_o
V^\*$.  It follows that $\bc'=\bc$ and $\sq(A',\bc) = \sq(A,\bc)$.

    The tangent space to a vector space $Q_\bc$ (i.e. to a fibre of $\ze$) coincides with  $Q_\bc$.

    The {\it vertical tangent bundle} to the vector bundle $\sQ(\zF^1_e(M);\sC M)$ is the space
        $$\sV Q = Q \fpr{(\ze,\ze)} Q=\{(q,\zd q) \in Q \times Q ;\; \ze(q) = \ze(\zd q)\}
                                                                                                        \tag \label{Fvp43}$$
    where $q= \sq(A,\bc)$ and $\zd q= \sq(\zd A,\bc)$ and $A \colon U \rightarrow \wedge^1_e V^\*$, $\zd A \colon U
\rightarrow \wedge^1_e V^\*$.  The {\it tangent projection} is the canonical projection
        $$\zt_Q \colon \sV Q \rightarrow Q \colon (q,\zd q) \mapsto q.
                                                                                                        \tag \label{Fvp44}$$
    There is no obvious choice of the bundle dual to $\sV Q$.

    We will use the fibre derivatives of functions  $k\in\sK(\zF^1_e(M);\sC M)$ as models of covectors.  The derivative
$\xD k$ of a function $k \colon Q \rightarrow \R$ is defined for each current $\bc$ separately.  It is evaluated on a pair
of vectors $q= \sq(A,\bc)\in Q_c$ and  $\zd q= \sq(\zd A,\bc)\in Q_c$. The result is the expression
        $$\align
        \xD k\left(\sq(A,\bc),\sq(\zd A,\bc)\right)
            &=\frac{\rd}{\rd s} k\left(\sq(A+s\zd A,\bc) \right)(0)\\
            &=\int_{\textstyle{\bc}} \left.\frac{\partial}{\partial s} \zk\circ \left(x, A+s\zd A,F+s\zd F
\right)\right|_{s=0},
                                                                                                        \tag \label{Fvp45}\endalign$$
    where  $F=\rd A$ and  $\zd F = \rd\zd A$.  The symbol $\rd k(\sq(A,\bc))$ will be used to denote the covector
characterized by the pairing
        $$\left\langle\rd k\left(\sq\left(A,\bc\right)\right),\sq(\zd A,\bc)\right\rangle=\xD k\left(\sq(A,\bc),\sq(\zd
A,\bc)\right)
                                                                                                        \tag \label{Fvp46}$$
    with vectors $\zd q= \sq(\zd A,\bc)\in Q_c$.

    So we need to calculate the expression
        $$\left.\frac{\partial}{\partial s} \zk\circ \left(x, A+s\zd A,F+s\zd F \right)\right|_{s=0}(x) = \xD
\zk_x(A(x),F(x), \zd A(x) \oplus \zd F(x)),
                                                                                                        \tag \label{Fvp47}$$
    for each $x\in U$. We are using the following known definition. If $V_1,V_2,W$ are vector spaces and
        $m\colon V_1\times V_2\rightarrow W$ is a smooth mapping, then its {\it derivative} is the mapping
        $$\xD m \colon V_1\times V_2\times \left(V_1\oplus V_2\right) \rightarrow W \colon (a,f,\zd a\oplus \zd f)
\mapsto\frac{\rd}{\rd s}\left(m(a+s\zd a,f+s\zd f)\right)(0).
                                                                                                        \tag \label{Fvp48}$$

    We define for each $x\in U$ the bilinear mappings
        $$\zl_x\colon \wedge^1_e V^\* \times  \wedge^1_e V^\* \rightarrow \wedge^4_o V^\*\colon (a,a')\mapsto
\zd^2\zk_x((a,0),(a',0)),
                                                                                                        \tag \label{Fvp49}$$
        $$\zm_x\colon \wedge^1_e V^\* \times  \wedge^2_e V^\* \rightarrow \wedge^4_o V^\*\colon (a,f)\mapsto
\zd^2\zk_x((a,0),(0,f))
                                                                                                        \tag \label{Fvp50}$$
    and
        $$\zn_x\colon \wedge^2_e V^\* \times  \wedge^2_e V^\* \rightarrow \wedge^4_o V^\*\colon (f,f')\mapsto
\zd^2\zk_x((0,f),(0,f')),
                                                                                                        \tag \label{Fvp51}$$
    obtaining the equality
        $$\zk_x(a,f) = \frac{1}{2} \zl_x(a,a) + \zm_x(a,f) + \frac{1}{2} \zn_x(f,f),
                                                                                                        \tag \label{Fvp52}$$
    for each $(a,f)\in\wedge^1_e V^\* \times \wedge^2_e V^\*$. The equality \Ref{Fvp52} in terms of the bilinear mappings
\Ref{Fvp49}, \Ref{Fvp50}, and \Ref{Fvp51} will be useful to calculate the expression \Ref{Fvp47}.

    We have the following lemmas.
    \claim \c{l}{Lemma}{}                                                       \label{L42}
    If $\zk_x \colon \wedge^1_e V^\* \times  \wedge^2_e V^\* \rightarrow  \wedge^4_o V^\*$ is a quadratic mapping and
$\zl_x,\zm_x,\zn_x$ are the mappings defined above, then
        $$\xD\zk_x(a,f,\zd a\oplus\zd f) = \frac{1}{2} \xD\zl_x(a,a,\zd a\oplus \zd a) + \xD\zm_x(a,f,\zd a \oplus \zd f)+
 \frac{1}{2}\xD\zn_x(f,f,\zd f\oplus \zd f),
                                                                                                        \tag \label{Fvp53}$$
    for each $(a,f,\zd a\oplus\zd f)\in \wedge_e^1 V^\* \times \wedge^2_e V^\* \times \wedge_e^1 V^\* \oplus \wedge^2_e
V^\*$.
    \endclaim
    \proof
        We have the equality
        $$\zk_x = \frac{1}{2}\zl_x\circ \zD\circ pr_1 + \zm_x + \frac{1}{2}\zn_x\circ \zD'\circ pr_2,
                                                                                                        \tag\label{Fvp54}$$
    where
        $$pr_1\colon \wedge_e^1 V^\* \times  \wedge^2_e V^\*\rightarrow  \wedge_e^1 V^\*\colon (a,f)\mapsto a,
                                                                                                        \tag \label{Fvp55}$$
        $$\zD\colon \wedge_e^1 V^\* \rightarrow  \wedge_e^1 V^\*\times  \wedge_e^1 V^\*\colon a\mapsto (a,a),
                                                                                                        \tag\label{Fvp56}$$
        $$pr_2\colon \wedge_e^1 V^\* \times  \wedge^2_e V^\*\rightarrow  \wedge_e^2 V^\*\colon (a,f)\mapsto f,
                                                                                                        \tag \label{Fvp57}$$
        $$\zD'\colon \wedge_e^2 V^\* \rightarrow  \wedge_e^2 V^\*\times  \wedge_e^2 V^\*\colon a\mapsto (a,a).
                                                                                                        \tag\label{Fvp58}$$
    The claim follows from the equality \Ref{Fvp54} by noting that
        $$\align
        \xD\zk_x(a,f,\zd a\oplus\zd f) &= \frac{1}{2} \xD\zl_x\circ (\zD,\xD\zD)\circ(pr_1,\xD pr_1)(a,f,\zd a\oplus\zd f)
+ \xD\zm_x(a,f,\zd a \oplus \zd f)\hskip19mm\\
                &\hskip30mm + \frac{1}{2} \xD\zn_x\circ (\zD',\xD\zD')\circ(pr_2,\xD pr_2)(a,f,\zd a\oplus\zd f)\\
            &= \frac{1}{2} \xD\zl_x\circ (\zD,\xD\zD)(a,\zd a) + \xD\zm_x(a,f,\zd a \oplus \zd f)\\
                &\hskip40mm + \frac{1}{2} \xD\zn_x\circ (\zD',\xD\zD')(f,\zd f)\\
            &= \frac{1}{2} \xD\zl_x(a,a,\zd a\oplus \zd a) + \xD\zm_x(a,f,\zd a \oplus \zd f)+
 \frac{1}{2}\xD\zn_x(f,f,\zd f\oplus \zd f).
                                                                                                        \tag \label{Fvp59}\endalign$$
    \endproof

    \claim \c{l}{Lemma}{}                                                       \label{L43}
    If $V_1,V_2,W$ are vector spaces and
        $$b\colon V_1\times V_2\rightarrow W
                                                                                                        \tag \label{Fvp60}$$
    is a bilinear mapping, then
        $$\xD b(a,f,\zd a\oplus \zd f) = b(a,\zd f) + b(\zd a,f),
                                                                                                        \tag \label{Fvp61}$$
    for each $(a,f,\zd a\oplus \zd f)\in V_1\times V_2\times \left(V_1\oplus V_2\right)$.
    \endclaim
    \proof
    From the definition \Ref{Fvp48} of the derivative it follows that
        $$\align
        \xD b(a,f,\zd a\oplus \zd f) &= \frac{\rd}{\rd s}\left(b(a,f+s\zd f)+sb(\zd a,f+s\zd f)\right)(0)\\
            &= \frac{\rd}{\rd s}\left(b(a,f)+sb(a,\zd f)+sb(\zd a,f)+s^2b(\zd a,\zd f)\right)(0)\\
            &= b(a,\zd f) + b(\zd a,f).
                                                                                                        \tag \label{Fvp62}\endalign$$
    \endproof

    In view of Lemmas \Ref{L42} and \Ref{L43} the integrand in right hand side of the equality \Ref{Fvp45} reduces to
        $$\align
        \xD \zk_x(A(x),F(x), \zd A(x) \oplus \zd F(x))&= \frac{1}{2}\xD\zl_x (A(x),A(x),\zd A(x)\oplus\zd A(x)) \\
            &\hskip5mm + \xD\zm_x (A(x), F(x),\zd A(x)\oplus\zd F(x))\hskip1mm\\
            &\hskip5mm + \frac{1}{2} \xD\zn_x (F(x),F(x), \zd F(x)\oplus\zd F(x))\\
            &= \frac{1}{2}\left(\zl_x (A(x),\zd A(x)) + \zl_x (\zd A(x), A(x))\right)\\
            &\hskip5mm + \zm_x (A(x),\zd F(x)) + \zm_x (\zd A(x),F(x)) \\
            &\hskip5mm+\frac{1}{2}\left( \zn_x (F(x),\zd F(x)) + \zn_x (\zd F(x),F(x))\right),
                                                                                                        \tag \label{Fvp63}\endalign$$
    or
        $$\align
            \xD \zk_x(A(x),F(x), \zd A(x) \oplus \zd F(x))&= \zl_x (A(x),\zd A(x)) +  \zm_x (A(x),\zd F(x)) \hskip20mm \\
            &\hskip 5mm  + \zm_x (\zd A(x),F(x)) +  \zn_x (F(x),\zd F(x)),
                                                                                                        \tag \label{Fvp64}\endalign$$
    since $\zl_x$ and $\zn_x$ are symmetric.  By using the Proposition \Ref{C4} contained in Section 6, this expression
reduces to
        $$\align
         \xD \zk_x(A(x),F(x), \zd A(x) \oplus \zd F(x)) =&\zd A(x) \wedge  \We_1(\ozl_x (A(x)) + \zd F(x) \wedge
\We_2(\ozm_x (A(x))\hskip25mm\\
        &+ \zd A(x) \wedge \We_1(\oozm_x (F(x)) + \zd F(x) \wedge \We_2(\ozn_x (F(x))\\
        =& \zd A(x) \wedge \left(\We_1(\ozl_x (A(x)) + \oozm_x (F(x)))\right)\\
            &+\zd F(x) \wedge \left(\We_2(\ozm_x (A(x)) + \ozn_x (F(x))) \right).
                                                                                                        \tag \label{Fvp65}\endalign$$
     where the linear mappings $\ozl_x$ and $\oozl_x$ are obtained from the bilinear mapping $\zl_x$ as prescribed in
Section 6, i.e.,
        $$\ozl_x \colon \wedge_e^1  V^\* \rightarrow \wedge_e^1 V \otimes \wedge_o^4  V^\* \colon a \mapsto
i_1(\zl_x(a,\cdot)),
                                                                                                        \tag \label{Fvp66}$$
        $$\oozl_x \colon \wedge_e^1  V^\* \rightarrow \wedge_e^1 V \otimes \wedge_o^4  V^\* \colon a \mapsto
i_1(\zl_x(\cdot, a)),
                                                                                                        \tag \label{Fvp67}$$
    for each $x\in U$.  The mappings $\ozm_x$, $\oozm_x$ are obtained from $\zm_x$, and $\ozn_x$, $\oozn_x$ are obtained
from $\zn_x$ in the same way.

    We define from $\ozl_x$, $\oozl_x$ the  mappings
        $$\ozl \colon  M \times \wedge_e^1 V^\* \rightarrow \wedge_e^1 V \otimes \wedge_o^4  V^\* \colon (x,a) \mapsto
\ozl_x(a),
                                                                                                        \tag \label{Fvp68}$$
        $$\oozl \colon  M \times \wedge_e^1 V^\* \rightarrow \wedge_e^1 V \otimes \wedge_o^4  V^\* \colon (x,a) \mapsto
\oozl_x(a).
                                                                                                        \tag \label{Fvp69}$$
    The mappings $\ozm$, $\oozm$, $\ozn$, $\oozn$ are defined respectively from  $\ozm_x$, $\oozm_x$, $\ozn_x$, $\oozn_x$
in the same way.

    By using the equalities \Ref{Fvp49} and \Ref{Fvp65} which hold for each  $x\in U$ we obtain the equality
        $$\align
        \left.\frac{\partial}{\partial s} \zk\circ \left(x, A+s\zd A,F+s\zd F \right)\right|_{s=0}&= -
\left(\We_1\circ\left(\ozl\circ (x,A) + \oozm\circ (x,F)\right)\right)\wedge \zd A \hskip15mm\\
        &\hskip15mm+ \left(\We_2\circ\left(\ozm\circ (x,A)+\ozn\circ (x,F) \right)\right)\wedge \zd F.
                                                                                                        \tag \label{Fvp70}\endalign$$
    We have also used the graded commutativity of the exterior product.

    Hence,
        $$\align
        \xD k\left(\sq(A,\bc),\sq(\zd A,\bc)\right)&=\int_{\textstyle{\bc}} \left.\frac{\partial}{\partial s} \zk\circ
\left(x, A+s\zd A,F+s\zd F \right)\right|_{s=0}\\
            &=-\int_{\textstyle{\bc}} \left(\We_1\circ\left(\ozl\circ (x,A) + \oozm\circ (x,F)\right)\right)\wedge \zd A\\
                &\hskip22mm + \int_{\textstyle{\bc}}\left(\We_2\circ\left(\ozm\circ (x,A)+\ozn\circ (x,F)
\right)\right)\wedge \zd F.
                                                                                                        \tag \label{Fvp71}\endalign$$

    Since $F=\rd A$ and $\zd F = \rd\zd A$, this equality can be converted to the form
        $$\align
        \xD k\left(\sq(A,\bc),\sq(\zd A,\bc)\right)&=-\int_{\textstyle{\bc}} \left(\We_1\circ\left(\ozl\circ (x,A) +
\oozm\circ (x,\rd A)\right)\right.\\
        &\hskip11mm+ \left.\rd\left(\We_2\circ\left(\ozm\circ (x,A)+\ozn\circ (x,\rd A) \right)\right)\right)\wedge \zd
A\hskip10mm\\
    &\hskip15mm+\int_{\textstyle{\bc}}\rd \left(\left(\We_2\circ\left(\ozm\circ (x,A)+\ozn\circ \rd A\right)\right)\wedge
\zd A\right).
                                                                                                        \tag \label{Fvp72}\endalign$$
    We note that the first integral contains the exterior product of $\zd A$ and an odd $3$-form, while the second integral
contains the exterior differential of the product of $\zd A$ and an odd $2$-form.

    The obtained expression suggests the representation of covectors as equivalence classes of elements of the sets
$\sX(\zF^2_o(M)\times \zF^3_o(M);\sC M)$ introduced below.

    An element of $\sX(\zF^2_o(M)\times \zF^3_o(M);\sC M)$ is a triple $(G,J,\bc)$ of a local odd differential $2$-form
$G\colon U\rightarrow \wedge_o^2 V^\*$, a local odd differential $3$-form $J\colon U\rightarrow \wedge_o^3 V^\*$  and a
current $\bc$ with support contained in $U$.  The odd $2$-form $G$ will be interpreted as the {\it electromagnetic
induction}, while the odd $3$-form $J$ will be interpreted as the {\it current}. A covector $p$ is an equivalence class
$\sp(G,J,\bc)$ of $(G,J,\bc) \in \sX(\zF^2_o(M)\times \zF^3_o(M);\sC M)$.  The equivalence relations in
$\sX(\zF^2_o(M)\times \zF^3_o(M);\sC M)$ are based on the expression
        $$\int_{\textstyle\bc}\left(\frac{1}{c^2} J \wedge \zd A - \frac{1}{4\zp c}\rd\left(G \wedge \zd A \right)\right).
                                                                                                        \tag \label{Fvp73}$$
    Elements $(G,J,\bc)$ and $(G',J',\bc')$ are equivalent if $\bc=\bc'$ and
        $$\int_{\textstyle\bc'}\left(\frac{1}{c^2} J' \wedge \zd A - \frac{1}{4\zp c}\rd\left(G' \wedge \zd A
\right)\right) = \int_{\textstyle\bc}\left(\frac{1}{c^2} J \wedge \zd A - \frac{1}{4\zp c}\rd\left(G \wedge \zd A
\right)\right),
                                                                                                        \tag
\label{Fvp74}$$ for each $(\zd A,\bc) \in Q_c$.

This equivalence relation is  related to the equivalence  relation \Ref{Fvp38}. They can be used in the construction of
dual objects. Indeed a (vertical) tangent vector might also (more generally) be defined as an equivalence class of pairs
$(\zd A, \bc)$ with respect to an equivalence relation similar to \Ref{Fvp38}. A convenient representation of such relation
is a relation similar to \Ref{Fvp74}, but with $(G,J,\bc)$ fixed, defining the equivalence between two pairs $(\zd A, \bc)$
and $(\zd A', \bc)$.  Such a construction is not needed in our case since each $Q_\bc$ is a vector space, so that its
tangent space is canonically identified with $Q_\bc$ itself.

    The vector space $\zP_\bc$ of covectors associated with the current $\bc$ will be used as the dual space of the space
$Q_{\bc}$ thought of as the space of vertical tangent vectors  with the pairing
        $$\blangle p, \zd q\brangle_\bc = \int_{\textstyle\bc}\left(\frac{1}{c^2} J \wedge \zd A - \frac{1}{4\zp
c}\rd\left(G \wedge \zd A \right)\right),
                                                                                                        \tag \label{Fvp75}$$
    where $\zd q = \sq(\zd A,\bc)\in Q_{\bc}$ and $p = \sp(G,J,\bc) \in \zP_\bc$. The space of all covectors is the union
        $$\zP = \bigcup_{\bc\in\sC M}\zP_{\bc}.
                                                                                                        \tag \label{Fvp76}$$
 There is a natural projection
        $$\ze' \colon \zP \rightarrow \sC M \colon \sp(G,J,\bc) \mapsto \bc.
                                                                                                        \tag \label{Fvp77}$$
    from the space of fields to the space $\sC M$ of currents in $M$. The {\it phase space} is the space
        $$\bPH =  Q \fpr{(\ze,\ze')} \zP=\{(q,p) \in Q \times \zP ;\; \ze(q) = \ze'(a)\}.
                                                                                                        \tag \label{Fvp78}$$
    The symbol $\bPH_{\bc}$ will denote the set $Q_{\bc} \times \zP_{\bc} \subset \bPH$.

\vskip10mm
        \sect{A virtual action principle for electrodynamics.}
    In this section a variational principle for electrodynamics similar to the virtual action principle of analytical
mechanics (see [3]) will be formulated. The construction is an example of similar constructions needed for the variational
formulation of a general field theory. The linearity of the theory and the choice of formulating it on the affine Minkowski
space makes our presentation simpler though containing the core of the general framework (which can be found in [1]).

    The {\it action} is the differentiable function
        $$W\colon \sQ(\zF^1_e(M);\sC M)\rightarrow \R\colon \sq(A,\bc)\mapsto \int_{\textstyle\bc}L\circ (A,\rd A)
                                                                                                        \tag \label{Fvp79}$$
    derived from the quadratic {\it Lagrangian density}
        $$L \colon  \wedge^1_e V^\*\times \wedge^2_e V^\* \rightarrow \wedge^4_o V^\* \colon (a,f) \mapsto - \frac{1}{8\zp
c}\langle f,\wedge_e^2 g^{-1}(f)\rangle \sqrt{|g|}.
                                                                                                        \tag \label{Fvp80}$$

    We are denoting by $\sqrt{|g|}$ the odd 4-covector defined by the Minkowski metric (see [5]). Later we will use the
same symbol to denote the constant 4-covector field constructed from it.

    The $1$-form $A$ is called the {\it potential}, the $2$-form $F=\rd A$ is called the {\it electromagnetic field}.

    We note that the Lagrangian is a quadratic mapping which depends only on its second argument $f$ and thus $\zl=\zm=0$
and $\zn$ does not depend on $x$ in the formula \Ref{Fvp52}.

    A phase $\bph=(\sq(A,c),\sp(G,J,\bc))$ satisfies the {\it virtual action principle} if the equality
        $$\langle \rd W(q), \zd q\rangle - \blangle p, \zd q\brangle_\bc = 0
                                                                                                        \tag \label{Fvp81}$$
    holds for each {\it virtual displacement} $\zd q = \sq(\zd A,\bc)\in Q_{\bc}$.
    For each current $\bc$ the {\it dynamics} associated with the current $\bc$ is the set $\bD_{\bc}\subset \bPH_{\bc}$ of
phases  which satisfy the virtual action principle. The {\it dynamics} is the subset
        $$\bD=\bigcup_{\bc\in\sC M}\bD_{\bc}
                                                                                                        \tag \label{Fvp82}$$
    of the {\it phase space} $\bPH$.

    A {\it phase space trajectory} is a triple of local differential forms
        $$(A,G,J) \colon U \rightarrow \wedge^1_e V^\* \times \wedge^2_o V^\*\times \wedge^3_o V^\*.
                                                                                                        \tag \label{Fvp83}$$
    The dynamics of a system can also be represented as a set $\bcD$ of phase space trajectories $(A,G,J)$ such that for
each current $\bc$ with support included in $U$ the phase  $\bph=(\sq(A,c),\sp(G,J,\bc))$ is in $\bD_{\bc}$.

    The equation \Ref{Fvp81} is too abstract to be used directly. A more concrete expression is given in the following
proposition.
        \claim \c{p}{Proposition}{}                                                                     \label{C6}
    A phase $\bph=(\sq(A,c),\sp(G,J,\bc))$ satisfies the virtual action principle if and only if the equality
        $$\align
            \frac{1}{4\zp c} \int_{\textstyle{\bc}} & \left( \rd \left( \left(\wedge_e^2 g^{-1} \circ \rd A\right) \lpr
\sqrt{|g|}\right) \wedge \zd A - \rd\left( \left(\left(\wedge_e^2 g^{-1} \circ \rd A\right) \lpr \sqrt{|g|}\right) \wedge
\zd A\right) \right)\\
                &= \int_{\textstyle\bc}\left(\frac{1}{c^2} J \wedge \zd A - \frac{1}{4\zp c}\rd\left(G \wedge \zd A
\right)\right)
                                                                                                        \tag \label{Fvp84}\endalign$$
    is satisfied for each {\it virtual displacement} $\sq(\zd A,\bc)$.
    \endclaim
    \proof
    By applying the formula \Ref{Fvp72} to the action $W$ we obtain that its variation is
        $$\langle \rd W(q), \zd q \rangle = \int_{\textstyle{\bc}} \left(- \rd (\We_2 \circ \ozn \circ \rd A) \wedge \zd A
+ \rd\left((\We_2 \circ \ozn \circ \rd A) \wedge \zd A\right) \right),
                                                                                                        \tag \label{Fvp85}$$
    since  $\zl=\zm=0$. Moreover
        $$\zn_x(f,f) = -\frac{1}{4\zp c}\langle f,\wedge_e^2 g^{-1}(f)\rangle \sqrt{|g|},
                                                                                                        \tag \label{Fvp86}$$
    from which it follows after some calculation that
        $$\ozn(x,f) = - \frac{1}{4\zp c}\wedge_e^2 g^{-1}(f)\otimes \sqrt{|g|},
                                                                                                        \tag \label{Fvp87}$$
    for each $f\in \wedge^2_e V^\*$. Hence,
        $$\We_2 \circ \ozn \circ (x,\rd A) =  - \frac{1}{4\zp c} \left(\left(\wedge_e^2 g^{-1} \circ \rd A\right) \lpr
\sqrt{|g|}\right)
                                                                                                        \tag \label{Fvp88}$$
    and the variation of the action reduces to
        $$\langle \rd W(q), \zd q \rangle = \frac{1}{4\zp c} \int_{\textstyle{\bc}} \left( \rd \left( \left(\wedge_e^2
g^{-1} \circ \rd A\right) \lpr \sqrt{|g|}\right) \wedge \zd A - \rd\left( \left(\left(\wedge_e^2 g^{-1} \circ \rd A\right)
\lpr \sqrt{|g|}\right) \wedge \zd A\right) \right).
                                                                                                        \tag \label{Fvp89}$$
    Recalling that, on the other hand
        $$\blangle p, \zd q\brangle_\bc = \int_{\textstyle\bc}\left(\frac{1}{c^2} J \wedge \zd A - \frac{1}{4\zp
c}\rd\left(G \wedge \zd A \right)\right),
                                                                                                        \tag \label{Fvp90}$$
    the claim follows.
    \endproof

     A phase space trajectory belongs to the dynamics $\bcD$, if and only if it satisfies the virtual action principle for
each current $\bc$ with support included in its domain of definition.  There is a characterization of the dynamics of phase
space trajectories in terms of differential equations. This is shown in the following propositions.

        \claim \c{p}{Theorem}{}                                                                     \label{C66}
    A phase space trajectory $(A,G,J)$ belongs to the dynamics $\bcD$ if and only if it satisfies the {\it Euler-Lagrange
equation}
        $$\rd\left(\left(\wedge_e^2 g^{-1} \circ \rd A\right) \lpr \sqrt{|g|}\right) = \frac{4\zp}{c}J
                                                                                                        \tag \label{Fvp91}$$
    and the {\it constitutive relation}
        $$G = \left(\wedge_e^2 g^{-1} \circ \rd A\right) \lpr \sqrt{|g|}.
                                                                                                        \tag\label{Fvp92}$$
    \endclaim
    \proof
    If  a phase space trajectory $(A,G,J)$ satisfies the Euler-Lagrange equation and the constitutive relation, then by
substituting the expressions \Ref{Fvp91} and \Ref{Fvp92} of $J$ and $G$ in terms of the electromagnetic field $F$ in the
virtual action principle \Ref{Fvp84} it follows that $(A,G,J)$ belongs to the dynamics $\bcD$.

    The inverse implication will be proved in the next section.
    \endproof
    The constitutive relation \Ref{Fvp92} produced by our variational principle corresponds to the momentum-velocity
relation of analytical mechanics.

        \claim \c{p}{Proposition}{}                                                                     \label{C77}
    A phase space trajectory $(A,G,J)$ satisfies the Euler-Lagrange equation and the constitutive relation if and only if
it satisfies the {\it Maxwell's equations}
        $$\rd G = \frac{4\zp}{c} J
                                                                                                        \tag \label{Fvp93}$$
    and the {\it constitutive relation}
        $$G = \left(\wedge_e^2 g^{-1} \circ F\right) \lpr \sqrt{|g|},
                                                                                                        \tag \label{Fvp94}$$
    with $F=\rd A$.
        \endclaim
    \proof
    The constitutive relation \Ref{Fvp93} is satisfied if and only if the equation \Ref{Fvp92} is satisfied for $F=\rd A$.

    If  a phase space trajectory $(A,G,J)$ satisfies the Maxwell's equations and the constitutive relation, then by
substituting  the expression \Ref{Fvp94} of the electromagnetic induction in the equation \Ref{Fvp93} we see that the
Euler-Lagrange equation is satisfied, since $F=\rd A$.

    Conversely if the Euler-Lagrange equation and the constitutive relation are satisfied, then, again by substitution, we
obtain that the Maxwell's equations \Ref{Fvp93} are satisfied.
    \endproof

We remark that since the virtual action principle \Ref {Fvp79} is more complete than the Hamilton principle, our
formulation leads to the Maxwell's equations \Ref{Fvp93} with external sources. The proposed variational principle also
permits the derivation of the constitutive relations \Ref{Fvp94} which are usually postulated separately since the
variations normally considered are not general enough to derive them from the variational principle.

\vskip10mm\penalty-400
        \sect{The Dynamics in a compact domain.}
    Let the current $\bc$ consist in integrating an odd $4$-form on a compact domain $K\subset M$ with smooth boundary
$\partial K$.  Field configurations, tangent vectors and covectors are equivalence classes of equivalence relations based
on the expressions
        $$ \int_{\textstyle K}\zk \circ (x,A,\rd A)
                                                                                                        \tag \label{Fvp95}$$
    and
        $$\int_{\textstyle K}\left(\frac{1}{c^2} J \wedge \zd A - \frac{1}{4\zp c}\rd\left(G \wedge \zd A \right)\right) =
\frac{1}{c^2} \int_{\textstyle K} J \wedge \zd A - \frac{1}{4\zp c} \int_{\textstyle \partial K} G \wedge \zd A.
                                                                                                        \tag \label{Fvp96}$$
    It follows that a field $q = \sq(A,K)$ is represented by the restriction
        $$A|K\colon K\rightarrow \wedge _e^1 V^\*
                                                                                                        \tag \label{Fvp97}$$
     of the potential $A$ to the
 the domain $K$.  A tangent vector $\zd q = \sq(\zd A,K)$ is represented by the restriction
        $$(\zd A)|K\colon K\rightarrow \wedge _e^1 V^\*
                                                                                                        \tag \label{Fvp98}$$
    of the variation $\zd A$ to the domain $K$.  A covector $p = \sp(G,J,K)$ is represented by the pair of the restriction
        $$G|\partial K\colon \partial K\rightarrow \wedge _o^2 V^\*
                                                                                                        \tag \label{Fvp99}$$
    of the electromagnetic induction $G$ to the boundary $\partial K$ of the domain $K$ and the restriction
        $$J|\iK\colon \iK\rightarrow \wedge _o^3 V^\*
                                                                                                        \tag \label{Fvp100}$$
    of the current $J$ to the interior $\iK$ of the domain $K$.

    A phase is a pair $(q,p)$ of a field $q = \sq(A,K)$ and a covector $p = \sp(G,J,K)$.

    The action is
        $$W\colon Q_K \rightarrow \R\colon \sq(A,K)\mapsto \int_{\textstyle K}L\circ (A,\rd A).
                                                                                                        \tag \label{Fvp101}$$
    The {\it virtual action principle} is the equality
        $$\langle \rd W(q), \zd q\rangle - \blangle p, \zd q\brangle_K = 0
                                                                                                        \tag \label{Fvp102}$$
    and the {\it dynamics} in the domain $K$ is the set $\bD_K\subset \bPH$ of phases satisfying the virtual action
principle.

        \claim \c{p}{Proposition}{}                                                                     \label{C65}
    A phase $\bph = (\sq(A,K),\sp(G,J,K))$, defined in a compact domain $K$, belongs to the dynamics $\bD_K$ if and only if
the Euler-Lagrange equation
        $$\rd\left(\left(\wedge_e^2 g^{-1} \circ \rd A\right) \lpr \sqrt{|g|}\right)|\iK = \frac{4\zp}{c}J|\iK
                                                                                                        \tag \label{Fvp103}$$
    and the constitutive relation
        $$G|\partial K = \left(\left(\wedge_e^2 g^{-1} \circ \rd A\right) \lpr \sqrt{|g|}\right)|\partial K
                                                                                                        \tag \label{Fvp104}$$
    are satisfied.
        \endclaim
    \proof
    If $q=\sq(A,K)$, $p = \sp(G,J,K)$ and $\zd q = \sq(\zd A,K)$, then
        $$\align
            \langle \rd W(q), \zd q \rangle &= \frac{1}{4\zp c} \int_{\textstyle{K}} \rd \left( \left(\wedge_e^2 g^{-1}
\circ \rd A\right) \lpr \sqrt{|g|}\right) \wedge \zd A\hskip10mm\\
        &\hskip17mm-\int_{\textstyle{K}} \rd\left( \left(\left(\wedge_e^2 g^{-1}\circ \rd A\right) \lpr \sqrt{|g|}\right)
\wedge \zd A\right)\\
            &= \frac{1}{4\zp c} \int_{\textstyle{K}}  \rd \left( \left(\wedge_e^2 g^{-1} \circ \rd A\right) \lpr
\sqrt{|g|}\right) \wedge \zd A\\
        &\hskip17mm - \frac{1}{4\zp c} \int_{\textstyle{\partial K}} \left(\left(\wedge_e^2 g^{-1} \circ \rd A\right) \lpr
\sqrt{|g|}\right) \wedge \zd A.\hskip12mm
                                                                                                        \tag \label{Fvp105}\endalign$$
    On the other hand
        $$\blangle p, \zd q\brangle_K = \frac{1}{c^2} \int_{\textstyle K} J \wedge \zd A - \frac{1}{4\zp c}
\int_{\textstyle \partial K} G \wedge \zd A.
                                                                                                        \tag \label{Fvp106}$$
    Thus the virtual action principle assumes the form
        $$\align
        &\int_{\textstyle{K}}  \rd \left( \left(\wedge_e^2 g^{-1} \circ \rd A\right) \lpr \sqrt{|g|}\right) \wedge \zd A -
\int_{\textstyle{\partial K}} \left(\left(\wedge_e^2 g^{-1} \circ \rd A\right) \lpr \sqrt{|g|}\right) \wedge \zd A
\hskip5mm\\
        &\hskip55mm = \frac{4\zp}{c} \int_{\textstyle K} J \wedge \zd A - \int_{\textstyle \partial K} G \wedge \zd A.
                                                                                                        \tag \label{Fvp107}\endalign$$
    By using variations with $(\zd A)| \partial K = 0$ we derive the Euler-Lagrange equation
        $$\rd\left(\left(\wedge_e^2 g^{-1} \circ \rd A\right) \lpr \sqrt{|g|}\right)|\iK = \frac{4\zp}{c}J|\iK.
                                                                                                        \tag \label{Fvp108}$$
    Assuming that this equation is satisfied and using arbitrary variations, the constitutive relation
        $$G|\partial K = \left(\left(\wedge_e^2 g^{-1} \circ \rd A\right) \lpr \sqrt{|g|}\right)|\partial K
                                                                                                        \tag \label{Fvp109}$$
    follows.
    \endproof

    The following Proposition completes the proof of the Theorem \Ref{C66}.
        \claim \c{p}{Proposition}{}                                                                     \label{C7}
    If a phase space trajectory
        $$(A,G,J) \colon U \rightarrow \wedge^1_e V^\* \times \wedge^2_o V^\*\times \wedge^3_o V^\*.
                                                                                                        \tag \label{Fvp110}$$
 belongs to the dynamics $\bcD$, then the Euler-Lagrange equation
        $$\rd\left(\left(\wedge_e^2 g^{-1} \circ \rd A\right) \lpr \sqrt{|g|}\right) = \frac{4\zp}{c}J
                                                                                                        \tag \label{Fvp111}$$
    and the constitutive relation
        $$G = \left(\wedge_e^2 g^{-1} \circ \rd A\right) \lpr \sqrt{|g|}
                                                                                                        \tag \label{Fvp112}$$
    are satisfied in $U$.
        \endclaim
    \proof
     If $(A,G,J)$ is a phase space trajectory, defined in the open set $U\subset M$, which belongs to the dynamics $\bcD$,
then the equation
        $$\rd\left(\left(\wedge_e^2 g^{-1} \circ \rd A\right) \lpr \sqrt{|g|}\right)|\iK = \frac{4\zp}{c}J|\iK
                                                                                                        \tag \label{Fvp113}$$
    and the boundary relation
        $$G|\partial K = \left(\left(\wedge_e^2 g^{-1} \circ \rd A\right) \lpr \sqrt{|g|}\right)|\partial K
                                                                                                        \tag \label{Fvp114}$$
    are satisfied for each compact domain $K\subset U$.  It follows that equations \Ref{Fvp111} and \Ref{Fvp112} are
satisfied in every $x\in U$.
        \endproof

\vskip10mm\penalty-400
        \sect{The Lagrangian formulation of electrodynamics.}
    The Lagrangian formulation of dynamics is the infinitesimal limit of the formulation in a compact domain with the
domain shrinking to a point.  A formal method  which greatly simplifies the passage to the infinitesimal limit is to
replace the compact domain --- which is used exclusively as domain of integration --- with the current $\bc = \zd(x)w$,
where $\zd(x)$ is the Dirac delta function in $x\in M$ and $w\in\wedge_o^4 V$ is an odd $4$-vector, with $w \neq 0$.  The
construction of infinitesimal fields, tangent vectors and covectors is based on the expressions
        $$ \int_{\zd(x)w} \zk \circ (x,A,\rd A) = \left\langle \zk \left(x,A(x),\rd A(x)\right), w\right\rangle
                                                                                                        \tag \label{Fvp115}$$
    and
        $$\int_{\zd(x)w} \left(\frac{1}{c^2} J \wedge \zd A - \frac{1}{4\zp c}\rd\left(G \wedge \zd A \right)\right) =
\left\langle \frac{1}{c^2}  J(x) \wedge \zd A(x) - \frac{1}{4\zp c} \rd(G \wedge \zd A)(x), w \right\rangle.
                                                                                                        \tag \label{Fvp116}$$
    Since $w\neq 0$ and $\wedge_o^4 V$ is one-dimensional, it follows from the first expression that a field $q =
\sq(A,\bc)$ is represented by the pair
        $$(A(x),F(x))\in \wedge_e^1 V^\*  \times \wedge_e^2 V^\*
                                                                                                        \tag
\label{Fvp117}$$ of an even $1$-covector $A(x)$ and an even $2$-covector  $F(x)$.

    The second expression reduces to
        $$- \frac{1}{4\zp c}\left\langle \left(\rd G(x)- \frac{4\zp}{c} J(x)\right) \wedge \zd A(x) + G(x) \wedge \zd F(x),
w\right\rangle,
                                                                                                        \tag \label{Fvp118}$$
    since $\rd\zd A(x) = \zd F(x)$.

    It follows that a tangent vector $\zd q = \sq(\zd A,\bc)$ is represented by the pair
        $$(\zd A(x),\zd F(x))\in \wedge_e^1 V^\*  \times \wedge_e^2 V^\*
                                                                                                        \tag \label{Fvp119}$$
    and a covector $p = \sp(G,J,\bc)$ is represented by the pair
        $$\left(G(x),\rd G(x)- \frac{4\zp}{c}J(x)\right) \in  \wedge_o^2 V^\*  \times \wedge_o^3 V^\* .
                                                                                                        \tag \label{Fvp120}$$
    The pairing
        $$\blangle p, \zd q\brangle_\bc = \int_{\textstyle\bc}\left(\frac{1}{c^2} J \wedge \zd A - \frac{1}{4\zp
c}\rd\left(G \wedge \zd A \right)\right)
                                                                                                        \tag \label{Fvp121}$$
    assumes the form
        $$\blangle p, \zd q\brangle^\ssL = - \frac{1}{4\zp c}\left\langle \left(\rd G(x)- \frac{4\zp}{c} J(x)\right) \wedge
\zd A(x) + G(x) \wedge \zd F(x), w \right\rangle.
                                                                                                        \tag \label{Fvp122}$$

    We have constructed the space of infinitesimal fields $Q_{\zd} = \wedge_e^1 V^\* \times\wedge_e^2 V^\*$ and the space
of infinitesimal covectors $\zP_{\zd} =  \wedge_o^2 V^\* \times  \wedge_o^3 V^\*$.   Hence, the infinitesimal phase space is
        $$\bPH_{\zd} = Q_{\zd} \times \zP_{\zd} = \wedge_e^1 V^\* \times  \wedge_e^2 V^\* \times \wedge_o^2 V^\* \times
\wedge_o^3 V^\*.
                                                                                                        \tag \label{Fvp123}$$
    The infinitesimal action is
        $$W(\sq(A,\zd(x)w))=\left\langle L(A(x), F(x)),w\right\rangle.
                                                                                                        \tag
\label{Fvp124}$$ The infinitesimal dynamics is the set
        $$\bD_{\zd} = \left\{(q,\zd q) \in \bPH_{\zd};\; \all{\zd q \in Q_{\zd}} \left\langle \rd W(q), \zd q\right\rangle
= \blangle p, \zd q \brangle^\ssL \right\},
                                                                                                        \tag
\label{Fvp125}$$ It is easy to verify that the infinitesimal dynamics $\bD_{\zd}$ admits also the following more explicit
expression
        $$\bD_{\zd} = \left\{(a,f,g,h) \in \bPH_{\zd};\;  \all{(\zd a,\zd f)\in \wedge_e^1 V^\*  \times \wedge_e^2 V^\* }
\xD L(a,f,\zd a,\zd f) = -\frac{1}{4\zp c}\left(h \wedge \zd a + g \wedge \zd f \right)\right\}.
                                                                                                        \tag \label{Fvp126}$$

    The infinitesimal dynamics $\bD_{\zd}$ is characterized by the following Proposition.

        \claim \c{p}{Proposition}{}                                                                     \label{C8}
    An infinitesimal phase $\bph = (\sq(A,\zd(x)w),\sp(G,J,\zd(x)w))$, with $w\neq0$, belongs to the infinitesimal dynamics
$\bD_{\zd}$ if and only if the equations
        $$G(x) = \left(\wedge_e^2 g^{-1} (F(x))\right) \lpr \sqrt{|g|}
                                                                                                        \tag \label{Fvp127}$$
    and
        $$\rd G(x) = \frac{4\zp}{c} J(x)
                                                                                                        \tag \label{Fvp128}$$
    are satisfied.
        \endclaim
    \proof
    If $q = \sq(A,\zd(x)w))$, $p = \sp(G,J,\zd(x)w)$, and $\zd q = \sq(\zd A,\zd(x)w)$, with $w\neq0$, then the variation
of the action \Ref{Fvp89}, which can also be expressed in the form
        $$\left\langle \rd W(q), \zd q \right\rangle = -\frac{1}{4\zp c} \int_{\textstyle{\bc}}  \left( \left(\wedge_e^2
g^{-1} \circ F\right) \lpr \sqrt{|g|}\right) \wedge \zd F,
                                                                                                        \tag \label{Fvp129}$$
    reduces to
        $$\left\langle \rd W(q), \zd q \right\rangle = - \frac{1}{4\zp c} \left\langle \left( \left(\wedge_e^2
g^{-1}(F(x))\right) \lpr \sqrt{|g|}\right) \wedge \zd F(x),w\right\rangle.
                                                                                                        \tag \label{Fvp130}$$

    Thus the virtual action principle
        $$\left\langle \rd W(q), \zd q \right\rangle = \blangle p, \zd q\brangle^\ssL
                                                                                                        \tag \label{Fvp131}$$
    has the explicit form
        $$\left\langle \left(\rd G(x)- \frac{4\zp}{c} J(x)\right) \wedge \zd A(x) + \left(G(x) - \left(\left(\wedge_e^2
g^{-1}(F(x))\right)\lpr \sqrt{|g|}\right)\right)\wedge \zd F(x), w \right\rangle  = 0.
                                                                                                        \tag \label{Fvp132}$$
    Therefore if $\bph$ satisfies the equations \Ref{Fvp127} and \Ref{Fvp128}, then \Ref{Fvp132} is satisfied and
$\bph\in\bD_{\zd}$.

    Conversely, if $\bph\in\bD_{\zd}$, then it satisfies the virtual action principle, which implies
        $$\left(\rd G(x)- \frac{4\zp}{c} J(x)\right) \wedge \zd A(x) + \left(G(x) - \left(\left(\wedge_e^2
g^{-1}(F(x))\right)\lpr \sqrt{|g|}\right)\right)\wedge \zd F(x) = 0,
                                                                                                        \tag \label{Fvp133}$$
    since $w\neq 0$ and $\wedge_o^4 V$ is one-dimensional.  The equations \Ref{Fvp127} and \Ref{Fvp128} follow, since $\zd
A(x)$ and $\zd F(x)$ are independent.
    \endproof

    In a preceding section we showed that the dynamics of phase space trajectories can also be characterized by the
Maxwell's equations and the constitutive relation. This fact can now be proved directly.

    Indeed, if a phase space trajectory $(A,G,J)$ belongs to the dynamics $\bcD$ and $x$ is a point in the domain of
definition of the trajectory, then the virtual action principle is satisfied for every infinitesimal current $\zd(x)w$.
Thus the Maxwell's equations and the constitutive relation are satisfied in $x$, thanks to the previous proposition.  It
follows that they are satisfied in the whole domain of definition of the trajectory.

    To prove the inverse implication, we observe that the virtual action principle \Ref{Fvp84} can also be expressed in the
form
        $$ -\frac{1}{4\zp c}\int_{\textstyle\bc} \left(\wedge_e^2 g^{-1} \circ F\right) \lpr \sqrt{|g|} \wedge \zd F =
\int_{\textstyle\bc}\left(\frac{1}{c^2} J \wedge \zd A - \frac{1}{4\zp c}\rd\left(G \wedge \zd A \right)\right)
                                                                                                        \tag \label{Fvp134}$$
    or
        $$ \int_{\textstyle\bc} \left(\left(G- \left(\wedge_e^2 g^{-1} \circ F\right) \lpr\sqrt{|g|}\right) \wedge \zd F +
\left(\rd G -\frac{4\zp}{c} J \right) \wedge \zd A \right) = 0,
                                                                                                        \tag \label{Fvp135}$$
    for each {\it virtual displacement} $\sq(\zd A,\bc)$.   Thus if a phase space trajectory $(A,G,J)$ satisfies the
Maxwell's equations, then it satisfies the virtual action principle for every current $\bc$ with support contained in the
domain of definition of the trajectory and hence it belongs to $\bcD$.

\vskip10mm
        \sect{The Hamiltonian formulation of electrodynamics.}
    We associate with the Lagrangian density
        $$L \colon  \wedge^1_e V^\*\times  \wedge^2_e V^\* \rightarrow \wedge^4_o V^\* \colon (a,f) \mapsto - \frac{1}{8\zp
c}\left\langle f,\wedge_e^2 g^{-1}(f)\right\rangle \sqrt{|g|},
                                                                                                        \tag \label{Fvp136}$$
    the {\it energy density}
        $$E  \colon  \wedge^1_e V^\*\times  \wedge^2_o V^\* \times  \wedge^2_e V^\* \rightarrow \wedge^4_o V^\*
                                                                                                        \tag \label{Fvp137}$$
    defined by
        $$\align
            E(a,g,f) &= -\frac{1}{4\zp c} g\wedge f - L(a,f)\\
                      &= -\frac{1}{4\zp c} g\wedge f + \frac{1}{8\zp c}\left\langle f,\wedge_e^2 g^{-1}(f)\right\rangle
\sqrt{|g|}\\
                       &= -\frac{1}{4\zp c} g\wedge f + \frac{1}{8\zp c}f\wedge \left(\wedge_e^2 g^{-1}(f)\lpr
\sqrt{|g|}\right)\\
                         &= -\frac{1}{8\zp c} \left( 2g - \wedge_e^2 g^{-1}(f) \lpr \sqrt{|g|}\right) \wedge f
                                                                                                        \tag \label{Fvp138}\endalign$$
    and treat this mapping as a family
        $$E(a,g,\cdot )  \colon  \wedge^2_e V^\* \rightarrow \wedge^4_o V^\*
                                                                                                        \tag \label{Fvp139}$$
    of mappings on the fibres of the projection
        $$pr_{\bP} \colon  \wedge^1_e V^\*\times  \wedge^2_o V^\* \times  \wedge^2_e V^\* \rightarrow \wedge^1_e V^\*\times
\wedge^2_o V^\*
                                                                                                        \tag \label{Fvp140}$$
    onto the  {\it field-momentum space} $\bP=\wedge^1_e V^\*\times \wedge^2_o V^\*$.  The set
        $$Cr(E,pr_{\bP}) =\left\{(a,g,\zl)\in \wedge^1_e V^\*\times  \wedge^2_o V^\* \times \wedge^2_e V^\*;  \all{\zd
\zl\in \wedge^2_e V^\*}\xD E(a,g,\zl,0,0,\zd\zl) = 0 \right\}
                                                                                                        \tag \label{Fvp141}$$
    is the {\it critical set} of the family.  The equality
        $$\align
        \xD E(a,g,\zl,\zd a,\zd g,\zd \zl)  &= -\frac{1}{4\zp c}\left(\zd g\wedge \zl + g\wedge  \zd \zl\right) - \xD
L(a,\zl,\zd a,\zd \zl)\\
            &= -\frac{1}{4\zp c}\left(\zd g\wedge \zl + g\wedge  \zd \zl\right) +\frac{1}{4\zp c}\zd \zl\wedge
\left(\wedge_e^2 g^{-1}(\zl)\lpr \sqrt{|g|}\right)\hskip8mm\\
            &= -\frac{1}{4\zp c}\left(\zd g\wedge \zl + \left(g - \left(\wedge_e^2 g^{-1}(\zl)\lpr
\sqrt{|g|}\right)\right)\wedge\zd \zl \right)
                                                                                                        \tag \label{Fvp142}\endalign$$
        implies
        $$\xD E(a,g,\zl,0,0,\zd \zl) =  -\frac{1}{4\zp c}  \left(g - \left(\wedge_e^2 g^{-1}(\zl)\lpr
\sqrt{|g|}\right)\right)\wedge\zd \zl.
                                                                                                        \tag \label{Fvp143}$$
        Thus we obtain the expression
        $$Cr(E,pr_{\bP}) = \left\{(a,g,\zl)\in  \wedge^1_e V^\*\times  \wedge^2_o V^\* \times \wedge^2_e V^\*;\,
g=\wedge_e^2 g^{-1}(\zl)\lpr \sqrt{|g|} \right\}.
                                                                                                        \tag \label{Fvp144}$$
        The critical set is the graph of the {\it Legendre mapping}
        $$\zL\colon \wedge^1_e V^\*\times  \wedge^2_e V^\* \rightarrow \wedge^2_o V^\*\colon (a,\zl)\mapsto \wedge_e^2
g^{-1}(\zl)\lpr \sqrt{|g|}.
                                                                                                        \tag \label{Fvp145}$$
    For each $a\in\wedge^1_e V^\*$, the mapping $\zL(a,\cdot)$ is invertible. Its inverse is the mapping
        $$\zL(a,\cdot)^{-1}\colon  \wedge^2_o V^\*\rightarrow\wedge^2_e V^\* \colon  g\mapsto \wedge_e^2
g\left(\sqrt{|g^{-1}|}\rpr g\right),
                                                                                                        \tag
\label{Fvp146}$$ where $\sqrt{|g^{-1}|}$ denotes the odd 4-vector characterized by $\langle \sqrt{|g|},
\sqrt{|g^{-1}|}\rangle = 1$. It follows that the critical set is the image of the section
        $$\zs \colon  \wedge^1_e V^\*\times  \wedge^2_o V^\*\rightarrow  \wedge^1_e V^\*\times  \wedge^2_o V^\* \times
\wedge^2_e V^\* \colon (a,g)\mapsto \left(a,g,\wedge_e^2 g\left(\sqrt{|g^{-1}|}\rpr g\right)\right)
                                                                                                        \tag \label{Fvp147}$$
    of the projection $pr_{\bP}$. The {\it Hamiltonian density} is the mapping
        $$H = E\circ \zs \colon  \wedge^1_e V^\*\times  \wedge^2_o V^\*\rightarrow \wedge^4_o V^\*,
                                                                                                        \tag \label{Fvp148}$$
    defined by the formula
        $$ H(a,g)= -\frac{1}{8\zp c} g\wedge\left(\wedge_e^2 g\left(\sqrt{|g^{-1}|}\rpr g\right)\right),
                                                                                                        \tag \label{Fvp149}$$
     for each $(a,g)\in\wedge^1_e V^\*\times  \wedge^2_o V^\*$.  The passage from the Lagrangian density $L$ to the
Hamiltonian density $H$ is the {\it Legendre transformation of electrodynamics}.

    We show that the energy density can be used to generate the infinitesimal dynamics $\bD_{\zd}$.  We consider the set
        $$\align
            \bD_E &= \left\{(a,f,g,r)\in\bPH_{\zd};\; \exi{\zl\in \wedge^2_e V^\*} \vphantom{\frac{1}{4\zp c}} \all{(\zd
a,\zd g,\zd\zl)\in \wedge^1_e V^\*\times \wedge^2_o V^\*\times \wedge^2_e V^\*}\right.\hskip35mm\\
            &\hskip46mm \left.\xD E(a,g,\zl,\zd a,\zd g,\zd\zl) =\frac{1}{4\zp c} \left(r \wedge \zd a - f\wedge \zd
g\right) \right\}.
                                                                                                        \tag \label{Fvp150}\endalign$$
    This set is obtained by projecting the set
        $$\align
            \widetilde{\bD_E}&=\left\{(a,f,g,r,\zl)\in\bPH_{\zd}\times  \wedge^2_e V^\*;\; \vphantom{\frac{1}{4\zp c}}
\all{(\zd a,\zd g,\zd\zl)\in \wedge^1_e V^\*\times \wedge^2_o V^\*\times \wedge^2_e V^\*} \right.\hskip33mm\\
            &\hskip46mm \left.\xD E(a,g,\zl,\zd a,\zd g,\zd\zl) =\frac{1}{4\zp c}
 \left(r \wedge \zd a - f\wedge \zd g\right) \right\}
                                                                                                        \tag \label{Fvp151}\endalign$$
    onto the phase space $\bPH_{\zd} =  \wedge_e^1 V^\* \times \wedge_e^2 V^\*  \times  \wedge_o^2 V^\* \times \wedge_o^3
V^\*$.

    It follows from \Ref{Fvp142} that $\zl=f$ and that the set $\widetilde{\bD_E} $ reduces to
        $$\align
            \widetilde{\bD_E}&=\left\{(a,f,g,r,\zl)\in\bPH_{\zd}\times  \wedge^2_e V^\*;\; \zl=f, \vphantom{\frac{1}{4\zp
c}}\right.\hskip68mm\\
            &\hskip8mm \left.\all{(\zd a,\zd g,\zd f)\in \wedge^1_e V^\*\times \wedge^2_o V^\*\times \wedge^2_e V^\*} \xD
L(a,f,\zd a,\zd f) =\frac{1}{4\zp c} \left(r \wedge\zd a + g\wedge \zd f\right) \right\}.
                                                                                                        \tag \label{Fvp152}\endalign$$
    It projects onto the infinitesimal dynamics
        $$\align
        \bD_{\zd} &= \left\{(a,f,g,r) \in \bPH_{\zd};\; \all{(\zd a,\zd f)\in  \wedge_e^1 V^\* \times \wedge_e^2 V^\*}
\vphantom{-\frac{1}{4\zp c}} \right.\hskip58mm\\
        & \hskip50mm \left.\xD L(a,f,\zd a,\zd f) = -\frac{1}{4\zp c}\left(r \wedge \zd a + g\wedge\zd f \right)\right\}.
                                                                                                        \tag \label{Fvp153}\endalign$$
    Hence, $\bD_E = \bD_{\zd}$.

    It is clear from the definition of the set $\widetilde{\bD_E}$ that this set is included in the set
        $$\left\{(a,f,g,r,\zl)\in\bPH_{\zd}\times  \wedge^2_e V^\*;\; (a,g,\zl) \in Cr(E,pr_{\bP})\right\}.
                                                                                                        \tag \label{Fvp154}$$
     The use of the mapping $\zs$ results in
        $$\align
            &\widetilde{\bD_E}=\left\{(a,f,g,r,\zl)\in\bPH_{\zd}\times \wedge^2_e V^\*;\; \zl=f, \vphantom{\frac{1}{4\zp
c}}\right.\hskip62mm \\
            & \hskip20mm \all{(\zd a,\zd g)\in \wedge^1_e V^\*\times \wedge^2_o V^\*}
             \left.\xD H(a,g,\zd a,\zd g) =\frac{1}{4\zp c} \left(r \wedge\zd a - f\wedge \zd g\right) \right\}.
                                                                                                        \tag \label{Fvp155}\endalign$$
    The {\it Hamiltonian description} of the dynamics
        $$\align
            \bD_{\zd}&=\left\{(a,f,g,r)\in\bPH_{\zd};\;\vphantom{\frac{1}{4\zp c}} \all{(\zd a,\zd g)\in \wedge^1_e
V^\*\times \wedge^2_o V^\*} \right.\hskip55mm\\
            &\hskip55mm \left.\xD H(a,g,\zd a,\zd g) =\frac{1}{4\zp c} \left(r \wedge \zd a - f\wedge \zd g\right)
\right\}\\
            &=\left\{(a,f,g,r)\in\bPH_{\zd};\; \wedge_e^2 g \left(\sqrt{|g{}^{-1}|}\rpr g\right) = f \;,\; r=0
\vphantom{\frac{1}{4\zp c}} \right\}.
                                                                                                        \tag \label{Fvp156}\endalign$$
    is obtained by projecting onto the phase space $\bPH_{\zd}$.

        \claim \c{p}{Proposition}{}                                                                     \label{C88}
    An infinitesimal phase $\bph = (\sq(A,\zd(x)w),\sp(G,J,\zd(x)w))$, with $w\neq0$, belongs to the infinitesimal dynamics
$\bD_{\zd}$ if and only if the equations
        $$F(x) = \wedge_e^2 g \circ \left(\sqrt{|g^{-1}|}\rpr G(x)\right)
                                                                                                        \tag \label{Fvp157}$$
    and
        $$\rd G(x) = \frac{4\zp}{c} J(x)
                                                                                                        \tag \label{Fvp158}$$
    are satisfied.
    \endclaim
    \proof
    We recall that an infinitesimal phase $\bph = (\sq(A,\zd(x)w),\sp(G,J,\zd(x)w))$, with $w\neq0$, is represented by
        $$\left(A(x),F(x),\rd G(x)- \frac{4\zp}{c}J(x),G(x)\right)
                                                                                                        \tag \label{Fvp159}$$
    and  use the Hamiltonian description \Ref{Fvp156} of infinitesimal dynamics $\bD_{\zd}$. The claim easily follows.
    \endproof

    The resulting equations for the phase space trajectories $(A,G,J)$
        $$F = \wedge_e^2 g \circ \left(\sqrt{|g^{-1}|}\rpr G\right)
                                                                                                        \tag \label{Fvp160}$$
    and
        $$\rd G = \frac{4\zp}{c} J
                                                                                                        \tag \label{Fvp161}$$
    are called {\it Hamilton's equations}. The equations \Ref{Fvp161} are again Maxwell's equations and the equation
\Ref{Fvp160} is the inverse of constitutive relation \Ref{Fvp94}.

\sect{References}

\noindent [1] W. M. Tulczyjew, {\it The origin of variational principles}, in the volume Classical and quantum integrability
 (Warsaw, 2001),  41--75, Banach Center Publ., {\bf 59}, Polish Acad. Sci., Warsaw, 2003. \vskip \baselineskip

\noindent [2] G. Marmo, W. M. Tulczyjew, P. Urba\'nski, {\it Dynamics of autonomous systems with external forces}, Acta
Physica Polonica B, {\bf 33} (2002), 1181--1240. \vskip \baselineskip

\noindent [3] A. De Nicola, W. M. Tulczyjew, {\it A variational formulation of analytical mechanics in an affine space},
Rep. Math. Phys., {\bf 58} (2006), 335--350. \vskip \baselineskip

\noindent [4] W. M. Tulczyjew, {\it A symplectic framework of linear field theories}, Annali di Matematica pura ed
applicata, {\bf 130 } (1982), 177--195. \vskip \baselineskip

\noindent [5] G. Marmo, E. Parasecoli, W. M. Tulczyjew, {\it Space-time orientations and Maxwell's equations}, Rep. Math.
Phys. {\bf 56} (2005), 209--248. \vskip \baselineskip

\noindent [6] G. Marmo, W. M. Tulczyjew, {\it Time reflection and
the dynamics of particles and antiparticles}, Rep. Math. Phys. {\bf
58} (2006), 147--164. \vskip \baselineskip

\noindent [7] A. De Nicola, W. M. Tulczyjew, {\it A Note on a
Variational Formulation of Electrodynamics}, Proc. XV Int. Workshop
on Geom. and Phys., Tenerife (Spain), 2006, Publ. de la RSME {\bf
11} (2007), 316--323.  \vskip \baselineskip

\noindent [8] J. A. Schouten, {\it Ricci calculus}, Springer, Berlin, 1955. \vskip \baselineskip

\noindent [9] J. A. Schouten, {\it Tensor Analysis for Physicists}, Oxford University Press, London, 1951. \vskip
\baselineskip

\noindent [10] G. de Rham, {\it Vari\'et\'es Differentiables}, Hermann, Paris, 1955. \vskip \baselineskip

\noindent [11] T. Frankel, {\it The Geometry of Physics: An Introduction}, Cambridge University Press, Cambridge, 1997.
\vskip \baselineskip

\noindent [12] F. W. Hehl and Yu. N. Obukhov, {\it Foundations of Classical Electrodynamics, Charge, Flux, and Metric},
Birkh\"auser, Boston, MA, 2003. \vskip \baselineskip

\noindent [13] I. V. Lindell, {\it Differential Forms in Electromagnetics}, IEEE Press, Piscataway, NJ, and
Wiley-Interscience, 2004. \vskip \baselineskip

\noindent [14] D. Freed, {\it Classical field theory and Supersimmetry}, IAS Park City Mathematics Series Vol. 11, IAS,
Princeton, 2001. \vskip \baselineskip \end